\documentclass[11pt,a4paper]{article}

\usepackage{graphicx}
\usepackage{amssymb}
\usepackage{textcomp}
\usepackage{amsmath}
\usepackage{geometry}                
\usepackage{graphicx}
\usepackage{amssymb}
\usepackage{epstopdf}
\usepackage{bm}
\usepackage{times}
\usepackage{epsfig}
\usepackage{color}
\usepackage{array,multirow}
\usepackage{jheppub}

\newcommand{\pmns}{{\mbox{\tiny PMNS}}}

\newcommand{\trix}[1]{\left(\begin{array}{#1}}
\newcommand{\notrix}{\end{array}\right)}
\newcommand{\comment}[1]{}

\def\beq{\begin{equation}}
\def\eeq{\end{equation}}
\def\bea{\begin{eqnarray}}
\def\eea{\end{eqnarray}}

\usepackage[T1]{fontenc}
\usepackage[utf8]{inputenc}
\usepackage{mathtools} 
\usepackage{amsfonts}
\usepackage{ amssymb }
\usepackage[toc,page,title]{appendix}
\usepackage{xcolor}
\usepackage[compat=1.1.0]{tikz-feynman}
\usepackage{graphicx}
\usepackage{subcaption}
\usepackage{float}
\usepackage{caption}
\usepackage{breqn}
\usepackage{amsmath}
\usepackage{etoolbox}
\makeatletter
\patchcmd{\maketitle}{\@fpheader}{}{}{}
\makeatother

\usepackage[compat=1.1.0]{tikz-feynman}
\numberwithin{equation}{section}
\title{\Large  {{\bf{$R$-parity Violating Decays of\\ 
Wino Chargino and Wino Neutralino\\
LSPs and NLSPs at the LHC}}}}
\author{{Sebastian Dumitru$^{1}$, Burt A.~Ovrut$^{1}$ and Austin Purves$^{2}$} \\[2mm
]
    {\it $^{1}$ Department of Physics and Astronomy, University of Pennsylvania} \\
   {\it Philadelphia, PA 19104--6396}\\
   {\it $^{2}$ Department of Physics, Manhattanville College}\\
   {\it Purchase, NY 10577} \\[4mm]
}
\date{\today}

{\footnotetext{\mbox{sdumitru@sas.upenn.edu, ~ovrut@elcapitan.hep.upenn.edu, ~austin.purves@mville.edu} }}
\abstract{
The $R$-parity violating decays of both Wino chargino and Wino neutralino LSPs are analyzed within the context of the $B-L$ MSSM ``heterotic standard model''. These LSPs correspond to statistically determined initial soft supersymmetry breaking parameters which, when evolved using the renormalization group equations, lead to an effective theory satisfying all phenomenological requirements; including the observed electroweak vector boson and Higgs masses. The explicit decay channels of these LSPs into standard model particles, the analytic and numerical decay rates and the associated branching ratios are presented. The decay lengths of these RPV interactions are discussed. It is shown that the vast majority of these decays are ``prompt'', although a small, but calculable, number correspond to ``displaced vertices'' of various lengths. It is demonstrated that for a Wino chargino LSP, the NLSP is the Wino neutralino with a mass only slightly higher than the LSP-- and vice-versa. As a consequence, we show that both the Wino chargino and Wino neutralino LSP/NLSP $R$-parity violating decays should be simultaneously observable at the CERN LHC. }

\begin{document}

\maketitle

\newpage

\newpage
%
\section{Introduction}

In a previous paper \cite{new} we presented, within the context of the $N=1$ supersymmetric $B-L$ MSSM,  the decay channels and the associated analytic decay rates for Wino/Higgsino chargino and arbitrary neutralino $R$-parity violating (RPV) decays to  standard model particles. These results were valid for any such charginos and neutralinos, regardless of their mass; that is, whether or not they are the lightest supersymmetric particles (LSPs). As was discussed in detail in that paper, the dimensionful soft supersymmetry breaking parameters were chosen statistically to lie in an interval that, although centered around a mass of several TeV, was wide enough to allow chargino and neutralino masses to be as low as $\sim 200~ {\rm GeV}$ and as heavy as $\sim 10 ~{\rm TeV}$. That is, the mass spectrum of these sparticles overlaps with the range potentially observable at the CERN LHC.

The $B-L$ MSSM is a minimal extension of the MSSM that arises as a vacuum state of heterotic M-theory \cite{Horava:1996ma,Lukas:1997fg,Lukas:1998ew,Lukas:1998yy,Lukas:1998tt}. It was shown using a series of papers \cite{Evans:1985vb,Donagi:2000fw,Donagi:2004ia,Braun:2005zv} that, when compactified to four-dimensions on a specific Calabi-Yau threefold \cite{Braun:2005nv}, this vacuum has exactly the particle spectrum of the MSSM-- that is, three families of quark and lepton chiral supermultiplets, a pair of Higgs-Higgs conjugate chiral superfields and three right-handed neutrino chiral supermultiplets, one per family. The associated gauge group is $SU(3)_{C} \times SU(2)_{L} \times U(1)_{Y}$ of the standard model augmented, however, by an additional gauged Abelian group factor, $U(1)_{B-L}$. Although derived from heterotic M-theory, this specific low energy effective field theory was unique and had not been introduced in any previous context. The $B-L$ MSSM was re-introduced, this time in the phenomenological literature, several years later in \cite{Barger:2008wn}. Importantly, it was first stated in \cite{Barger:2008wn} that the gauged $U(1)_{B-L}$ symmetry could, in principle, be spontaneously broken by at least one of the right-handed sneutrino scalars obtaining a non-vanishing VEV. They emphasized this point by presenting the associated sneutrino vacuum expectation value in terms of specific soft supersymmetry breaking parameters. Be that as it may, it is essential that it be proven that this VEV could dynamically occur via radiative breaking in the $B-L$ MSSM. This was first proven using a full RG analysis in \cite{Ambroso:2009jd}. This extra gauge factor contains $R$-parity and sufficiently suppresses both proton decay and lepton number violating decays, even after it is spontaneously broken \cite{Marshall:2014cwa,new}. As discussed in many of the previously referenced papers, as well as in \cite{new}, when supersymmetry is softly broken in an interval of order the TeV scale, and then run to lower mass scales using the renormalization group (RG), we find--for a large number of statistically scattered and uncorrelated initial conditions--that the $B-L$ MSSM is consistent with all present experimental bounds. Specifically, it was shown that 1) the gauged $U(1)_{B-L}$  symmetry is radiatively broken by the third family right-handed sneutrino acquiring a vacuum expectation value (VEV). This VEV is sufficiently large that the associated vector boson mass exceeds the present experimental bound. The process of $B-L$ breaking also yields a natural explanation for Majorana neutrino masses via a seesaw mechanism \cite{Barger:2010iv, Mohapatra:1986aw}. 2) Electroweak symmetry is radiatively broken by the neutral Higgs fields acquiring radiative VEVs. The associated $W^{\pm}$ and $Z^{0}$ vector bosons have precisely their experimentally measured values. 3) All supersymmetric sparticles exceed their present experimental lower bounds. 4) Finally, and remarkably, the Higgs boson mass satisfies the three sigma bound established at the LHC.

Furthermore, several important ``stringy'' theoretic aspects of the $B-L$ MSSM are potentially amenable to calculation. First, it has been demonstrated that, in principle, the potential energy functions for the geometric, vector bundle and five-brane moduli can be calculated and the vacuum state of these moduli stabilized \cite{Anderson:2011cza,Buchbinder:2002pr,Donagi:1999jp,Lima:2001jc}. Second, it has been shown \cite{Anderson:2009nt,Anderson:2009ge,Blesneag:2015pvz,Blesneag:2016yag} that the Yukawa couplings are, in principle, directly calculable from the harmonic representatives of the sheaf cohomology classes \cite{Braun:2006me}. Similarly, gauge couplings are potentially calculable from string unification threshold corrections \cite{Deen:2016vyh,Kaplunovsky:1992vs,Kaplunovsky:1995jw,Mayr:1993kn,Dienes:1995sq,Dienes:1996du,Nilles:1997vk,Ghilencea:2001qq,deAlwis:2012bm,Bailin:2014nna}. Finally, gauging the $N=1$ supersymmetry couples the $B-L$ MSSM directly to $N=1$ supergravitation. This allows both theories of inflation \cite{Deen:2016zfr,Cai:2018ljy,Ferrara:2015cwa,Linde:2016bcz} and ``bouncing universe'' cosmologies \cite{Koehn:2012ar,Battarra:2014tga} to naturally arise within this context. We note in passing that for the $B-L$ MSSM inflation theory to be consistent with the present cosmological data, its soft supersymmetry breaking scale must be raised to the order of $10^{13}$ GeV. Be that as it may, it was shown in \cite{Deen:2016zfr} that the low energy theory can remain completely consistent with all phenomenological bounds listed above.

 For all of these reasons, {\it the $B-L$ MSSM appears to be the simplest possible phenomenologically realistic theory of heterotic superstring/M-theory; being exactly the MSSM with right-handed neutrino chiral supermultiplets and spontaneously broken R-parity}. We would like to point out that, although the $B-L$ MSSM was originally derived from the ``top-down'' point of view of heterotic M-theory, it was also constructed from a low energy, ``bottom-up'' approach in \cite{FileviezPerez:2008sx,Barger:2008wn,FP:2009gr,Everett:2009vy,FileviezPerez:2012mj,Perez:2013kla}. For all of these reasons, it would seem to be 
to be a rich arena to study the phenomenological predictions of the $B-L$ MSSM at energies low enough to be observable by the ATLAS detector at the LHC at CERN. This requires taking the interval of soft supersymmetry breaking parameters to be in the range discussed in detail in our recent paper \cite{new}. We will do this, henceforth. However, the generic supersymmetric interactions of the $B-L$ MSSM are extremely complicated for arbitrary mass sparticles, with the RP conserving processes being much larger than, and, hence, potentially making  unobservable, the RPV decays calculated in \cite{new}.
However, there is one very clean and obvious window where experimental observation of supersymmetric interactions becomes vastly simplified. That window is for the so-called lightest supersymmetric particle--the LSP. By definition, in an $R$-parity conserving theory, the LSP cannot further decay, either to other sparticles or to standard model particles. However, in a theory in which $R$-parity is spontaneously broken, the LSP, while still unable to decay via RP conserving interactions, can now decay through RPV processes to standard model particles. In the $B-L$ MSSM, these decay channels, their decay rates and the associated branching fractions can be explicitly calculated. 
We propose, therefore, that the RPV decays of the LSPs of the $B-L$ MSSM be searched for experimentally, and the results compared to the theoretical predictions. Any positive result obtained in this regard could be a first indication of the existence of $N=1$ supersymmetry, as well as a potential confirmation of the $B-L$ MSSM theory.

This program has already been carried out for the lightest admixture stop, which was shown to be one of the LSPs of the $B-L$ MSSM. The branching ratios for the RPV decay of the lightest stop LSP to a bottom quark and a charged lepton, the dominant decay mode, along with the relationship of these decays to the neutrino mass hierarchy and the $\theta_{23}$ neutrino mixing angle, were presented in \cite{Marshall:2014cwa,Marshall:2014kea}. The admixture stop LSP was chosen for two reasons. First, it carries both electric and color charge and, therefore, is ``exotic''; in the sense that in RP conserving theories such an LSP would contribute to ``dark matter'' which must be gauge neutral. Second, it has a high production cross section at the LHC. Based on the results of these two papers, a search for stop LSP decays in the recent ATLAS LHC data was carried out in \cite{Aaboud:2017opj,ATLAS:2017hbw,Jackson:2015lmj,ATLAS:2015jla}. No direct detection was observed. However, the lower bounds on the stop LSP mass were significantly strengthened. Be that as it may, as was discussed in \cite{Ovrut:2014rba,Ovrut:2015uea,new} and will be described in the next section, the number of physically realistic initial conditions leading to a stop LSP are relatively small compared to other sparticles. Therefore, in a series of papers, we will pursue this program focussing, however, on other sparticles that occur more frequently as LSPs of the $B-L$ MSSM.

 Specifically, in this paper, we will explore the decay channels, the decay rates and calculate the branching ratios to standard model particles for Wino chargino and Wino neutralino LSPs, using the explicit results for generic chargino and neutralino sparticles presented in \cite{new}. The Wino charginos/neutralinos occur with much more frequency as LSPs of realistic initial conditions of the $B-L$ MSSM. As in the case of the stop LSP, we will discuss the relationship of their decays to standard model particles to the neutrino mass hierarchy and the $\theta_{23}$ neutrino mixing angle. Finally, we find that for a Wino chargino LSP, the NLSP is the Wino neutralino, and vice versa. Furthermore, the mass splitting between them is very small, on the order of several hundred MeV. It follows that a) the RPC decays of the Wino NLSP are highly suppressed relative to its RPV decays and b) that, in addition to the RPV decays of the Wino LSP,  the RPV decays of the Wino NLSP should be observable in the detector at the LHC as well. For clarity, in a series of Appendices, we give a summary of our notation, present the precise definitions of the Wino chargino and Wino neutralino and give the analytic expressions for both Wino chargino and Wino neutralino decay rates first derived in \cite{new}.

Finally, we want to make three important statements concerning the computations in, and the context of, this paper.  These are:

\begin{enumerate}

\item All calculations in this paper, as well as those in previous analyses of the $B-L$ MSSM such as \cite{new}, are carried out using the {\it one-loop 
corrected} $\beta$ and $\gamma$ renormalization group functions associated with the dimensionless and dimensionful parameters of the theory. However, we systematically {\it ignore all higher-loop corrections to the RGEs as well as any finite one-loop and higher-loop corrections} to the effective Lagrangian. For the purposes of this paper this is sufficient, since our goal is to present the allowed RPV decay channels of Wino chargino and Wino neutralino LSPs in the $B-L$ MSSM theory and to give their {\it leading order} decay rates, branching ratios and the relationship of these to the neutrino mass hierarchy. However, we are well aware that some of these processes can be substantially effected by higher-loop corrections, both in the RG running of the parameters  and in finite quantities, such as particle masses. For example, in the Higgs mass calculation two-loop RGEs and higher-loop finite corrections could indeed be very important. We conclude that the calculations presented here could, and depending on specific experimental searches being performed to verify them should, be carried out to higher precision than the results presented in this paper. This would put the $B-L$ MSSM computations on the same footing as the the more commonly studied MSSM. Indeed, the computational tools required to extend our work to finite one-loop and higher-loop RG and finite corrections already exist in the literature, such as in ISAJET \cite{Paige:2003mg}, FlexibleSUSY \cite{Athron:2014yba},
NMSPEC  \cite{Ellwanger:2006rn}, SUSPECT \cite{Djouadi:2002ze}, SARAH \cite{Staub:2008uz}, SPHENO \cite{Porod:2003um}, SUSEFLAV \cite{Chowdhury:2011zr} and the latest version of SOFTSUSY \cite {Allanach:2016rxd}. We will carry out these higher-loop RG and finite corrections to the $B-L$ MSSM in future publications.

\item In this paper, as well as our previous papers \cite{Ovrut:2014rba, Ovrut:2015uea, new}, the initial soft supersymmetry breaking parameters are selected statistically using a {\it ``log-uniform''} distribution over a mass range compatible with LHC energies. As discussed Section 2 of this paper, this is the {\it standard} distribution used in analyzing  such initial conditions. We are well aware, however, that one could choose other statistical distributions for the initial parameters--such as a uniform distribution. However, the justification for which distribution to use depends on the choice of the explicit mechanism for spontaneous supersymmetry breaking. In this paper, as well as the series of papers  \cite{Ovrut:2014rba, Ovrut:2015uea, new} on which it is based, the analysis is restricted to the low energy phenomenology of the observable sector only, and does not specify the mechanism of supersymmetry breaking. This could be due, for example, to various non-vanishing F-terms, D-terms or gaugino condensation in the hidden sector of the theory, and is far from unique. Therefore, in this paper and \cite{Ovrut:2014rba, Ovrut:2015uea, new}, we simply add to the effective Lagrangian the most general allowed soft supersymmetry breaking terms and choose the values of their parameters statistically. Since the log-uniform distribution is the {\it standard} distribution, as justified in Section 2, we will employ it uniquely. 
Furthermore, as we now discuss, choosing a log-uniform distribution is sufficient for the purposes of this paper.

To begin, we are simply seeking a set of ``viable'' initial parameters that, when scaled using the RG to lower energy, are completely consistent with all present phenomenological requirements. Using this log-uniform distribution, we show that there are indeed a very large number of such viable initial conditions. It is then demonstrated that, within this context, there is a subset of such parameters that lead to Wino chargino and Wino neutralino LSPs.
Were one to use a different initial distribution of parameters, one would find a potentially different set of viable points, some presumably already contained within the log-uniform distribution and, perhaps, some new viable points. However, any such new viable points will not greatly effect the calculations and conclusions of this paper. The first important example of this is the following. It is of some interest to ask, within the context of a log-uniform distribution, what the set of all allowed LSPs is and, furthermore, what percentage of  the viable initial points correspond to Wino charginos and Wino neutralinos. This information is presented in the histogram in Section 2. We find, within the log-uniform context, that the percentage of valid initial points with Wino chargino and Wino neutralino LSPs is relatively large. Now, it is indeed possible that choosing a different initial distribution would change the percentages for the individual LSPs in this histogram. 
That being said, the actual content of this paper is {\it independent} of whether or not these specific LSPs are statistically prominent. Rather, as was the case for the stop LSPs discussed in previous work \cite{Marshall:2014cwa,Marshall:2014kea}--which are statistically minimal in the histogram in Section 2--we focus on Wino chargino and Wino neutralino LSPs because their RPV decays are {\it readily observable at the ATLAS detector at the LHC}. 

Secondly, it is obvious that the explicit decay channels, the {\it analytic expressions} for the the decay rates and, hence, the LSP lifetimes, as well as the {\it analytic expressions} for the associated branching ratios--both summing and not summing over the families of final states--are completely {\it independent} of the choice of initial parameters and, hence, the choice of the initial distribution. That being said, the statistical plots for the branching ratios and decay lifetimes of Wino charginos and Wino neutralinos, presented in Sections 3 and 4 respectively, as well as the statistical decay rates of the RPC versus RPV processes for the NLSP presented in Section 5 are, like the histogram in Section 2, all explicitly calculated using a {\it log-uniform} initial distribution only. The choice of some different initial distribution of parameters, while reproducing much of these plots, can be expected to alter them somewhat. However, exactly as with the histogram in Section 2, these statistical plots are presented to give a concrete representation of what decay channels of the Wino charginos and Wino neutralinos should be observed at the ATLAS detector at the LHC, whether or not an individual such decay can be ``prompt'', has a ``displaced vertex'' or occurs outside the ATLAS detector, and what the relative probability is of observing a given decay channel as opposed to another channel. If one simply adds the new viable points of a different distribution to the log-uniform points, all of the LHC observational conclusions drawn from the log-uniform priors will remain, essentially, unchanged. However, were one to repeat the entire analysis using a completely different initial distribution--{\it not including the log-uniform priors}--then, although the explicit decay channels will remain the same, their decay lifetimes and relative branching ratios could be altered. However, there is no physical reason to expect the viable points of the log-uniform distribution to be excluded. Furthermore, the choice of a {\it ``non-standard''} initial distribution, not including the viable log-uniform priors, would require a physical and mathematical analysis of the exact mechanism of spontaneous supersymmetry  breaking which, as discussed above, is beyond the scope of the present paper. To conclude, the present paper uses the standard log-prior distribution of initial points. The observational LHC conclusions will only be minimally altered if the viable initial points of additional distributions are added.

\item Finally, there is a long literature discussing RPV decays within a vast variety of contexts. Reference \cite{Barbier:2004ez} reviews the theoretical aspects of RPV violation with both bilinear and trilinear RPV couplings added in the superpotential. Relevant to the content of our present paper, this review discussed both explicit and, more briefly, spontaneous RPV due to both left- and right-chiral sneutrinos developing VEVs. More recently, the subject was reviewed in 2015 \cite{Mohapatra:2015fua}. This discussed explicit RPV in the MSSM but, in particular, focused on spontaneous breaking of $R$-parity in theories where the standard model symmetry is extended by a gauged $U(1)_{B-L}$.  More recently, there was a comprehensive paper \cite{Dercks:2017lfq} 
investigating the phenomenology of the MSSM extended by a single trilinear RPV coupling
at the unification scale. It goes on to discuss the RPV decay of some of the LSPs; specifically the Bino neutralino and the stau sparticle, within the context of the RPV-CMSSM. The mechanism of generating Majorana neutrino masses through RPV bilinear terms is treated in \cite {Hirsch:2008ur, Kayla2013, Mitsou:2015eka, Mitsou:2015kpa}. This set of papers also studies the decay modes of some LSPs, with emphasis on the decay modes of the lightest neutralino. There are papers such as \cite{Bomark:2014rra, Csaki:2015uza, Dercks:2017lfq}, which study the RPV decay signatures of chargino, stop, gluinos and charged and neutral Higgsinos, using parameter scans in agreement with the existent experimental bounds. However, they work in different, more general theoretical contexts than our own.

The RPV decays of the Wino chargino and Wino neutralino LSPs presented in this paper share many of the concepts and techniques contained in these papers, such as RG evolution, the associated LSP calculations and their RPV decays, relationship to neutrino masses and so on. However, the purpose of our present paper is to discuss the RPV decays of Wino charginos and Wino neutralinos precisely within the context of the $B-L$ MSSM; a minimal and specific extension of the MSSM with spontaneously broken $R$-parity. Furthermore, the initial conditions of this theory are chosen so as to be {\it completely consistent with all phenomenological requirements}, a property not shared by much of the previous literature. Our analysis is performed so as to predict RPV LSP decays amenable to observation at the LHC and arising from a minimal, realistic, $N=1$ supersymmetric theory. The calculation of the leading order RPV decays of the Wino chargino and Wino neutralino LSPs in this specific context have not previously appeared in the literature.

\end{enumerate}


\section{The LSPs of the $B-L$ MSSM}

A review of the $B-L$ MSSM, including the structure of its $R$-parity violating interactions, was presented in \cite{new}. In that paper, we also discussed the relationship between the Majorana neutrino masses, for both the normal and inverted neutrino hierarchies, and the RPV parameters $\epsilon_{i}$ and ${\nu_{L}}_{i}$, $i=1,2,3$ using the most up-to-date neutrino data. The basic input parameters of our RG computer code were also presented. We refer the reader to \cite{new} for details. Here, we want to emphasize that the interval over which all 24 dimensionful soft supersymmetry breaking parameters are statistically scattered was chosen in \cite{new} to be
\begin{equation}
\big[~\frac{M}{f},Mf~\big] \quad {\rm where}~~~M=1.5~{\rm TeV}~, ~f=6.7 \ .
\label{burt1}
\end{equation}
This guarantees, as mentioned above, that all mass parameters in the theory lie approximately in the range 
\begin{equation}
\big[200~{\rm GeV},10~{\rm TeV}\big]\ .
\label{burt2}
\end{equation}
The values of $M$ and $f$ were chosen to maximize the number of points that are of phenomenological interest --- that is, compatible with current LHC bounds while also being potentially amenable to observation at the LHC. Since this mass interval is ultimately used our analysis, it is important that our results do not significantly depend on these values. Varying $M$ and $f$ changes the range that the massive soft supersymmetry breaking parameters can take and, hence, effects how spread out the low energy mass spectrum can be.  However, the primary focus of this paper is the branching ratios and decay lengths of Wino LSPs and NLSPs. We have tested our code with substantially different values of $M$ and $f$ and found that our primary results, that is, the branching ratios and decay lengths of Wino LSPs and NLSPs, indeed do not significantly depend on the choice of $M$ and $f$.

The soft supersymmetry breaking parameters are statistically scattered in the range \eqref{burt1} with a log-uniform distribution. This is a {\it standard choice} of prior distribution. For examples and discussion see \cite{Athron:2017fxj,Fichet:2012sn,Fundira:2017vip,Bomark:2014rra}. There are at least three reasons for choosing a log-uniform distribution. First, it has the intuitive property of scattering masses evenly around the value $M$. That is, 50\% of the scattered masses will be above $M$ and 50\% will be below. A uniform distribution does not have this property. See \cite{Ovrut:2015uea} for further discussion. Second, using a log-uniform distribution addresses ambiguities in how the soft supersymmetry breaking parameters are scattered. For example, should we scatter the mass or the mass squared? The log-uniform distribution addresses this because it is actually invariant with respect to such choices. This ensures that our results are independent of these choices. See \cite{Fichet:2012sn,Fundira:2017vip} for discussion of this. Third, the statistical inference literature identifies the log-uniform distribution as a more objective one to use because it is non-informative in a formal sense \cite{Kass}. In short, we use the log-uniform distribution because it a standard choice and there are multiple good reasons for this choice.

In addition to a discussion of the input parameters for the computer code, reference \cite{new} also presented all of the phenomenological requirements that must be fulfilled for the low energy $B-L$ MSSM vacuum to be physically realistic. Whether or not these requirements are satisfied depends entirely on the initial soft supersymmetry breaking parameters that are chosen. As discussed in detail in a number of previous papers, we choose these initial conditions statistically, randomly throwing all 24 soft breaking parameters over the interval \eqref{burt1}. A detailed analysis of this was presented in \cite{new}. Here, we simply present the relevant results. Out of 100 million initial statistical data points, we found that 65,576 satisfied all phenomenological requirements when scaled to low energy using the RGEs. These were called ``valid black points''. 

\begin{figure}[t]
   \centering

   \begin{subfigure}[b]{0.8\textwidth}
\includegraphics[width=1.\textwidth]{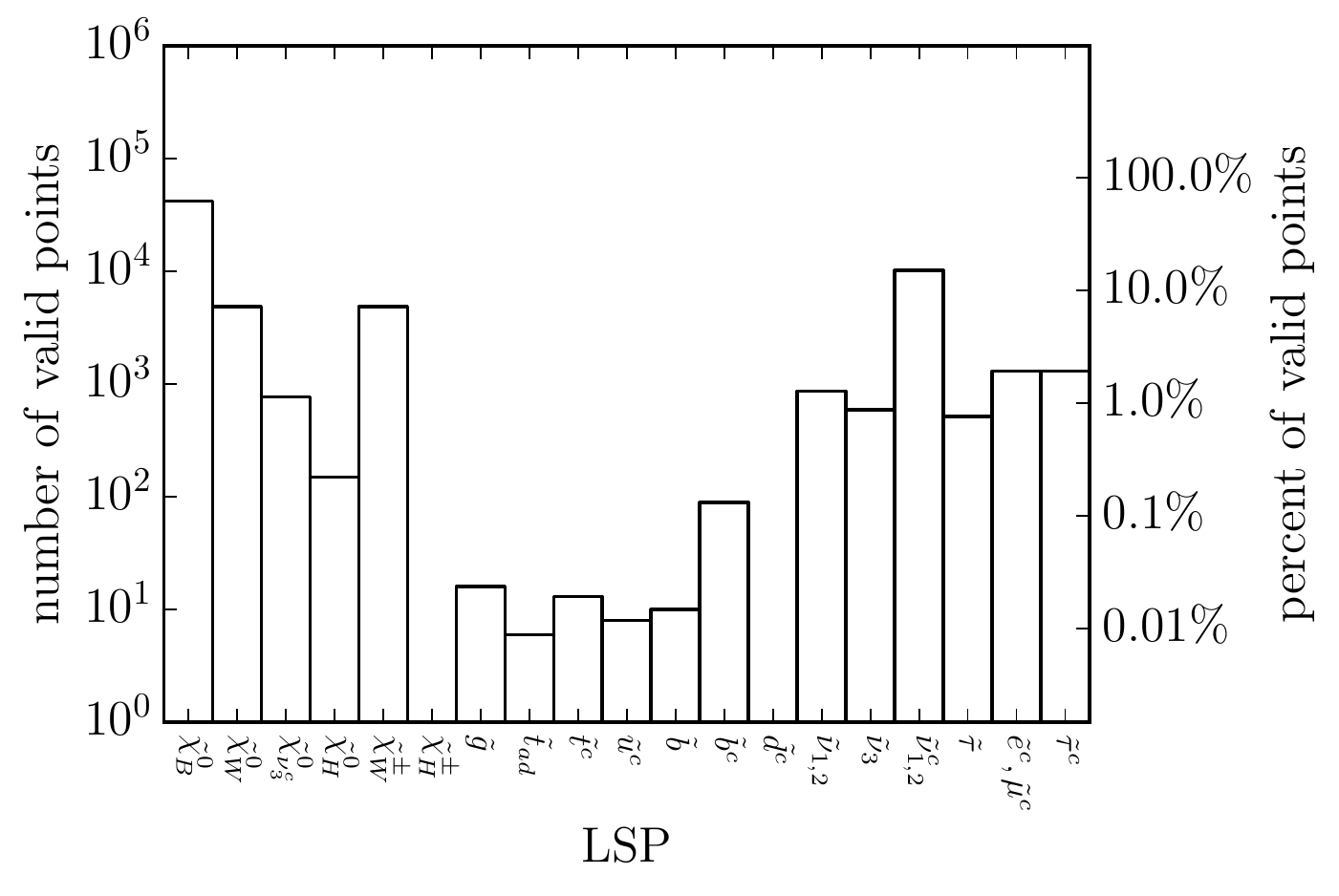}
\end{subfigure}
\caption{A histogram of the LSPs associated with a random scan of 100 million initial data points, showing the percentage of valid black points with a given LSP. Sparticles which did not appear as LSPs are omitted. The y-axis has a log scale. The notation and discussion of the sparticle symbols on the x-axis were presented in \cite{new}.}
\label{fig:LSP_Hist}
\end{figure}

Although each such black point satisfies all physical requirements, they can have different LSPs. A statistical study of the LSPs associated with the 65,576 black points was carried out in \cite{new}. The results are reproduced below in the histogram in Figure \ref{fig:LSP_Hist}. First, notice that, despite having a large production cross section, the percentage of physically realistic vacua having an admixture stop LSP is, as mentioned above, relatively small-- of the order of 0.01\%. Much more statistically favored are the ``Wino chargino'', ${\tilde{\chi}}^{\pm}_{W}$, and the associated ``Wino neutralino'', ${\tilde{\chi}}^{0}_{W}$. A generic chargino is an $R$-parity conserving mixture of a charged Wino, ${\tilde{W}}^{\pm}$, and a charged Higgsino, ${\tilde{H}}^{\pm}$, along with a small RPV  component of charged leptons, $e^{c}_{i}, e_{i}$,~ $i=1,2,3$.  A Wino chargino is a chargino which is predominantly the charged Wino. A generic neutralino is an $R$-parity conserving linear combination of six neutralino sparticles, including  the neutral Wino, ${\tilde{W}}^{0}$, along with a small RPV  component of neutrinos, $\nu_{i}$,~ $i=1,2,3$. We refer the reader to \cite{new} for details. A Wino neutralino is a neutralino that is predominantly the neutral Wino. Our statistical analysis shows that the Wino chargino, ${\tilde{\chi}}^{\pm}_{W}$, and the Wino neutralino, ${\tilde{\chi}}^{0}_{W}$, arise from 4,858 and 4,869 valid black points respectively; that is, each occurring approximately 7.40\% of the time. Therefore, even though each has a lower production cross section than the stop LSP, they clearly would play a significant role in any experimental search for RPV decays in the $B-L$ MSSM. It is also of interest, and important, to note that the percentage of ``Higgsino chargino'' LSPs, ${\tilde{\chi}}_{H}^{\pm}$, that is, a chargino which is predominantly a charged Higgsino, is approximately zero. In fact, out of the 65,576 valid black points, only 1 was found to have a Higgsino chargino LSP. The reason for this paucity of charged Higgsinos was explained in \cite{new}. Hence, in this and future publications we will ignore the Higgsino chargino LSP. 
It is clear from the histogram in Figure \ref{fig:LSP_Hist} that, in addition to the Wino chargino and Wino neutralino, there are other possible LSPs of the $B-L$ MSSM that are potentially of interest experimentally. For example, the RPV decays of ${\tilde{\chi}}_{B}^{0}$ and ${\tilde{\chi}}_{H}^{0}$, with 42,039 and 105 valid black points respectively, are possibly observable at the LHC. However, in this paper, we will confine our analysis to Wino charginos and Wino neutralinos only, returning to other LSP RPV decays in future publications.

It will be helpful in our analysis to be more explicit about the properties of Wino chargino and Wino neutralino mass eigenstates. We refer the reader to Appendices B and C for a detailed definition of these states respectively. Here, we simply point out that the RPV terms in the definition of both of these sparticles are always very small compared to the $R$-parity conserving terms. It follows that, although essential in the discussion of RPV decays to standard model particles in the following sections, these RPV terms give negligible contributions to the mass eigenvalues. Hence, in this section, where we are discussing their LSP properties, we will consider only the $R$-parity conserving components of the Wino chargino and Wino neutralino eigenstates.

First consider the Wino chargino. As discussed in Appendix B, after diagonalizing the chargino mass mixing matrix, one obtains two chargino mass eigenstates, $\tilde \chi_1^\pm$ and $\tilde \chi_2^\pm$, labeled such that $\tilde \chi_1^\pm$ is lighter than $\tilde \chi_2^\pm$.
If the dimensionful parameters $M_{2}$ and $\mu$ in the $B-L$ MSSM satisfy $|M_{2}|<|\mu|$, then (ignoring the $R$-parity violating components) the lighter state $\tilde \chi^\pm_1$ is given by 
\begin{equation}
\tilde \chi^\pm_1=\cos \phi_\pm \tilde W^\pm+\sin \phi_\pm \tilde H^\pm \ ,
\label{car1}
\end{equation}
%
where $W^\pm$ and $\tilde H^\pm$ are the pure charged Wino and charged Higgsino respectively. The angles $\phi_{\pm}$ are defined in equations \eqref{bernard1} and \eqref{bernard2}. As discussed in Appendix B, the 4,858 viable initial points leading to an observable chargino LSP will naturally tend to require $|M_{2}|<|\mu|$ and, hence, satisfy \eqref{car1}.
Generically, one finds $|M_{2}|$ to be of order of several hundred GeV to ensure that the associated LSP is observable at the LHC, whereas $\mu$ is much larger, of order a few TeV, to solve the ``little hierarchy problem''.
Importantly, these mass scales, along with one other important input, allow one to estimate the sizes of the $\phi_{\pm}$ angles. As given in Appendix B, these angles are defined by

%
%
%
\begin{equation}
\tan 2\phi_-=2\sqrt{2}M_{W^\pm}\frac{\mu \cos \beta +M_2 \sin \beta}{\mu^2-M_2^2-2M_{W^\pm}^2
\cos 2\beta}
\label{bernard1}
\end{equation}
\begin{equation}
\tan 2\phi_+=2\sqrt{2}M_{W^\pm}\frac{\mu \sin \beta +M_2 \cos \beta}{\mu^2-M_2^2+2M_{W^\pm}^2
\cos 2\beta} \ .
\label{bernard2}
\end{equation}

\begin{figure}[t]
   \centering
   \begin{subfigure}[b]{0.5\textwidth}
\includegraphics[width=1.0\textwidth]{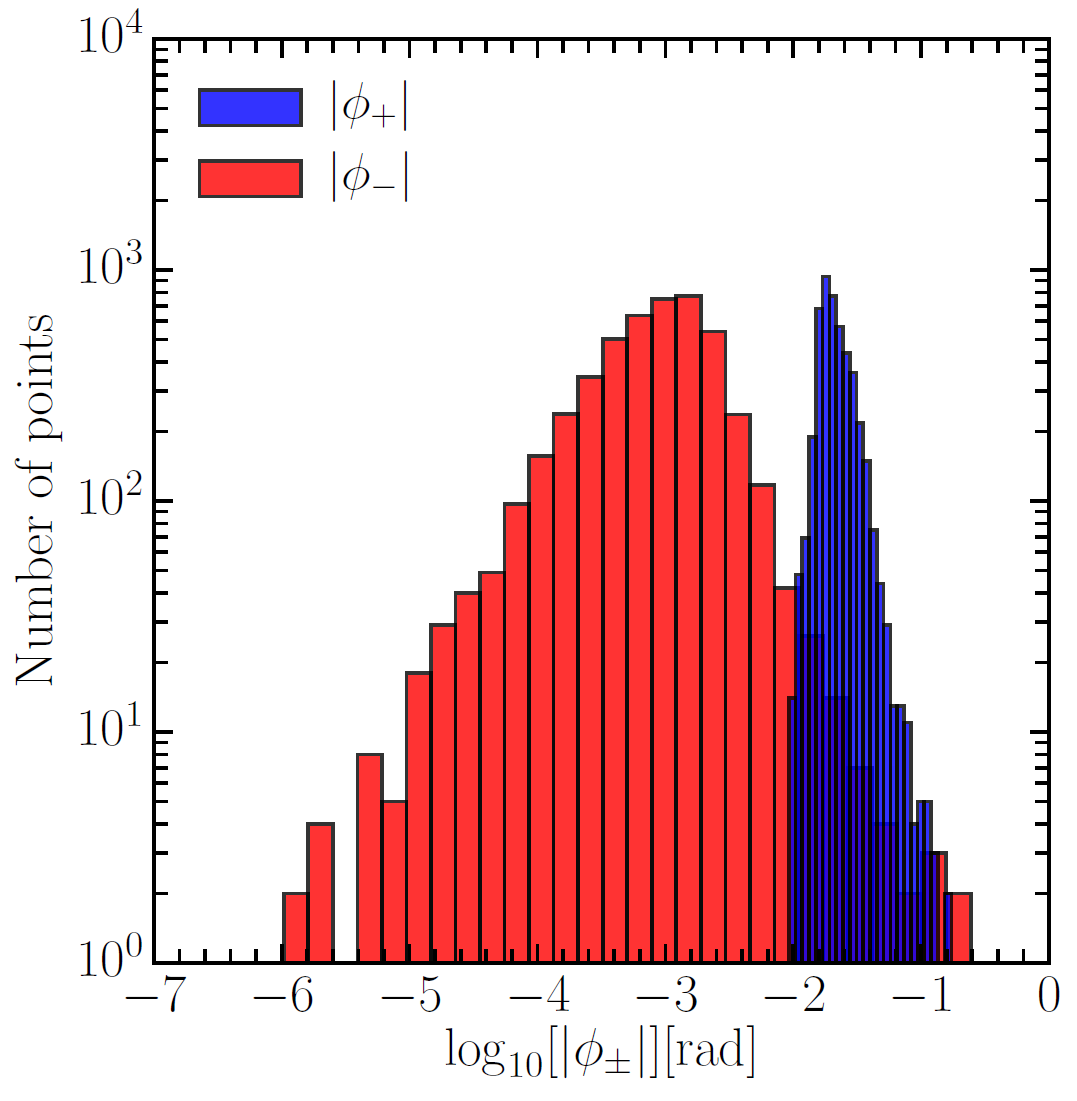}
\end{subfigure}
\caption{ The mixing angles $|\phi_{+}|$ and $|\phi_{-}|$ plotted against the 4,858 valid black points leading to a Wino chargino LSP's. It is clear that both angles are extremely small and, hence $\cos \phi_{\pm} \approx 1$ and $\sin \phi_{\pm} \approx 0$ for any such LSP.}
\label{fig:phi_scatter}
\end{figure}

\begin{figure}[t]
   \centering
   \begin{subfigure}[b]{0.49\textwidth}
\includegraphics[width=1.0\textwidth]{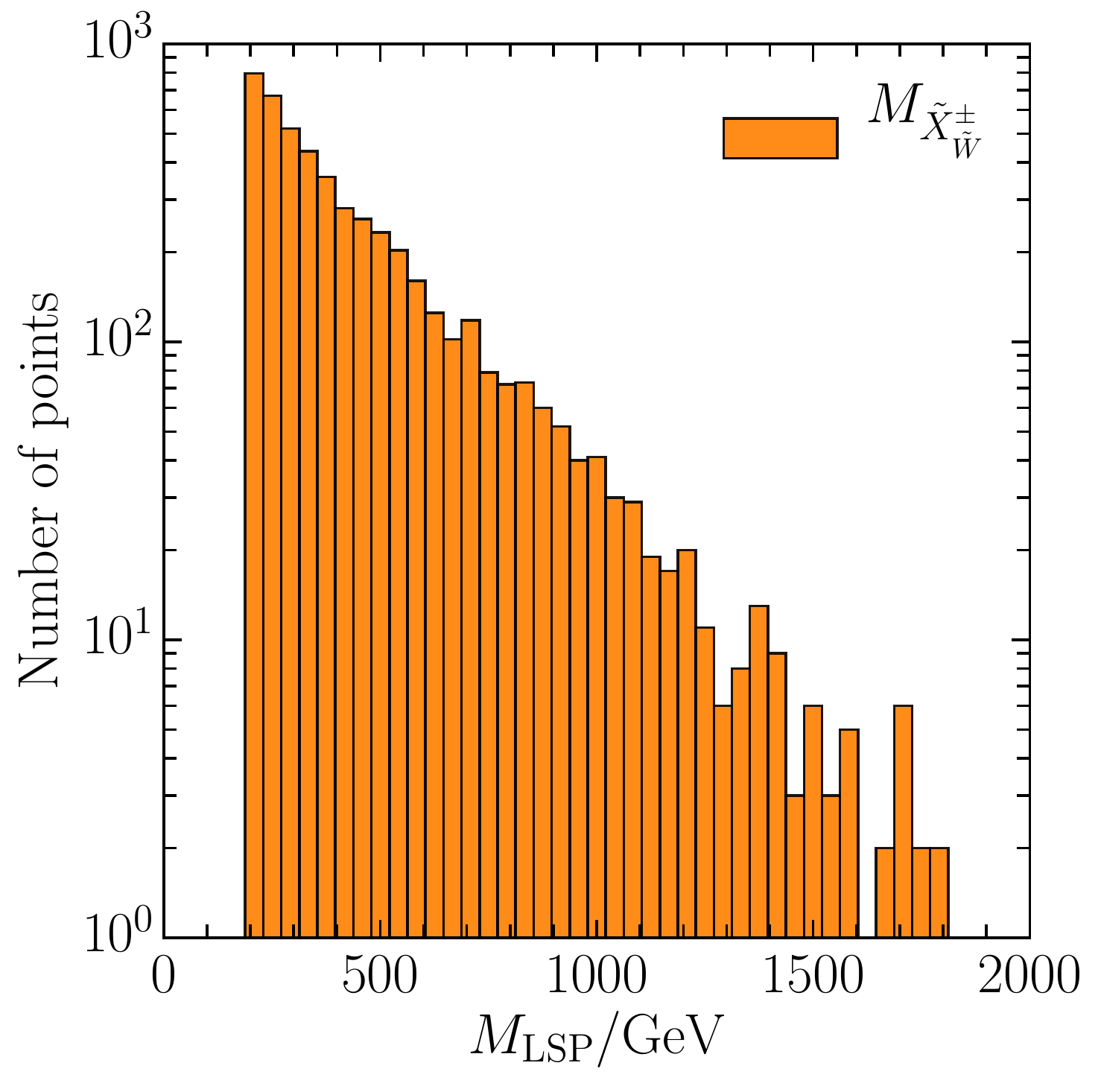}
\label{fig:mass_hist2}
\end{subfigure}
   \begin{subfigure}[b]{0.49\textwidth}
\includegraphics[width=1.0\textwidth]{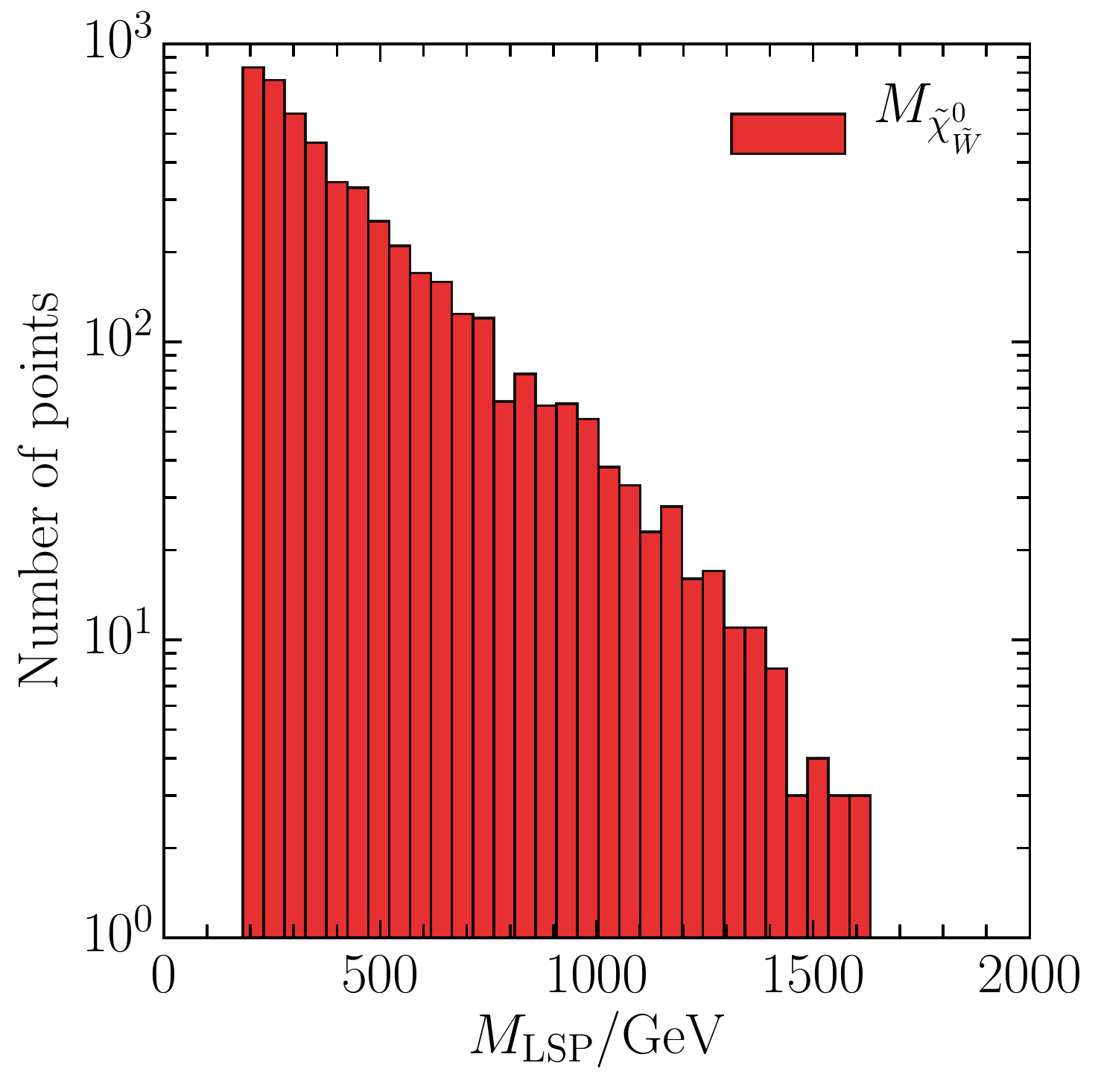}
\label{fig:mass_hist1}
\end{subfigure}
\caption{ a) Mass distribution of the Wino chargino LSP's for the 4,858 valid black points. The masses range from 200 GeV to 1820 GeV, peaking towards the low mass end. b) Mass distribution of the Wino neutralino LSP's for the 4,869 valid black points. The masses range from 200 GeV to 1734 GeV, peaking towards the low mass end.}
\label{fig:mass_hist}
\end{figure}

\noindent Clearly, the final required input is an estimate of the size of ${\rm tan}~ \beta$. 
Although we sample $\tan \beta$ with a uniform prior between 1 and 65, we find that the 4,858 valid black points, subject to all low energy phenomenological constraints, tend to prefer larger values of $\tan \beta$ over the smaller ones. That is, for most of the black points with chargino LSPs, $\sin \beta \approx 1$ and $\cos \beta \ll 1$.
With these insights, we expect 
\begin{equation}
\phi_+ \approx \frac{M_{W^\pm}}{\mu} \sim \mathcal{O} \left( 10^{-2}, 10^{-1} \right)
\label{late2}
\end{equation}
 and 
 \begin{equation}
 \phi_-\approx \frac{M_{W^\pm}}{\mu}\left( \cos \beta+\frac{M_2}{\mu} \right) \sim \mathcal{O} \left(10^{-4}, 10^{-2}  \right),
 \label{late3}
 \end{equation}
where $M_{W^\pm}=80.379$ GeV is the measured mass of the $W^{\pm}$ weak gauge bosons. 
These angles $|\phi_\pm|$ can be evaluated numerically for each of the 4,858 associated black points. The results are shown as a histogram in Figure \ref{fig:phi_scatter}. It is clear that both angles are extremely small for any such black points,
%
%
which is in agreement with our estimates in \eqref{late2} and \eqref{late3}.
It then follows from \eqref{car1} that
\begin{equation}
\tilde \chi_{1}^{\pm} \simeq \tilde W^\pm 
\label{book1}
\end{equation}
Therefore, we will denote
\begin{equation}
\tilde{\chi}_{1}^{\pm} \equiv \tilde \chi^\pm_W 
\label{late1}
\end{equation}
and refer to $\tilde \chi^\pm_W$ as the Wino chargino.
The mass of a Wino chargino was discussed in \cite{new}. An analytic expression for the $R$-parity preserving part of the mass eigenvalue was presented, as was the method for numerically computing the RPV extension. Using these results, the Wino chargino mass can be evaluated numerically for each of the 4.858 
 associated black points. The results were presented in \cite{new} and, for clarity, are reproduced here in Figure \ref{fig:mass_hist}.
The reason that the Wino chargino mass distribution peaks toward the low mass values was discussed in detail in \cite{new}, and we refer the reader to that paper for details.

Let us now consider the Wino neutralino. As discussed in \cite{new}, ignoring the very small RPV corrections, there are 6 neutralino mass eigenstates, each a complicated linear combination of the neutral gauge eigenstates. Here, we will only consider one of them, the Wino neutralino LSP , ${\tilde {\chi}}^{0}_W$, and the 4,869 valid black points associated with it. As discussed in \cite{new}, a numerical calculation allows us to compute, for each valid black point, the coefficients of the linear combination of neutral gauge eigenstates comprising the Wino neutralino. 
Here, we will simply state the result that the coefficient of the neutral Wino, $W^{0}$, component is largely predominant, whereas all other coefficients are very small. See Appendix \ref{appendix:Neutralinos} for a detailed discussion.
Hence, to a high degree of approximation,
\begin{equation}
{\tilde {\chi}}^{0}_W \simeq W^{0} \ .
\label{book2}
\end{equation}

That is, the Wino neutralino LSP mass eigenstate is almost the neutral Wino. The methods for numerically evaluating both the $R$-parity conserving  and the RPV contributions to the Wino neutralino mass were discussed in \cite{new}.  Using these methods, the Wino neutralino mass can be evaluated numerically for each of the 4.869 associated valid black points.  The results were presented in \cite{new} and, for clarity, are reproduced here in Figure \ref{fig:mass_hist}. Again, the reason that the Wino neutralino mass distribution peaks toward the low mass values was discussed in detail in \cite{new}, and we refer the reader to that paper for details.

\section{Wino Chargino LSP Decays}

The minimal B-L extension of the MSSM, that is, the $B-L$ MSSM, introduces RPV vertices that allow LSPs to decay directly into SM particles. In this section, we  will investigate the RPV decays of a Wino chargino LSP. As discussed in \cite{new}, a generic chargino mass eigenstate is a superposition of a charged Wino, a charged Higgsino and charged lepton gauge eigenstates. The $R$-parity conserving component of the Wino chargino is given by the linear combination of a charged Wino and charged Higgsino presented in \eqref{car1}, where the charged Wino component dominates. The smaller RPV contribution to the Wino chargino was presented in subsection 5.1 of \cite{new}. For the case at hand, where $|M_{2}|<|\mu|$, this was found to be
\begin{equation}
\mathcal{V}_{1\>2+i} e^{c}_{i} \quad {\rm where} \quad \mathcal{V}_{1\>2+i}=-\cos \phi_+ \frac{g_2 \tan \beta m_{e_i}}{\sqrt{2}M_2\mu}v_{L_i}+\sin \phi_+\frac{m_{e_i}}{\mu v_d}v_{L_i}
\label{wait1}
\end{equation}
for ${\tilde{\chi}}^{+}_{W}$ and 
\begin{equation}
\mathcal{U}_{1\>2+i} e_{i}  \quad {\rm where} \quad \mathcal{U}_{1\>2+i}=-\cos \phi_- \frac{g_2 v_d}{\sqrt{2}M_2\mu}\epsilon_i^*+\sin \phi_-\frac{\epsilon_i^*}{\mu}
\label{wait2}
\end{equation}
for ${\tilde{\chi}}^{-}_{W}$. We sum \eqref{wait1}
 and \eqref{wait2} over $ i=1,2,3$.

\begin{figure}[t]
 \begin{minipage}{1.0\textwidth}
     \centering
   \begin{subfigure}[b]{0.24\linewidth}
   \centering
       \includegraphics[width=1.0\textwidth]{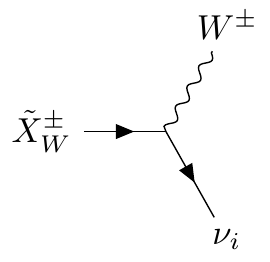} 
\caption*{(a) ~${\tilde X}^\pm_W\rightarrow W^\pm \nu_{i}$}
       \label{fig:table2}
   \end{subfigure} 
   \hfill
   \centering
   \begin{subfigure}[b]{0.24\linewidth}
   \centering
         \includegraphics[width=1.0\textwidth]{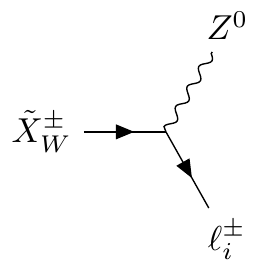} 
\caption*{(b) ~${\tilde X}^\pm_W\rightarrow Z^0 \ell_{i}^\pm$}
       \label{fig:table2}
\end{subfigure}\hfill
   \centering
   \begin{subfigure}[b]{0.24\textwidth}
   \centering
         \includegraphics[width=1.0\textwidth]{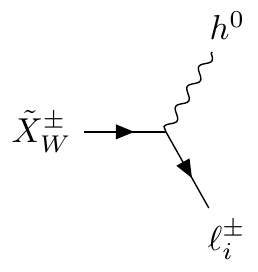} 
\caption*{(d) ~${\tilde X}^\pm_W \rightarrow h^0 \ell_{i}^\pm$}
\end{subfigure}
\end{minipage}
\caption{RPV decays of a general massive chargino state $X_W^\pm$. There are three possible channels, each with $i=1,2,3$, that allow for Wino chargino LSP decays. The decay rates into each individual channel were calculated for generic charginos in our previous paper, and are reproduced here in Appendix \ref{appendix:A}}\label{Figure4}
\end{figure}

One of the goals of of this paper is to predict the possible signals produced by the RPV decays  of Wino chargino LSPs, were such particles to exist and be light enough to be detected at the LHC. In our previous paper \cite{new}, we analyzed RPV decay channels using 4-component spinor notation for the mass eigenstates. For example, the Dirac spinor associated with the  Weyl fermions ${\tilde{\chi}}^{\pm}_{W}$  is defined to be 

\begin{equation}
\tilde X^\pm_W=
\left(
\begin{matrix}
\tilde \chi_W^\pm\\
\tilde \chi_W^{\mp \dag}
\end{matrix}
\right) \ .
\end{equation}
We found that ${\tilde{ X}}^\pm_W$ can decay into standard model particles via three RPV channels. These are shown in Figure \ref{Figure4}.
\noindent  Each of these three channels have different properties concerning their potential experimental detection.
The  ${\tilde X}^\pm_W\rightarrow Z^0 \ell_i^\pm$
process is the Wino chargino decays most easily observed at the LHC. On the other hand, the left handed neutrinos produced during ${\tilde X}^\pm_W\rightarrow W^\pm \nu_i$ can only be detected as missing energy, while the Higgs boson $h^0$ resulting from the decay ${\tilde X}^\pm_W\rightarrow h^0 \ell_i^\pm$ couples to both quarks and leptons, leading to traces in the detector that are harder to interpret. 
In the following, we will explicitly compute the decay rates and branching ratios for all three channels. Sufficiently large probabilities for the process  ${\tilde X}^\pm_W\rightarrow Z^0 \ell_i^\pm$  may indicate favorable prospects for detecting Wino chargino LSPs at the LHC, whereas finding that this channel is subdominant would then make these experimental efforts more difficult. 

\subsection{Branching ratios of the decay channels}

The decay rates into each individual channel were calculated analytically for generic charginos in our previous paper and are reproduced in Appendix \ref{appendix:A}. {\it For a fixed lepton family $i$, these decay rates depend explicitly on the choice of the neutrino hierarchy and the value of $\theta_{23}$}\footnote{As shown in \cite{new}, each of the measured values of $\theta_{23}$ in both the normal and inverted hierarchies have small uncertainties around a central value. These uncertainties are incorporated into our computer code in all calculations. However, for simplicity of notation, when we refer to the value of $\theta_{23}$ in the text of this paper, we will suppress these error intervals and indicate the central values only.}. We will discuss this in detail at the end of this section. However, for the present, we will confine ourselves to a calculation of the overall branching ratio for a given type of decay process, which explicitly involves a sum $\sum^{3}_{i=1}$ over the lepton families. The relative prevalence of each channel type is determined by its associated branching ratio. {\it A statistical analysis shows that, for any given decay channel, the sum over the three lepton families makes the branching ratio approximately independent of both the choice of the neutrino hierarchy and the value of $\theta_{23}$}. We will now evaluate these branching ratios for each of the 4,858 valid black points associated with a Wino chargino LSP, separating the data into statistically relevant bins of both $\tan \beta$ and the Wino chargino mass. To begin with, these calculations will be carried out assuming a normal hierarchy with $\theta_{23}=0.597$. Later on in this section, we will  discuss the small statistical differences that would occur had we chosen one of the other possible neutrino data sets. To make this process transparent, we now present our explicit calculational procedure.

For specificity, let us first discuss ${\tilde X}^\pm_W\rightarrow Z^0 \ell_i^\pm$ for any $i=1,2,3$. For this decay channel, the branching ratio is defined by

\begin{equation}\label{eq:Branching1}
\text{Br}_{{\tilde X}^\pm_W\rightarrow Z^0 \ell^\pm}=\frac{\sum_{i=1}^{3} \Gamma_{{\tilde X}^\pm_W\rightarrow Z^0 \ell_i^\pm}}{\sum_{i=1}^{3} \Big( \Gamma_{{\tilde X}^\pm_W\rightarrow W^\pm \nu_{i}}+\Gamma_{{\tilde X}^\pm_W\rightarrow Z^0 \ell_i^\pm} +\Gamma_{{\tilde X}^\pm_W\rightarrow h^0 \ell_i^\pm}\Big)} \ .
\end{equation}

\noindent We now proceed to evaluate \eqref{eq:Branching1} for each of the valid black points associated with a Wino chargino LSP. Since there will be 4,858 different values of $\text{Br}_{{\tilde X}^\pm_W\rightarrow Z^0 \ell^\pm}$, we find it convenient to divide up this data into separate bins. Specifically, we will do the following. First, recall from Figure 3 that the physical mass of a Wino chargino is much more likely to be small, on the order of 200 GeV, and approximately $10^{-2}$ times less likely to be on the order of 1 TeV. This leads us to divide the Wino chargino LSP mass range into three bins given by
 \begin{equation}
 M_{{\tilde X}_W^\pm} \in [200, 300],\>[300,600],\> [600,1820]~ \text{GeV} \ .
 \label{bin1}
 \end{equation}
\begin{figure}[t]
   \centering
\begin{subfigure}[b]{1.0\textwidth}
\includegraphics[width=1.0\textwidth]{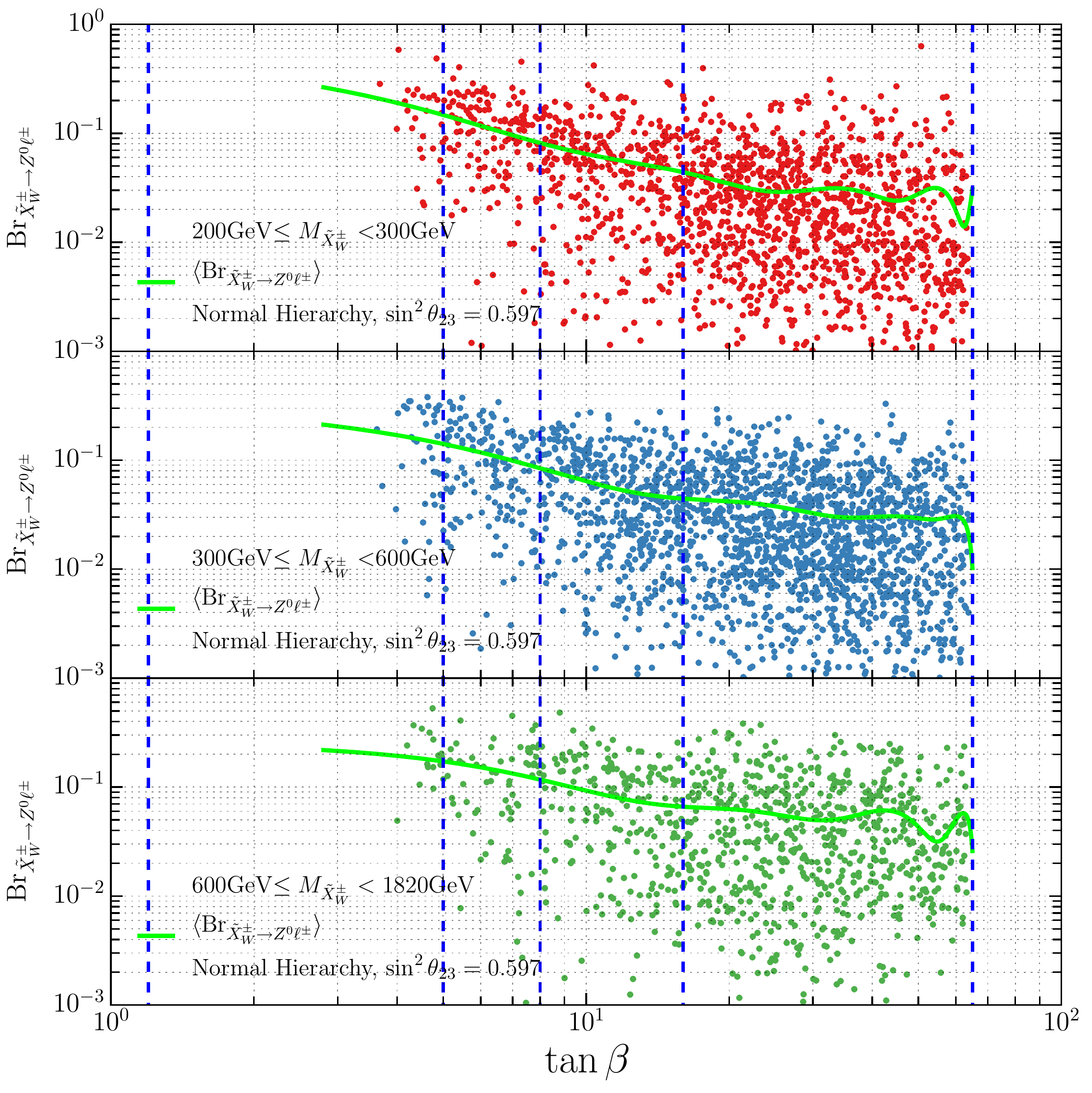}
\end{subfigure}
\caption{A scatter plot of all 4,858 branching ratios, $\text{Br}_{{\tilde X}^\pm_W\rightarrow Z^0 \ell^\pm}$, associated with a Wino chargino LSP versus $\tan \beta$. The plot is broken up into the three  $M_{{\tilde X}_W^\pm}$ mass bins given in \eqref{bin1}. In each plot, the values of the branching fractions are highly scattered around the a green curve which represents the ``best fit'' to the data. The vertical blue lines mark the boundaries of the four regions where the behavior of the best fit lines are approximately identical.}\label{fig:ScatterGammaRatios}

\end{figure}
The range of each bin is chosen so that each contains approximately a third of the 4,858 valid black points. Second, as we will see below, the value of $\tan \beta$ plays a significant role in the relative sizes of the branching ratios of the three decay channels. With this in mind, we plot the values of $\text{Br}_{{\tilde X}^\pm_W\rightarrow Z^0 \ell^\pm}$ against $\tan \beta$ for each of the three mass bins in \eqref{bin1}. In each case, we present the ``best fit'' to the data as a green curve. We further partition each of these plots into bins-- represented by the vertical, dashed blue lines --where the best fit curves in each plot behave similarly. The results are presented in Figure \ref{fig:ScatterGammaRatios}. \noindent We see from these plots that the range of $\tan \beta$ is naturally broken into four regions approximately given by
\begin{equation}
 \tan \beta \in [1.2,5],\> [5,8],\>[8,16], \>[16,65] \ .
 \end{equation}

\begin{figure}[H]
   \begin{subfigure}[b]{1.0\textwidth}
\includegraphics[width=1.0\textwidth]{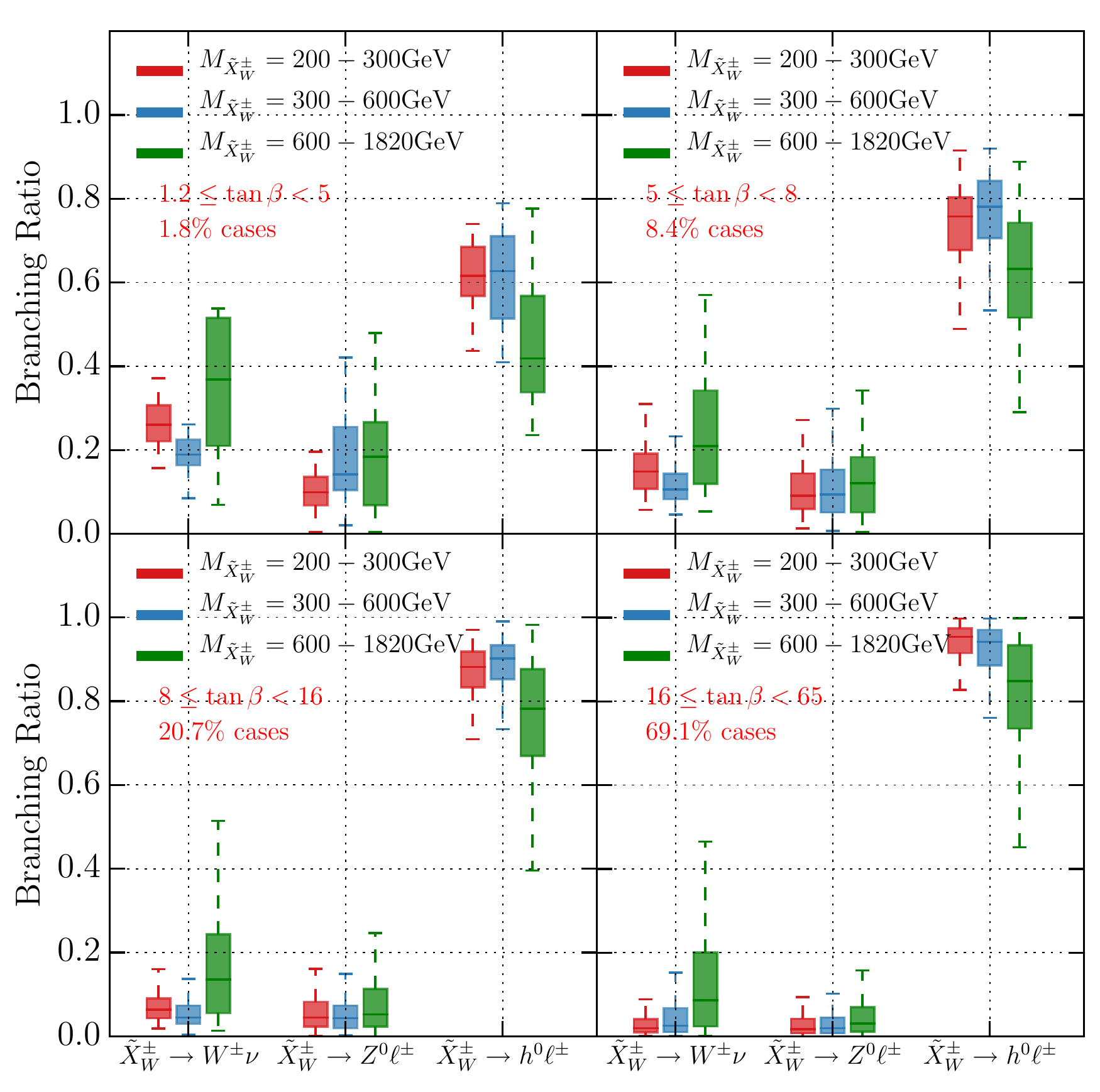}
\end{subfigure}\\
    \caption{Branching ratios for the four possible decay channels of the Wino chargino LSP, presented for the three $M_{{\tilde X}_W^\pm}$ mass bins and four $\tan \beta$ regions. The colored horizontal line inside each box indicate the median value of the branching fraction in that bin, the colored box indicates the interquartile range in that bin, while the dashed error bars show the range between the maximum and the minimum values of the branching ratio for that bin. The case percentage indicate what percentage of the valid initial points have $\tan \beta$ values within the range indicated. For each channel, we sum over all three families of possible leptons.
Note that ${\tilde X}_W^\pm\rightarrow h^0 \ell^\pm$ is strongly favored-- except perhaps in the $1.2 < \tan\beta < 5$ bin. The calculations were performed assuming a normal neutrino hierarchy, with $\theta_{23}=0.597$.}
\label{BarPlotChargino.pdf}
\end{figure}

Having broken up the ranges of $M_{{\tilde X}_W^\pm}$ and $\tan \beta$ into 3 and 4 bins respectively, we now calculate the median, the interquartile range and the maximum and the minimum values\footnote{To make these terms explicit-- a) the ``median'' is the value of a quantity for which 50\% of that quantity have larger values and 50\% are smaller and b) the ``interquartile'' range is the interval of that quantity which contains 25\% of all values that lie above the median and 25\% that lie below it. 3) The meaning of the ``maximum'' and ``minimum'' values is self-evident} of the branching ratio $\text{Br}_{{\tilde X}^\pm_W\rightarrow Z^0 \ell^\pm} $ for the decay channel ${\tilde X}^\pm_W\rightarrow Z^0 \ell_i^\pm$ in each of the $3\times 4$ data bins. Using an identical procedure, one can compute the same quantities for the remaining two branching ratios ${ \rm Br}_{{\tilde X}^\pm_W\rightarrow W^\pm \nu} $ and $ {\rm Br}_{{\tilde X}^\pm_W\rightarrow h^0 \ell^\pm}$ as well. The results are displayed in Figure \ref{BarPlotChargino.pdf}. 

For example, for the valid black points with Wino chargino LSP mass between $200$ and $300$ GeV and with $\tan \beta$ between 8 and 16, Figure \ref{BarPlotChargino.pdf} can be interpreted in the following way: 
\begin{itemize}
\item{The branching ratios have median values of $0.064$ for the $\tilde X_W^\pm \rightarrow W^\pm \nu $ channel, 
$0.0445$ for the $\tilde X_W^\pm \rightarrow Z^0 \ell^\pm$ channel, and $0.882$  for the $\tilde X_W^\pm \rightarrow h^0 \ell^\pm $ channel. We therefore expect $h^{0}$ Higgs boson production via chargino RPV decays to dominate for these ranges of mass and $\tan \beta$.}

\item{After solving for the RPV couplings and the decay rates, we generally obtain different branching ratio values for different viable initial conditions in our simulation. The data is scattered around the ``best fit'' values, as shown in Figure \ref{fig:ScatterGammaRatios} . However, the branching ratios take values only within certain ranges, allowing for theoretical predictions for the decay patterns of the Wino Chargino LSPs.  The dashed error bars in Figure \ref{BarPlotChargino.pdf} indicate the full range of values that the branching ratios take. For example, in our chosen bin, the branching ratios for the $\tilde X_W^\pm \rightarrow Z^0 \ell^\pm $ channel  are not higher than approximately $0.18$ while they can be very close to $0$.  At the same time, the branching ratios for the $\tilde X_W^\pm \rightarrow h^0 \ell^\pm $ channel  are approximately between  $ 0.72$ and $0.95$. }
\item{The boxes show the interquartile ranges, within which $50\%$ of the points lie around the median value. In our chosen bin we learn, for example, that while the branching ratios for the $\tilde X_W^\pm \rightarrow Z^0 \ell^\pm $ channel can take any value between approximately $0.18$ and $0$, they tend to accumulate in the more restricted interval $0.03-0.08$.}

\end{itemize}

Generically, for all three mass ranges and all four $\tan\beta$ bins in Figure \ref{BarPlotChargino.pdf}, one can conclude the following. It is clear that the ${\tilde X}^\pm_W\rightarrow h^0 \ell^\pm$ channel is the most abundant and becomes increasingly so for higher values of $\tan \beta$.  
. For $\tan \beta<5$, the medians of the branching fractions for the most experimentally visible channel,
${\tilde X}^\pm_W\rightarrow Z^0 \ell^\pm$, lie between 0.20-0.65, depending on the mass bin. However, there are very few such cases in our simulation-only $1.8\%$. Much more likely is a scenario in which $\tan \beta$ is large. We find that $69.1\%$ of the total number of points have $\tan \beta>16$. For this parameter region, however, the branching fraction of ${\tilde X}^\pm_W\rightarrow Z^0 \ell^\pm$ drops between $0.05-0.18$, and the prospects of detecting it become slimmer.

The results in Figure \ref{BarPlotChargino.pdf} were calculated using numerical inputs into the complicated expressions for the decay rates given in Appendix \ref{appendix:A}. Hence, the origin of the physical trends displayed in that Figure is obscure.
However, the formulas for the decay rates can, under certain assumptions, be simplified-- allowing for a physical interpretation for the observed relationships between the three decay channels. To do this, we note the following:

\begin{itemize}

\item{ As shown in Figure 2 above, the values for the angles $|\phi_{\pm}|$ are very small for each of the 4,858 black points associated with the Wino chargino; with $|\phi_{-}|$ being generically smaller than $|\phi_{+}|$. Hence, to a high degree of approximation, one can set $\cos \phi_{-}=1$ and $\sin \phi_{- }= 0$. However, due to the fact that the values for $|\phi_{+}|$, although very small, tend to be somewhat larger than $|\phi_{-}|$, we can only take $\cos \phi_{+}\approx 1$ and $\sin \phi_{+}\approx 0$.}

\item{ The lepton masses, $m_{\ell_i}$, are insignificant compared to the other masses in the expressions for the decay rates and, hence, the terms containing them can be neglected. Note that all occurences of the angle $\phi_{+}$ are contained in these terms. This facilitates our simplification even further, since the slightly larger values of $|\phi_{+}|$ no longer enter the approximation of the decay rates.}

\end{itemize}

Using these two approximations, we obtain simplified expressions for the decay rates of each of the three decay channels. They are

\begin{multline}\label{eq:decay_1}
\Gamma_{{\tilde X}_W^\pm\rightarrow W^\pm \nu_{i}} \approx \frac{g_2^4}{64\pi}
\Big(  \frac{ v_d}{\sqrt{2}M_2\mu}\epsilon_i-\frac{M_{BL} v_u}{M_1 v_R^2} \epsilon_j
 \left[V_{\text{PMNS}}\right]_{ji}  \Big)^2 \frac{M_{{\tilde X^\pm}_W}^3}{M_{W^\pm}^2}\left(1-\frac{M_{W^\pm}^2}{M_{{\tilde X}_W^\pm}^2}\right)^2
\left(1+2\frac{M_{W^\pm}^2}{M_{{\tilde X}_W^\pm}^2}\right) \ ,
\end{multline}

\begin{multline}\label{eq:decay_2}
\Gamma_{{\tilde X}_W^\pm\rightarrow Z^0 l_i^\pm}\approx\frac{g_{2}^4}{64\pi}
\Bigg( \frac{ \frac{1}{\sqrt 2} c_W(v_d\epsilon_i+\mu v_{L_i}^*)+
\frac{1}{c_W}\left( \frac{1}{2}-s_W^2 \right)v_d\epsilon_i}{M_2\mu} \Bigg)^2\times \\ 
\frac{M_{{\tilde X}_W^\pm}^3}{M_{Z^0}^2}\left(1-\frac{M_{Z^0}^2}{M_{{\tilde X}_W^\pm}^2}\right)^2
\left(1+2\frac{M_{Z^0}^2}{M_{{\tilde X}_W^\pm}^2}\right) \ ,
\end{multline}

\begin{equation}\label{eq:decay_4}
\Gamma_{{\tilde X}_W^\pm\rightarrow h^0 l_i^\pm}\approx\frac{g_2^2}{64\pi}\sin^2 \alpha
\Bigg(\frac{\epsilon_i}{2\mu} \Bigg)^2 M_{{\tilde X}_W^\pm}
\left(1-\frac{{M_{h^0}^2}}{M_{{\tilde X}_W^\pm}^2}\right)^2 \ .
\end{equation}

We refer the reader to Appendix \ref{appendix:notation} to understand all the parameters in this expressions. By examining \eqref{eq:decay_1}-\eqref{eq:decay_4}, we understand why the ${\tilde X}^\pm_W\rightarrow h^0 \ell^\pm$ channel dominates, being directly proportional to $\epsilon/\mu$, without the suppression $v_d/M_2$ that is present in the other decay channels for similar terms. However, the $v_d/M_2$ suppression becomes less pronounced for small $\tan \beta$ values, since $v_d=174~ \text{GeV}/(1+\tan \beta)$ increases. Therefore, channels of interest such as ${\tilde X}^\pm_W\rightarrow Z^0 \ell^\pm$ become increasingly more significant towards smaller $\tan \beta$ values.
The Goldstone equivalence theorem tells us that the first two channels are amplified by the longitudinal degrees of freedom of the massive $Z_\mu^0$ and $W_\mu^\pm$ bosons, so the traces of these two decays become more apparent in scenarios with more massive LSP's.

\subsection{Choice of neutrino data}

The neutrino mass hierarchy can be normal or inverted. Furthermore, for each of those possible hierarchies, two different values of the neutrino mixing angle $\theta_{23}$\footnote{As discussed above, each of the measured values of $\theta_{23}$ in both the normal and inverted hierarchies have small uncertainties around a central value-- which are incorporated into our computer code. However, in the text of this paper, for notational simplicity, we will ignore these uncertainties and write the central values only.} fit the existing data. See \cite{PDG,Capozzi:2018ubv}. For the normal hierarchy, the angle $\theta_{23}$ can be 0.597 or 0.417, while for the inverted one, $\theta_{23}$ can be 0.529 or 0.421. So far, out of the four possibilities, we have chosen a normal neutrino hierarchy with  $\sin \theta_{23}=0.597$ to compute the branching ratios-- each summed over all three families of leptons --and their relative properties for each decay channel.
The results were shown in Figures \ref{fig:ScatterGammaRatios} and \ref{BarPlotChargino.pdf}. Can choosing the other neutrino hierarchy and/or different values of $\theta_{23}$ 
modify those predictions? To explore this question, we begin by repeating the calculations of Subsection 3.1 leading to Figure \ref{BarPlotChargino.pdf}, but this time for an inverted hierarchy with $\theta_{23}=0.529$. We find that the new median values of the branching ratios change, but are never outside the interquartile ranges displayed in Figure \ref{BarPlotChargino.pdf}. 
Furthermore, we find that switching between the two possible values of the angle $\theta_{23}$ while keeping the hierarchy the same has no impact on the results-- for either the normal or the inverted hierarchy. 

\begin{figure}[t]
   \centering

   \begin{subfigure}[c]{0.495\textwidth}
\includegraphics[width=1.0\textwidth]{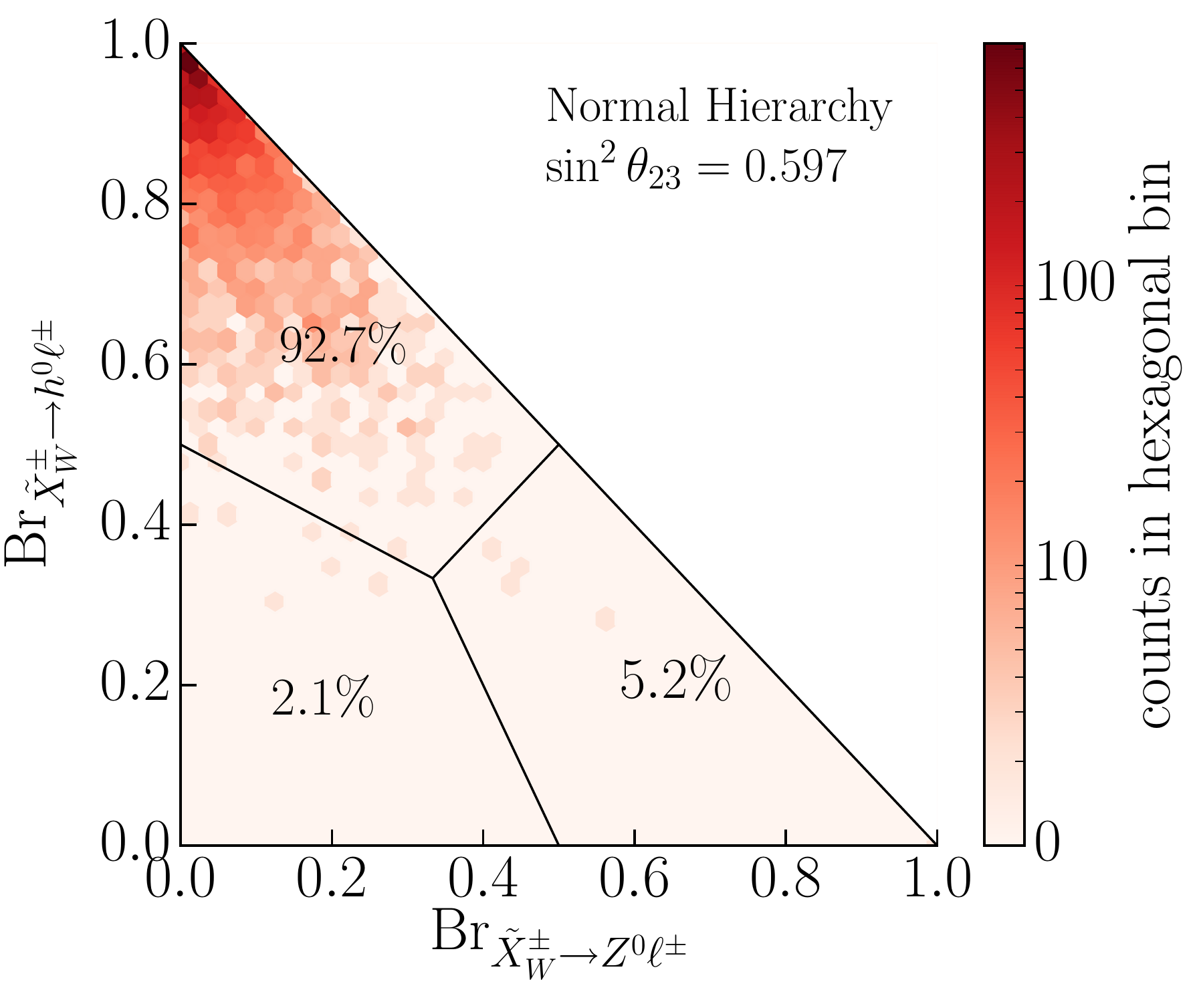}
\label{fig:ScatterHZ_normal}
\end{subfigure}
   \begin{subfigure}[c]{0.495\textwidth}
\includegraphics[width=1.0\textwidth]{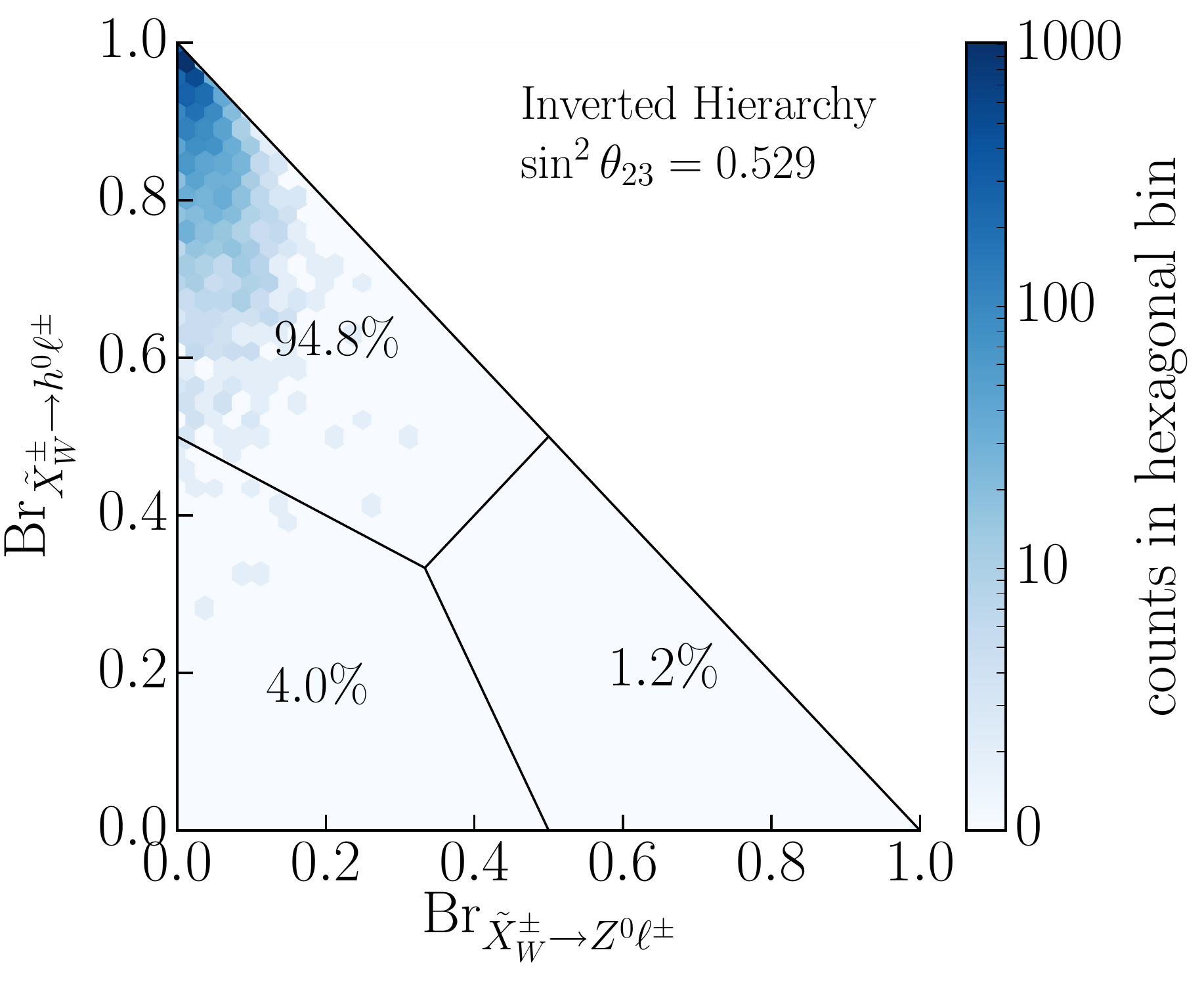}
\label{fig:ScatterHZ_normal}
\end{subfigure}
   \begin{subfigure}[c]{0.495\textwidth}
\includegraphics[width=1.0\textwidth]{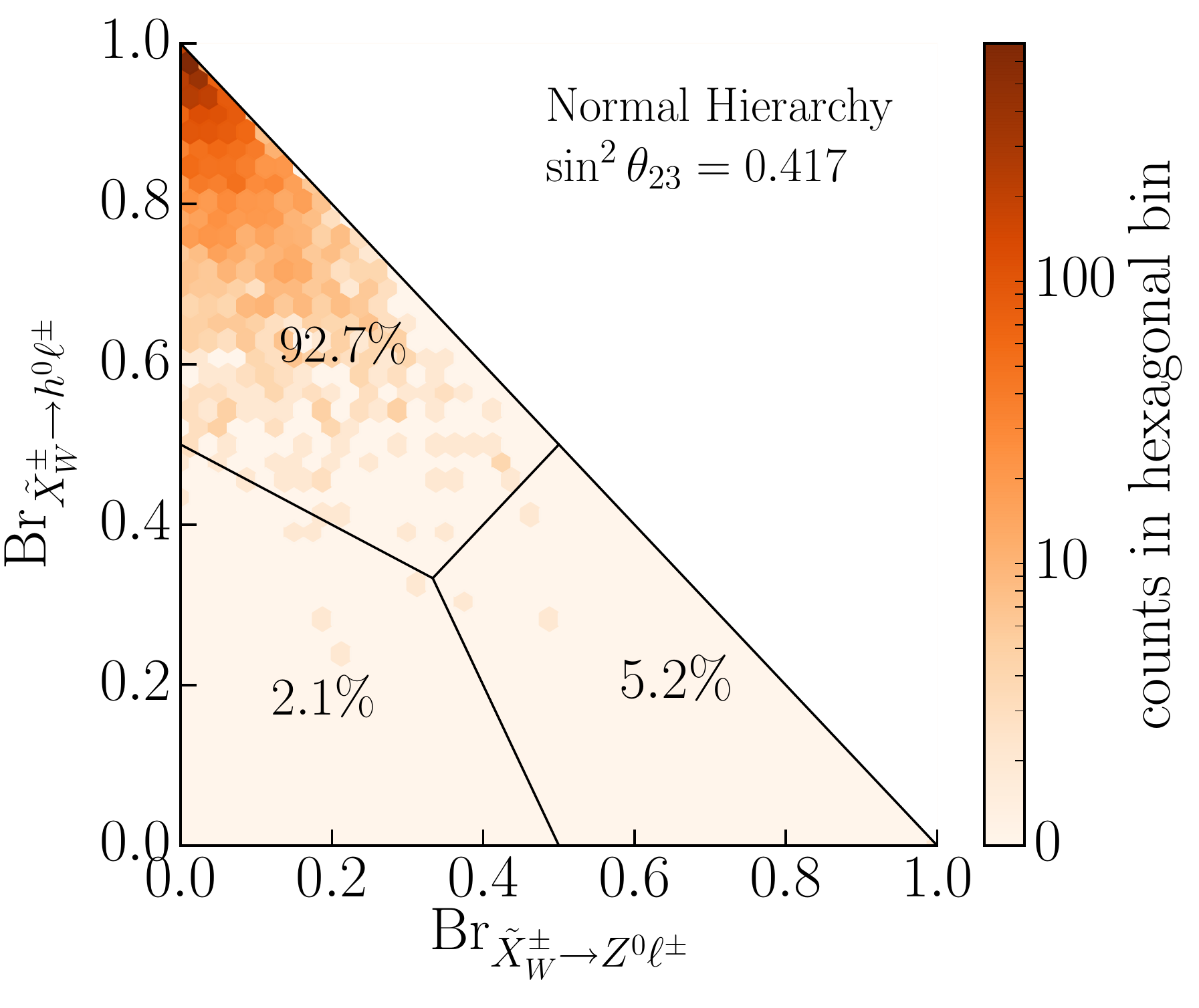}
\label{fig:ScatterHZ_normal}
\end{subfigure}
   \begin{subfigure}[c]{0.495\textwidth}
\includegraphics[width=1.0\textwidth]{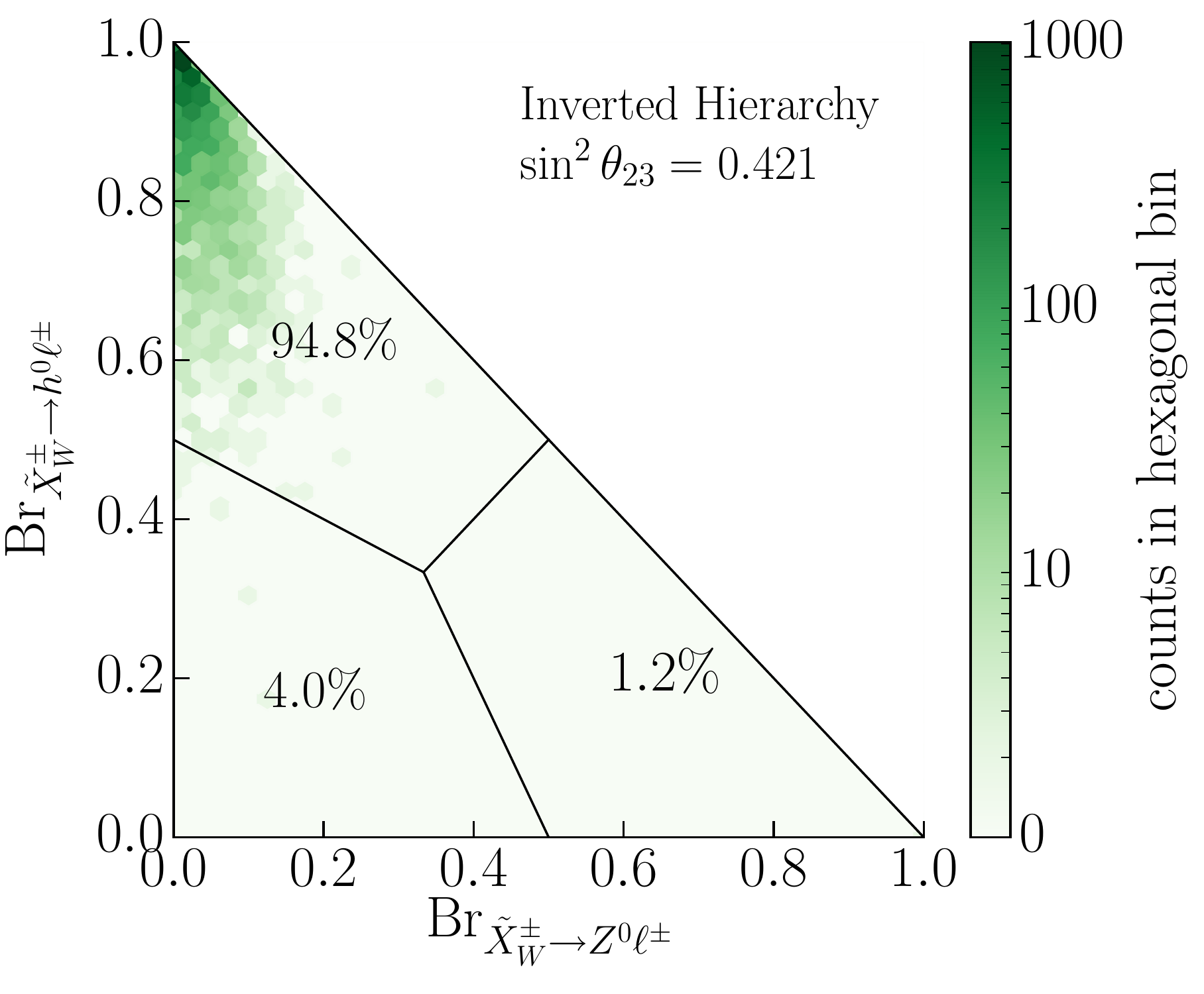}
\label{fig:ScatterHZ_normal}
\end{subfigure}

\caption{ Branching ratio to $h^0\ell^\pm$ versus branching ratio to $Z^0\ell^\pm$ for Wino chargino LSP decays, for both normal and inverted hierarchy. Wino chargino LSP decays via the  ${\tilde X}^\pm_W\rightarrow Z^0 \ell^\pm$ channel tend to be more abundant for a normal hierarchy. The choice of the angle $\theta_{23}$ has no impact on the statistics of these decays, for any of the two possible hierarchies. The percentages indicate what proportion of the points is contained within each third of the four plots.}
\label{fig:ScatterHZ}

\end{figure}

These results are best illustrated by plotting the branching ratios (summed over all mass and $\tan \beta$ bins) for each decay channel against the other two channels-- and doing this for each of the four choices of neutrino input data. Each such plot is simplified by using the fact 
\begin{equation}
\text{Br}_{{\tilde X}^\pm_W\rightarrow W^\pm \nu}+\text{Br}_{{\tilde X}^\pm_W\rightarrow Z^0 \ell^\pm}+\text{Br}_{{\tilde X}^\pm_W\rightarrow h^0 \ell^\pm}=1 \ .
\label{red1}
\end{equation}
We have demonstrated \eqref{red1} explicitly for the normal hierarchy with $\theta_{23}=0.597$, and have numerically shown that it remains true for the other three neutrino input possibilities. It follows that $\text{Br}_{{\tilde X}^\pm_W\rightarrow W^\pm \nu}$ can be determined, using \eqref{red1}, from the remaining two decay channels. Hence, one can plot 2D histograms associated with all 4,858 valid black points associated with a Wino chargino LSP for each of the four possible neutrino input scenarios. These are presented in Figure \ref{fig:ScatterHZ}.

\noindent The most obvious fact that one learns by comparing the top and bottom plots for each individual neutrino hierarchy in Figure \ref{fig:ScatterHZ} is that the $\theta_{23}$ angles play no role in determining the branching ratios-- as stated above. The reason for this is the following. First, note that the simplified expressions \eqref{eq:decay_1} - \eqref{eq:decay_4}, although originally presented for the normal hierarchy with $\theta_{23}=0.597$, remain valid for the other three sets of neutrino data as well. When we sum over the three lepton families in these expressions, the decay rates for each individual channel are proportional to the squared amplitudes of the RPV couplings. Changing the value of the $\theta_{23}$ angle results in a different unitary $V_{\text{PMNS}}$ matrix, which rotates the $\epsilon_i$ and $v_{L_i}, \>i=1,2,3$ components differently, but does not change the squared amplitudes of these couplings to produce a statistically observable effect. For this reason, switching between different $\theta_{23}$ values inside any hierarchy doesn't result in different data patterns, as clearly shown in Figure \ref{fig:ScatterHZ}. This is why using only one value of the angle ( for example $\theta_{23}=0.597$ for the normal hierarchy and $\theta_{23}=0.529$ for the inverted hierarchy) is sufficient to make experimental predictions. Note, however, that if one does {\it not} sum over the three lepton families, this argument is no longer valid, and the value of $\theta_{23}$ can play a substantial role. We will explore this scenario in Subsection 3.4.

The second fact that one learns from comparing the left-hand and right-hand plots of Figure \ref{fig:ScatterHZ} is that there {\it is} a difference in the distribution of branching ratios between the normal and the inverted neutrino hierarchies. This is because, in our theory, the three generations of left handed neutrinos have Majorana masses, directly proportional to the squared amplitudes of these RPV couplings. In the normal hierarchy
\begin{equation}
	m_1 = 0, \quad m_2 = (8.68 \pm 0.10) \times 10^{-3} ~{\rm eV},\quad m_3 = (50.84 \pm 0.50) \times 10^{-3}~{\rm eV}
\label{b2}
\end{equation}
while in the inverted one
\begin{equation}
	m_1 = (49.84 \pm 0.40) \times 10^{-3} ~{\rm eV}, \quad m_2 = (50.01 \pm 0.40) \times 10^{-3} ~{\rm eV},\quad m_3 = 0.
\label{b3}
\end{equation} 
We expect, therefore, that the amplitudes of the couplings will change with the choice of neutrino hierarchy--  leading to the differences in the branching ratios that we observe in Figure \ref{fig:ScatterHZ}. Note, however, from the distribution of points-- plotted as percentages --in the subsections of each plot, that the difference in branching ratios between the normal and inverted hierarchies is relatively small, on the order of a few percent. This is consistent with our statement above that the ``new median values of the branching ratios (for the inverted hierarchy) change, but are never outside the interquartile ranges displayed in Figure 6 (the normal neutrino hierarchy).''
Moreover, in the next section we show that the chargino decay lengths are generally smaller when we assume the inverted hierarchy, compared to when we assume a normal one.

Finally, from Figure \ref{fig:ScatterHZ} we learn that the Wino chargino decays via the  ${\tilde X}^\pm_W\rightarrow h^0 \ell^\pm$ channel tend to be slightly more abundant for a normal hierarchy. However, the incremental difference is relatively small, since the bulk of the points lie in the top left corner, where the decay to $h^0\ell^\pm$ dominates. Although the effect is too small to be statistically distinguishable, it is of interest to note how the choice of neutrino hierarchy can have small influence over the decay rates.

\subsection{Decay Length}

There is one more issue to be discussed; that is, are the decays of the Wino chargino ``prompt''-- defined to be decays where the {\it overall} decay length $L$, defined in \eqref{alan1},  satisfies  $L$~<~1mm? The key to this problem lies in the magnitudes of the RPV parameters, $\epsilon_i$ and $v_{L_i}$. We find that for prompt decays, at least one of the couplings $\epsilon_i$ needs to be larger than $10^{-4}$ GeV. The overall scale of neutrino masses guarantees that this is well satisfied. Putting the lower limit of this interval any lower would not change our results significantly. The upper limit of this interval eliminates the problem of unphysical finely tuned cancellations in the neutrino mass matrices. See \cite {new} for details. 
\begin{figure}[t]
\begin{subfigure}[t]{0.495\textwidth}
\includegraphics[width=1.\textwidth]{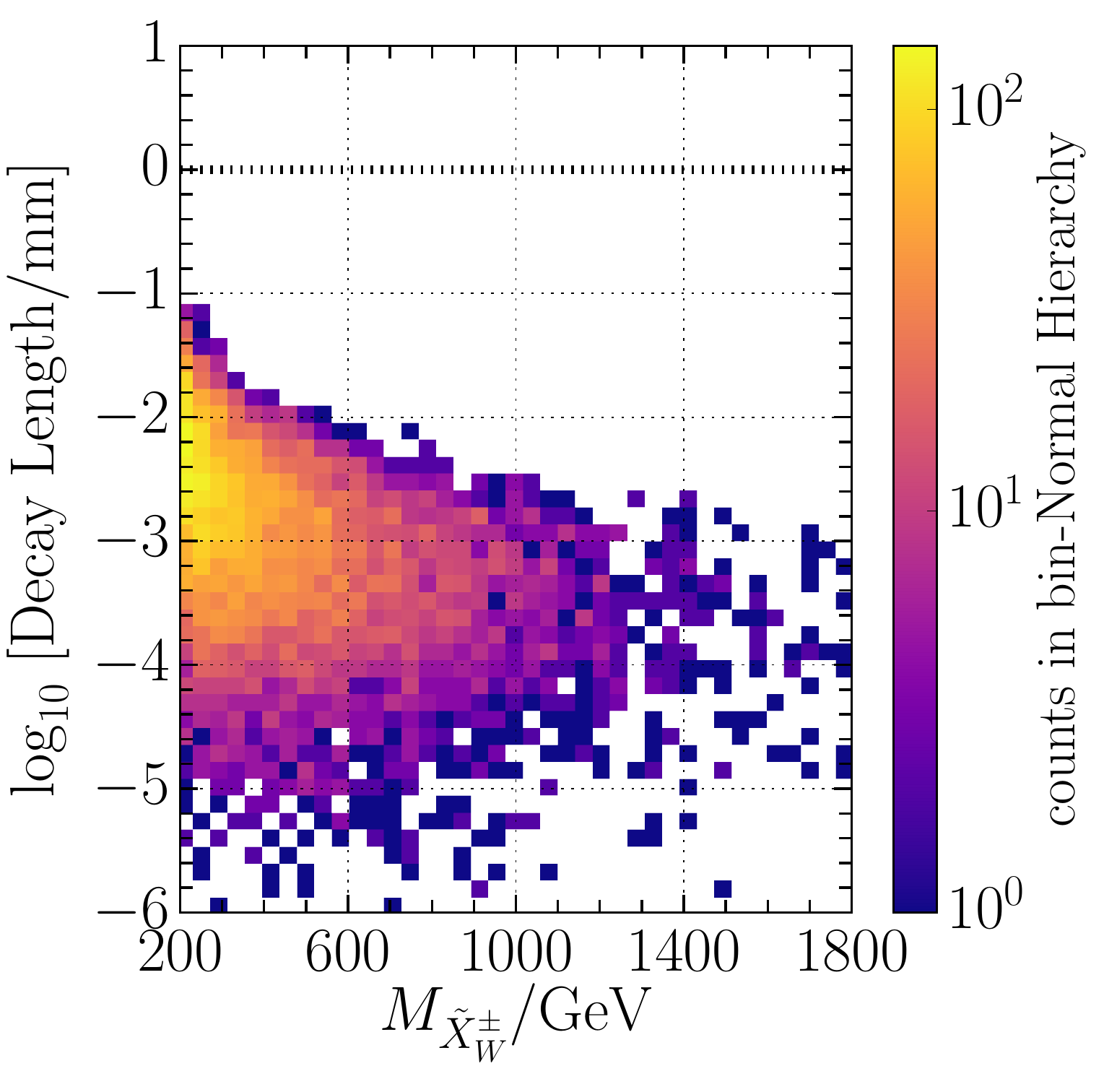}
\end{subfigure}
\begin{subfigure}[b]{0.495\textwidth}
\includegraphics[width=1.\textwidth]{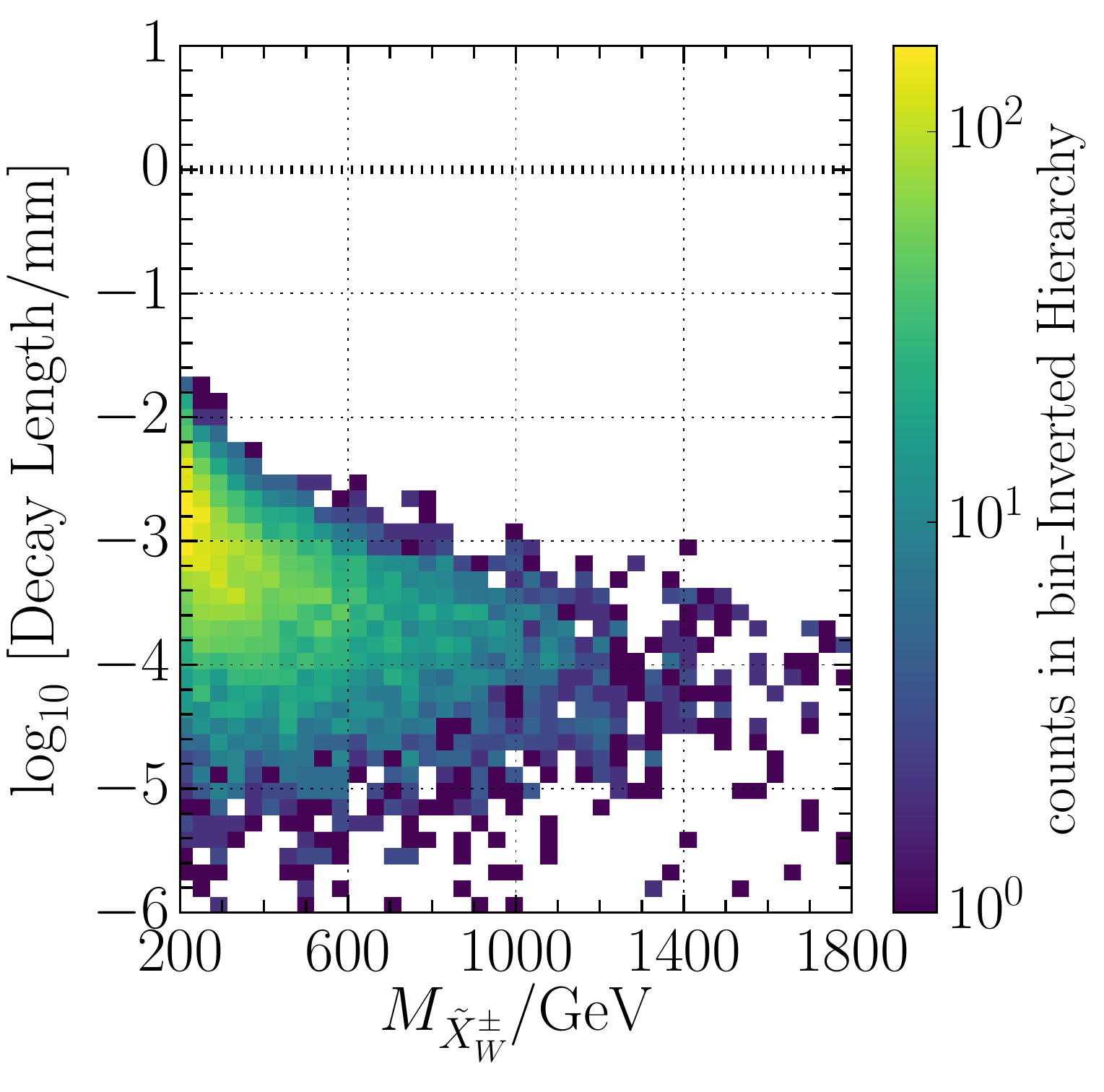}
\end{subfigure}
\caption{Wino Chargino LSP decay length in millimeters, for the normal and inverted hierarchies, summing over all three decay channels. The average decay length $L=c\times \frac{1}{\Gamma}$ decreases for larger values of $M_{\tilde X^\pm_W}$, since the decay rates are amplified because of the longitudinal degrees of freedom of the massive bosons produced. We have chosen $\theta_{23}=0.597$ for the normal neutrino hierarchy and $\theta_{23}=0.529$ for the inverted hierarchy. However, the choice of $\theta_{23}$ has no impact on the decay length.}
\label{fig:LSPprompt1}
\end{figure}
In Figure \ref{fig:LSPprompt1}, we present two scatter plots-- one for the normal and one for the inverted neutrino hierarchy --of the decay length 
\begin{equation}
L=c\times \frac{1}{\Gamma} \ , \quad  \Gamma= \sum_{i=1}^{3} \Big( \Gamma_{{\tilde {X}}_W^\pm\rightarrow W^\pm \nu_{i}} + \Gamma_{{\tilde {X}}^\pm_W \rightarrow Z^0 \ell^\pm_{i}}  + \Gamma_{{\tilde X}^\pm_W\rightarrow h^0 \ell^\pm_{i}} \Big) 
\label{alan1}
\end{equation}
against the Wino chargino LSP mass for all of the 4,858 valid black points with a Wino chargino LSP. The parameter $c$ is the speed of light. Since the {\it overall} decay rate involves a sum over the three lepton families, it follows from the results of the previous section that the value of $\theta_{23}$ plays no role for either hierarchy. We find that the viable Wino chargino LSPs in our simulation decay promptly and produce prompt vertices in the detector for both neutrino hierarchies. 
\noindent However, we note that the decay lengths tend to be slightly smaller in the case of the inverted hierarchy.
\noindent This follows from the fact that the masses of the neutrinos are, overall, slightly larger in the inverted case. Hence, the RPV couplings will be somewhat larger as well-- resulting in a tiny increase in the decay rates and, therefore, smaller decay lengths in the inverted hierarchy.

Although Figure \ref{fig:LSPprompt1} shows that Wino chargino LSPs decay promptly for all viable initial points, their decay rates are strongly dominated by the ${\tilde X}^\pm_W\rightarrow h^0 \ell^\pm$ channel in general. Recall that the notion of ``prompt'' used above involved a sum over all three separate channels. This stimulates us to study the ``promptness'' of each individual decay channel independently-- although we continue to sum over the three lepton families. 

For example, the decay length of ${\tilde X}^\pm_W\rightarrow Z^0 \ell^\pm$  is given by
\begin{equation}
L_{{\tilde X}^\pm_W\rightarrow Z^0 \ell^\pm}=c\times \frac{1}{\sum_{i=1}^{3}\Gamma_{{\tilde X}^\pm_W\rightarrow Z^0 \ell^\pm_{i}}}.
\end{equation}
  \begin{figure}[t]
\begin{subfigure}[b]{1.\textwidth}
\includegraphics[width=1.0\textwidth]{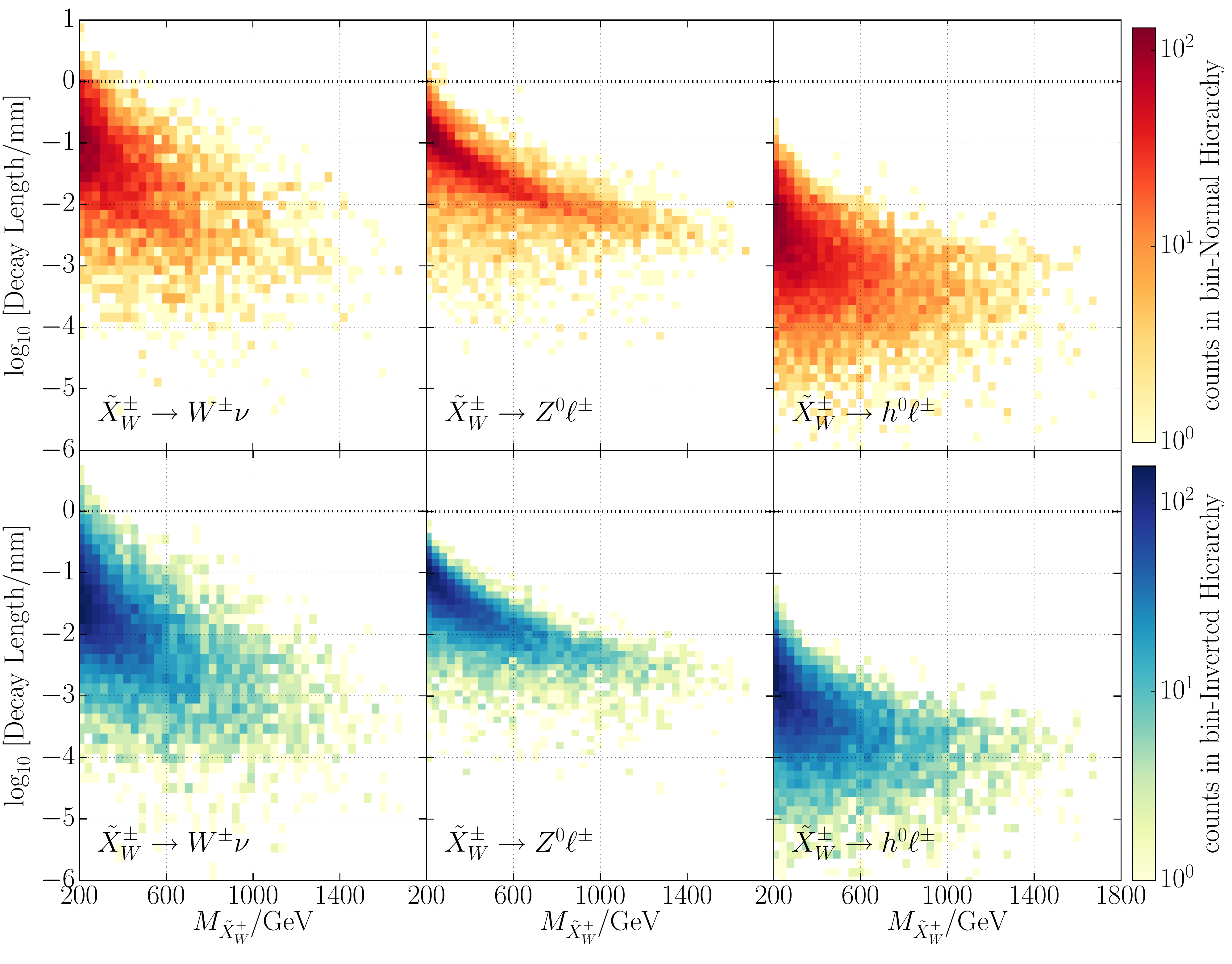}
\end{subfigure}
\caption{Wino Chargino LSP decay length in milimeters, for individual decay channels, for both normal and inverted hierarchies.  We have chosen $\theta_{23}=0.597$ for the normal neutrino hierarchy and $\theta_{23}=0.529$ for the inverted hierarchy. The choice of $\theta_{23}$ has no impact on the decay lengths. All individual channels have decay lengths $<1$mm }
\label{fig:LSPprompt3}
\end{figure}
 In Figure \ref{fig:LSPprompt3} we show that the Wino chargino LSP has decay lengths smaller than 1mm when decaying via any of the channels ${\tilde X}^\pm_W\rightarrow Z^0 \ell^\pm$, ${\tilde X}^\pm_W\rightarrow h^0 \ell^\pm$ and ${\tilde X}^\pm_W\rightarrow W^\pm\nu$.

\subsection{Lepton family production}

As discussed above, for any one of the three generic decay channels, the branching ratio for the decay into an single lepton family can, in principal, depend on the choice of the neutrino hierarchy and the value of $\theta_{23}$ used in determining the values of the $\epsilon_{i}$ and $v_{L_{i}}$ parameters. Using the available neutrino data with $3\sigma$ errors for the neutrino masses, along with the $V_{\text{PMNS}}$ rotation matrix angles and CP violating phases-- see \cite{new} for details --one can calculate, for any valid black point associated with a Wino chargino LSP, the decay rate into each individual lepton family for a given  decay channel. Clearly, the value of the decay rate will depend explicitly on the choice of neutrino hierarchy-- either normal or inverted --and, for a given hierarchy, on the choice of the two allowed values of $\theta_{23}$. 
For example, to quantify the probability to observe an electron $e^\pm$ in the generic decay process ${\tilde X}^\pm_W\rightarrow Z^0 \ell^\pm$, we compute

 \begin{figure}[t]
   \centering

   \begin{subfigure}[b]{0.49\textwidth}
\includegraphics[width=1.0\textwidth]{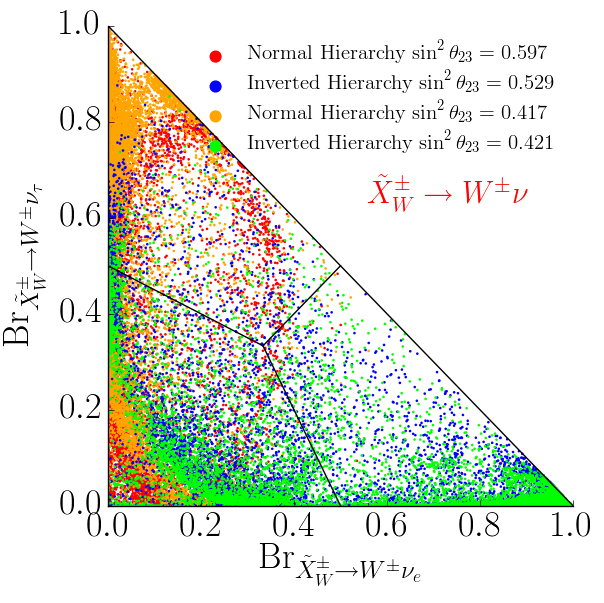}
\end{subfigure}
   \begin{subfigure}[b]{0.49\textwidth}
\includegraphics[width=1.0\textwidth]{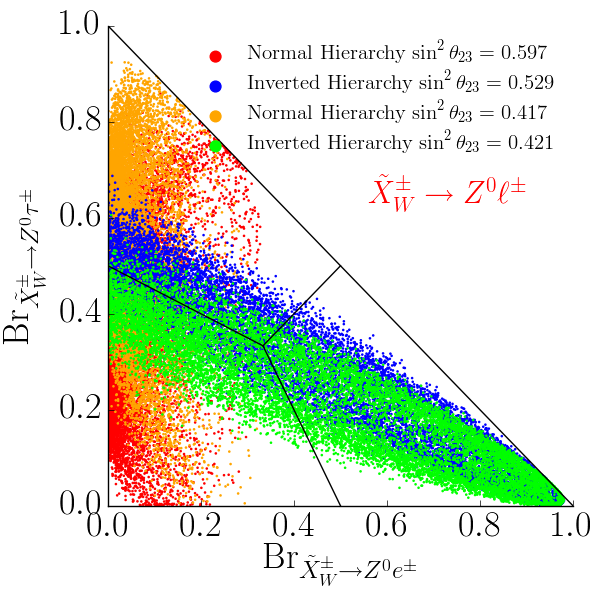}
\end{subfigure}\\
   \begin{subfigure}[b]{0.49\textwidth}
\includegraphics[width=1.0\textwidth]{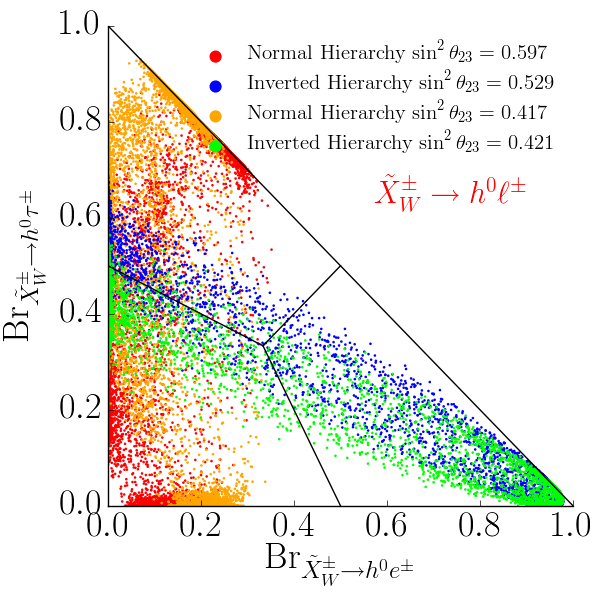}
\end{subfigure}
    \caption{Branching ratios into the three lepton families,
for each of the three main decay channels of a Wino chargino LSP. The associated neutrino hierarchy and the value of $\theta_{23}$ is specified by the color of the associated data point.}\label{fig:lepton_branching}
\end{figure}

\begin{equation}
\text{Br}_{{\tilde X}^\pm_W\rightarrow Z^0 e^\pm}=\frac{\Gamma_{{\tilde X}^\pm_W\rightarrow Z^0 e^\pm}}{\Gamma_{{\tilde X}^\pm_W\rightarrow Z^0 e^\pm}+\Gamma_{{\tilde X}^\pm_W\rightarrow Z^0 \mu^\pm} + \Gamma_{{\tilde X}^\pm_W\rightarrow Z^0 \tau^\pm} } \ ,
\end{equation}
and similarly for a muon, $\mu^{\pm}$, and a tauon, $\tau^{\pm}$, final state. Using this result, we proceed to quantify the branching ratios for each of the 3 decay processes ${\tilde X}^\pm_W\rightarrow W^\pm \nu_i$, ${\tilde X}^\pm_W\rightarrow Z^0 \ell_i^\pm$ and ${\tilde X}^\pm_W\rightarrow h^0 \ell_i^\pm$ into their individual lepton families. The results are shown in Figure \ref{fig:lepton_branching}.
Each subgraph in Figure \ref{fig:lepton_branching} has the following characteristics.
For a point near the top left corner of each subgraph, the branching ratio into a third family lepton is the largest, whereas for a point near 
the bottom right corner, the branching ratio into a first family lepton is the largest. Finally, using the fact that  
\begin{equation}
{\rm Br}_{{\tilde X}^\pm_W\rightarrow Z^0 e^\pm}+{\rm Br}_{{\tilde X}^\pm_W\rightarrow Z^0 \mu^\pm} + {\rm Br}_{{\tilde X}^\pm_W\rightarrow Z^0 \tau^\pm} = 1 \ ,
\label{cup1}
\end{equation}
it follows that for a point near the the bottom left corner, the branching ratio into a second family lepton is the largest.
Perhaps the most striking feature of each such graph is the connection between the Wino chargino decays, the neutrino hierarchy and the $\theta_{23}$ angle. Should experimental observation measure these branching ratios with sufficient precision, that could help shed light on the neutrino hierarchy and the value of $\theta_{23}$. For each neutrino hierarchy, there are two sets of points of different color, since the present experimental data allows for two values of $\theta_{23}$.

For example, let us consider the subgraph associated with the ${\tilde X}^\pm_W\rightarrow Z^0 \ell^\pm$ decay channels. If experimental observation finds that electrons are predominant after the Wino chargino LSP decays, then the hierarchy is inverted. Depending on whether the experimental result is a green or a blue point, implies that $\theta_{23}$ will be $0.421$ or $0.529$ respectively. However, if the branching ratios to either the second or third family leptons are highly dominant, then the hierarchy will be normal, with $\theta_{23}$ given, most likely, by $0.597$ and $0.417$ respectively. That is, with sufficiently precise measured branching ratios one could determine the type of neutrino hierarchy and the value of the $\theta_{23}$ mixing angle from the color of the associated data point.

\section{Wino Neutralino LSP Decays}

\begin{figure}[t]
 \begin{minipage}{1.0\textwidth}
     \centering
   \begin{subfigure}[b]{0.245\linewidth}
   \centering
       \includegraphics[width=1.0\textwidth]{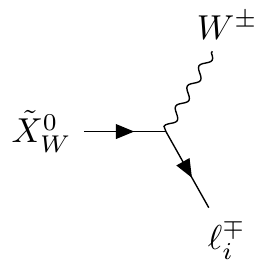} 
\caption*{${\tilde X}^0_W\rightarrow W^\pm \ell^\mp_{i}$}
       \label{fig:table2}
   \end{subfigure} 
   \hfill
   \centering
   \begin{subfigure}[b]{0.245\linewidth}
   \centering
      \includegraphics[width=1.0\textwidth]{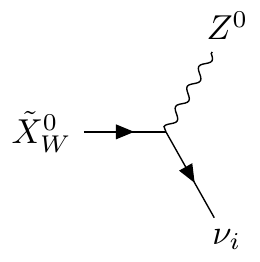} 
\caption*{ ${\tilde X}^0_W\rightarrow Z^0 \nu_{i}$}
       \label{fig:table2}
\end{subfigure}
\hfill
   \centering
     \begin{subfigure}[b]{0.245\textwidth}
   \centering
       \includegraphics[width=1.0\textwidth]{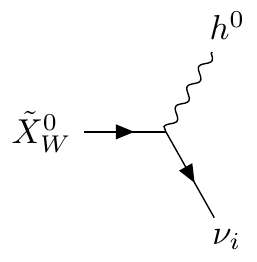} 
\caption*{ ${\tilde X}^0_W\rightarrow h^0 \nu_{i}$}
\end{subfigure}
\end{minipage}
\caption{RPV decays of a general massive Wino neutralino $X_W^0$. There are three possible channels, each with $i=1,2,3$, that allow for Wino neutralino LSP decays. The decay rates into each individual channel were calculated analytically in our previous paper and are reproduced in Appendix \ref{appendix:B}.} \label{fig:NeutralinoDecays}
\end{figure}

 In this section, we analyze the RPV decay signatures of the Wino neutralino LSPs. Written in 4-component spinor notation, the Wino neutralino Weyl spinor, $\tilde \chi_W^0$, becomes
\begin{equation}
\tilde X^0_W=
\left(
\begin{matrix}
\tilde \chi_W^0\\
\tilde \chi_W^{0 \dag}
\end{matrix}
\right) \ ,
\end{equation}
which is a Majorana spinor. In our previous paper \cite{new}, we analyzed the RPV decay channels using 4-component spinor notation for all neutralino mass eigenstates. These are presented, for specificity, in Appendix \ref{appendix:B} of this paper. The Wino neutralino corresponds to the case where n=2.
Unlike the Wino chargino, the Wino neutralino has only three possible decay channels, reproduced here in Figure \ref{fig:NeutralinoDecays}.\\

\subsection{Branching ratios of the decay channels}

The  ${\tilde X}^0_W\rightarrow W^\pm \ell^\mp_{i}$
processes is the most favored for detection at the LHC. Similarly to the Wino chargino decay products, the left handed neutrinos produced during ${\tilde X}^0_W\rightarrow Z^0 \nu_i$ decays can only be detected as missing energy, while the Higgs boson $h^0$ arising from ${\tilde X}^0_W\rightarrow h^0 \nu_i$ couples to both quarks and leptons, leading to decay remnants in the detector that are harder to interpret. Hence, the most interesting decay experimentally appears to be the Wino neutralino decay into a $W^\pm$ massive boson and a charged lepton. The decay rates into each individual channel were calculated in our previous paper and are reproduced in Appendix \ref{appendix:B}. The abundance of each channel is proportional to its branching ratio. For example, for the process ${\tilde X}^0_W\rightarrow W^\pm \ell^\mp$ the branching ratio is defined to be
\begin{equation}
\text{Br}_{{\tilde X}^0_W\rightarrow W^\pm \ell^\mp}=\frac{\sum_{i=1}^{3}  \Gamma_{ {\tilde X}^0_W\rightarrow W^\pm \ell^\mp_{i}}}{\sum_{i=1}^3 \Big( \Gamma_{{\tilde X}^0_W\rightarrow Z^0 \nu_i}+  \Gamma_{{\tilde X}^0_W\rightarrow W^\pm \ell^\mp_i}+\Gamma_{{\tilde X}^0_W\rightarrow h^0 \nu_i}\Big)}  \ .
\end{equation}

\begin{figure}[t]
   \centering
   \begin{subfigure}[b]{1.\textwidth}
\includegraphics[width=1.0\textwidth]{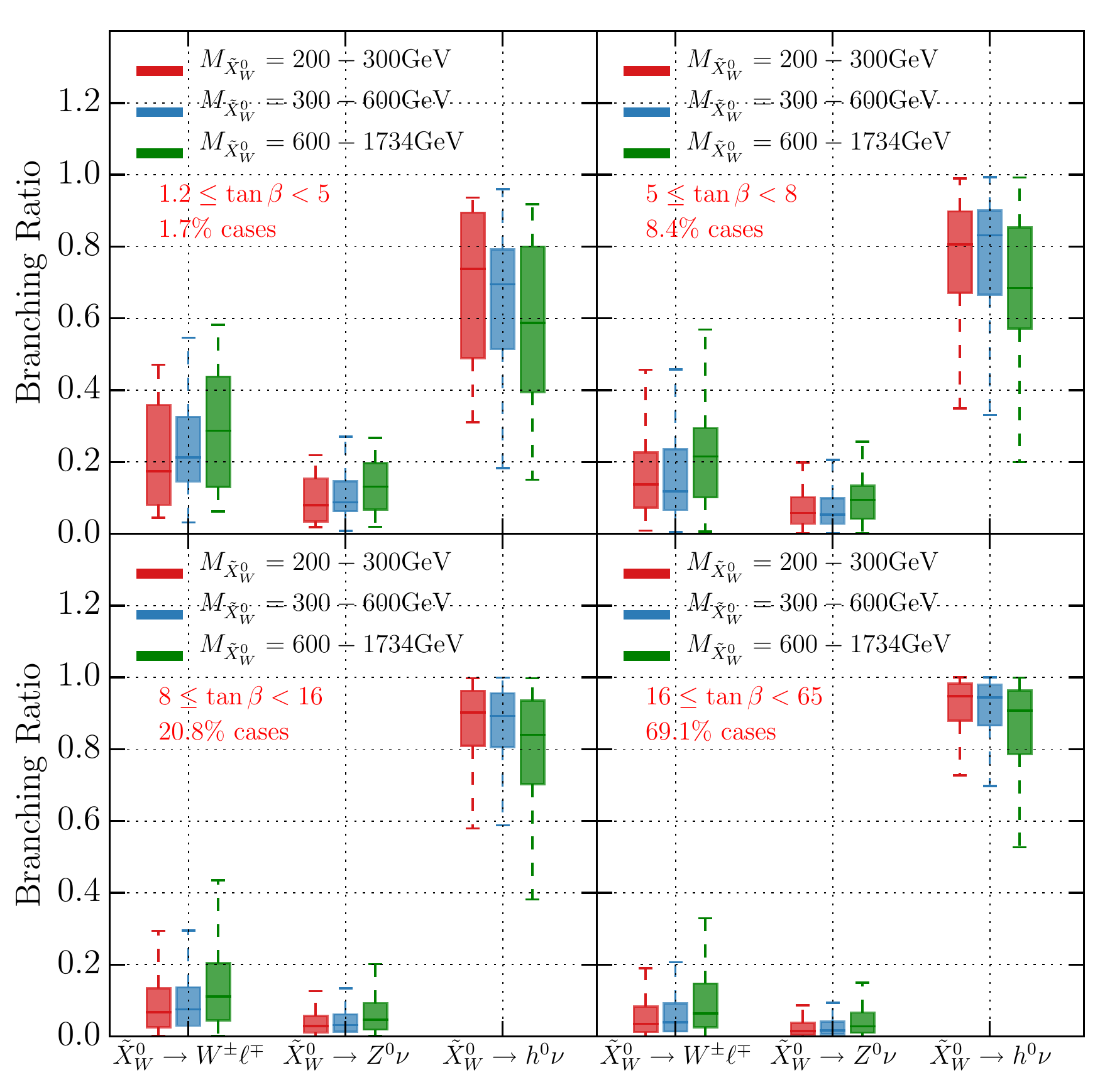}
\end{subfigure}
 
    \caption{Branching ratios for the three possible decay channels of a Wino neutralino LSP divided over three mass bins and four $\tan \beta$ regions. The colored horizontal lines inside the boxes indicate the median values of the branching fraction in each bin, the boxes indicate the interquartile range, while the dashed error bars show the range between the maximum and the minimum values of the branching fractions. The case percentage indicate what percentage of the physical mass spectra have $\tan \beta$ values within the range indicated.  We assumed a normal neutrino hierarchy, with $\theta_{23}=0.597$.}
\label{fig:bar_plot2}
\end{figure}

We now study the decay patterns and branching ratios for each for the 3 decay channels of the Wino neutralino. There are 4,869 valid black points associated with Wino neutralino LSPs. For each of these, we compute the decay rates via RPV processes, using the expressions \eqref{Neutralino_Decay_Rate1}-\eqref{Neutralino_Decay_Rate3} with $n=2$ given in Appendix \ref{appendix:B}. The branching ratios of the main channels take different values for different valid points in our simulation. These values are scattered around the median values of these quantities. We compute the median values, interquartile ranges and the minimum and maximum values of the branching fractions in the same ``bins'' of the parameter space as we used in the study of the Wino chargino LSP decay channels. That is,
we sample the average branching fractions in the three bins for the LSP mass $M_{{\tilde X}_W^0} \in [200, 300],\>[300,600],\>[600,1734]\footnote{Note that the highest mass for a Wino neutralino is somewhat smaller than that for a Wino chargino.}$ GeV and in the four intervals for $\tan \beta \in [1.2,5],\> [5,8],\>[8,16], \>[16,65]$. The results are presented in Figure \ref{fig:bar_plot2}. To carry out the explicit calculations, we have chosen a normal neutrino hierarchy with $\theta_{23}=0.597$. We again find that assuming an inverted neutrino hierarchy changes these results only slightly, while the exact value of $\theta_{23}$ is statistically irrelevant.

 Note that the ${\tilde X}^0_W\rightarrow h^0 \nu$ is dominant in all regions of the parameter space. The ${\tilde X}^0_W\rightarrow W^\pm \ell^\mp$ process has relatively high occurrence, especially for spectra characterized by small $\tan \beta$ values. Just as for charginos, the equations for the decay rates are complicated  and do not allow a simple explanation of the relative results. Furthermore, unlike for charginos, the rotation matrices involved are much more complicated since there are six neutralino species, while only two chargino species. Nevertheless, simplifying assumptions can be made. One such assumption is that the soft breaking terms have much larger magnitudes than the electroweak scale. This renders the Wino neutralino to be almost purely neutral Wino. Furthermore, using the fact that the charged lepton masses are much smaller than the soft breaking parameters further simplifies the equations. Using these approximations in the expressions in Appendix \ref{appendix:B}, one obtains the following simplified formulas for the decay rates. They are given by

\begin{equation}\label{eq:decay_neut1}
\Gamma_{{\tilde X}^0_W\rightarrow Z^0\nu_{i}} \approx
\frac{1}{64\pi}\Big({ \frac{g_2^2}{{2c_W}M_2\mu}(v_d\epsilon_i+\mu v_{L_i}^*)\left[V_{\text{PMNS}}\right]_{ij}^\dag  }\Big)^2
\frac{M_{{\tilde X}_W^0}^3}{M_{Z^0}^2}\left(1-\frac{M_{Z^0}^2}{M_{{\tilde X}_W^0}^2}\right)^2
\left(1+2\frac{M_{Z^0}^2}{M_{{\tilde X}^0_W}^2}\right) \ ,
\end{equation}

\begin{multline}
\Gamma_{{\tilde X}^0_W\rightarrow W^\mp \ell_i^\pm} \approx\frac{1}{64\pi}\Big(
\frac{g_2^2}{2M_2\mu}(v_d\epsilon^*_i+\mu v_{L_i}))
\Big)^2\times\\ \times
\frac{M_{{\tilde X}_W^0}^3}{M_{W^\pm}^2}\left(1-\frac{M_{W^\pm}^2}{M_{{\tilde X}_W^0}^2}\right)^2
\left(1+2\frac{M_{W^\pm}^2}{M_{{\tilde X}_W^0}^2}\right) \ ,
\end{multline}

\begin{equation}\label{eq:decay_neut4}
\Gamma_{{\tilde X}^0_W\rightarrow h^0\nu_{i}} \approx\frac{1}{64\pi}\Big({\frac{g_2}{{2}}\left[V_{\text{PMNS}}\right]_{ij}^\dag\Big(\sin \alpha \frac{\epsilon_j^*}{\mu}\Big)     }\Big)^2
M_{{\tilde X}_W^0}\left(1-\frac{M_{h^0}^2}{M_{{\tilde X}_W^0}^2}\right)^2 \ .
\end{equation}

We refer the reader to Appendix \ref{appendix:notation} to understand all the parameters in this expressions. Unlike the approximate expressions for the decay rates of Wino charginos in eqs. \eqref{eq:decay_1}-\eqref{eq:decay_4}, the above expressions are less exact. The neutralino mass matrix contains a significantly larger number of soft mass parameters which can take values of a few GeV, close to the electroweak breaking scale, where the approximation breaks down. Nevertheless, the above expressions still provide valuable insights into which decay channel is expected to dominate in the chosen regions of parameter space. Analyzing \eqref{eq:decay_neut1}-\eqref{eq:decay_neut4}, we expect the decay channels to have comparable contributions. Interestingly, the channels ${\tilde X}^0_W\rightarrow W^\pm \ell_i^\mp$ and 
 ${\tilde X}^0_W\rightarrow Z^0 \nu_{i}$ receive a suppression proportional to $\frac{v_d}{M_2}=\frac{174 \text{GeV}}{M_2\sqrt{1+\tan^2 \beta}}$. Therefore, for large values of $\tan \beta$, the channel involving the Higgs boson, $h^{0}$, dominates for Wino neutralino decays, just as the Higgs channel dominated the Wino chargino LSP decays for this range of $\tan \beta$.


\subsection{Decay length}

\begin{figure}[t]
\begin{subfigure}[t]{0.49\textwidth}
\includegraphics[width=1.\textwidth]{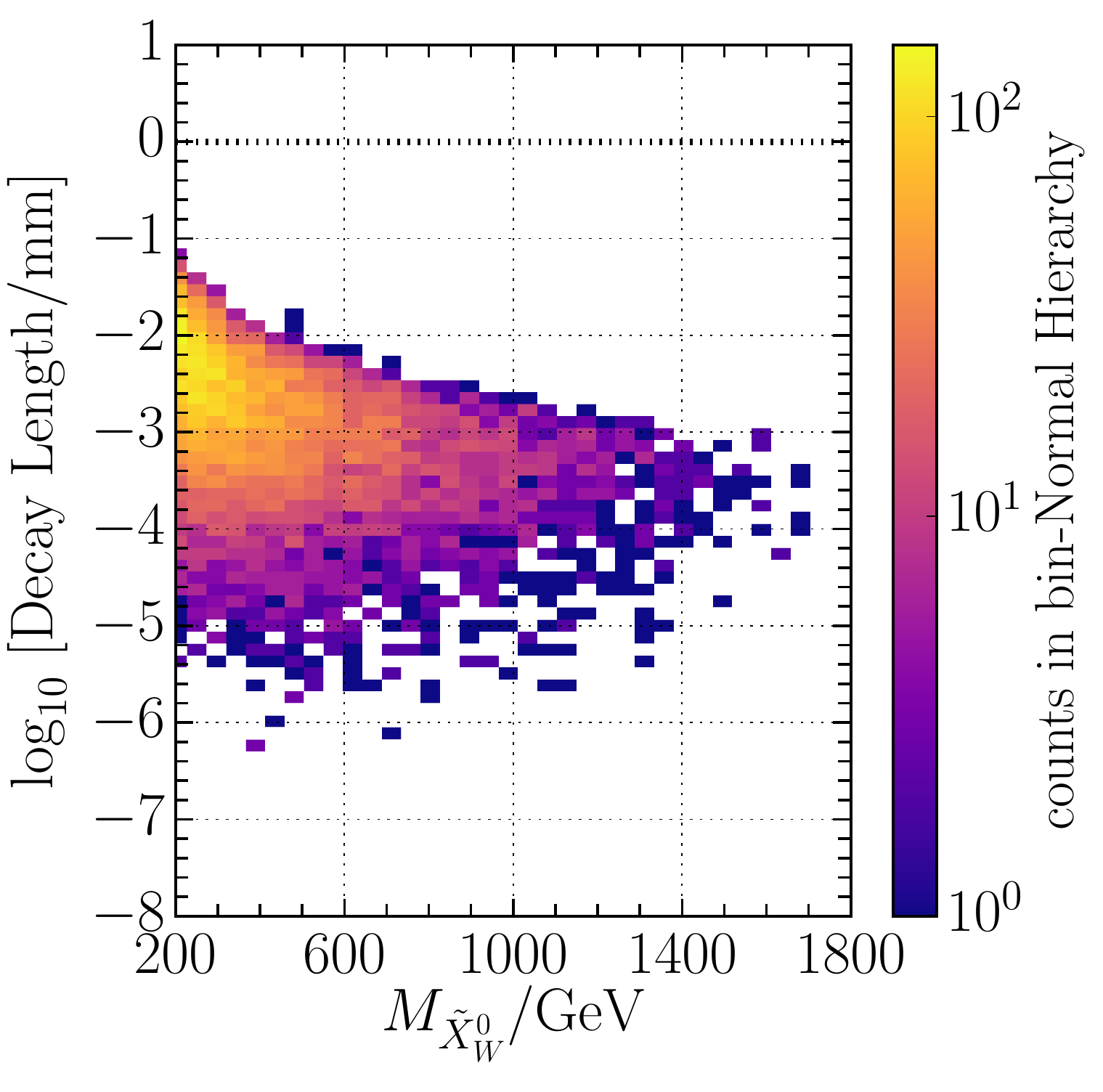}
\end{subfigure}
\begin{subfigure}[b]{0.49\textwidth}
\includegraphics[width=1.\textwidth]{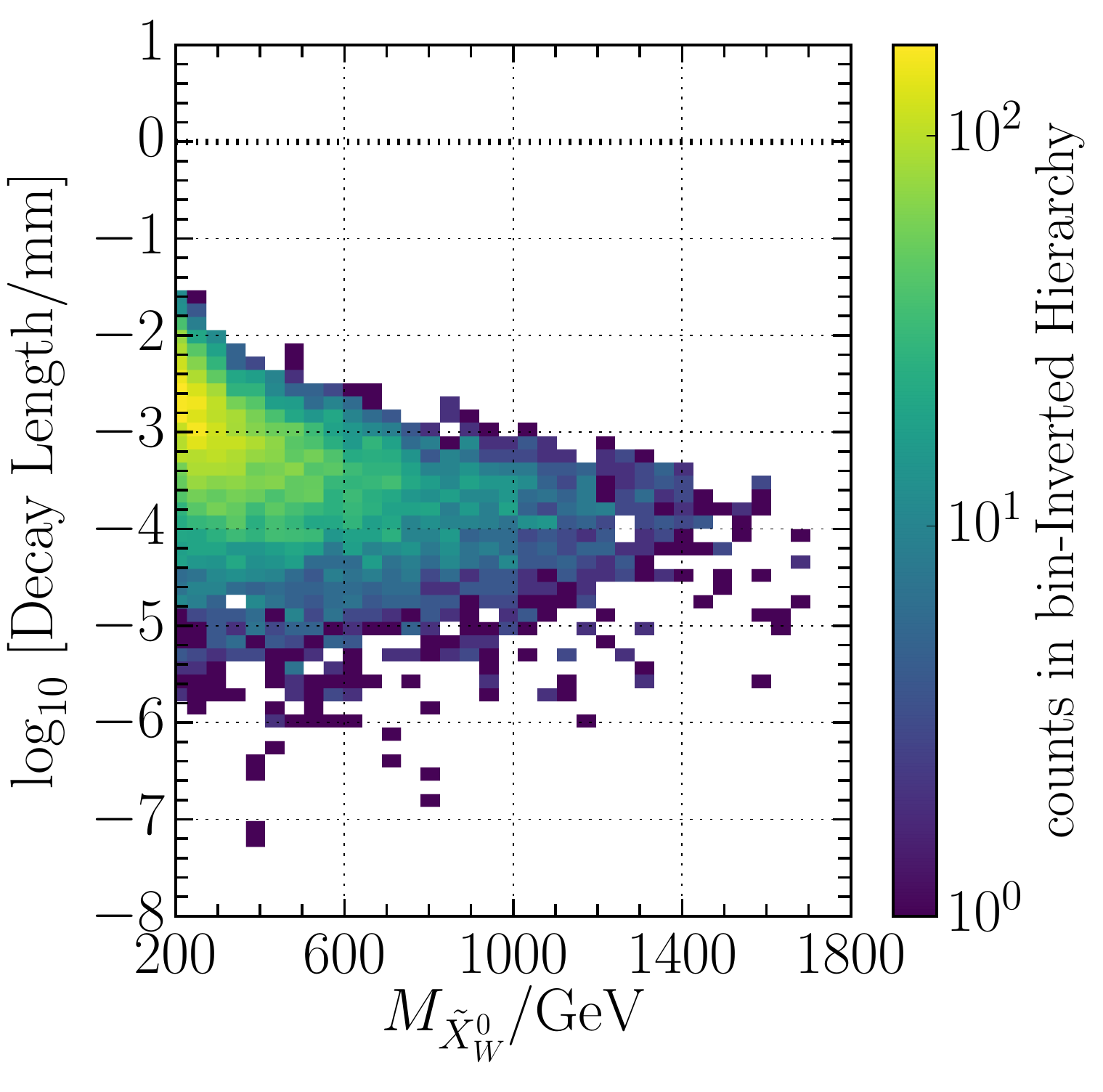}
\end{subfigure}
\caption{Wino neutralino LSP decay length in millimeters, for the normal and inverted hierarchies summed over all three channels. The average decay length $L=c\times \frac{1}{\Gamma}$ decreases for larger values of $M_{\tilde X^0_W}$, since the decay rates are amplified because of the longitudinal degrees of freedom of the massive bosons produced. We have chosen $\theta_{23}=0.597$ for the normal neutrino hierarchy and $\theta_{23}=0.529$ for the inverted hierarchy. However, the choice of $\theta_{23}$ has no impact on the decay length. The dotted line represents the 1mm line, below which all decays are considered prompt.}\label{fig:LSPprompt_neut}
\end{figure}

Figure \ref{fig:LSPprompt_neut} shows that Wino neutralino LSP decays are prompt-- that is, the {\it overall} decay length $L$ is less than 1mm --just as it is for Wino chargino LSP decays. Therefore, signals of both Wino chargino and Wino neutralino LSP decays produce point-like vertices. This insight is particularly useful when considering that the NLSPs of these two sparticle species (Wino neutralino NLSP for Wino chargino  LSP and Wino chargino NSLP for Wino neutralino LSP) are almost degenerate in mass with the LSPs. 
We observe that in the case of the inverted hierarchy, the decay lengths are generally a little smaller, since the values of the RPV couplings are somewhat larger, as we explained in the previous section. 

\begin{figure}[t]
\centering
\begin{subfigure}[b]{1.\textwidth}
\includegraphics[width=1.0\textwidth]{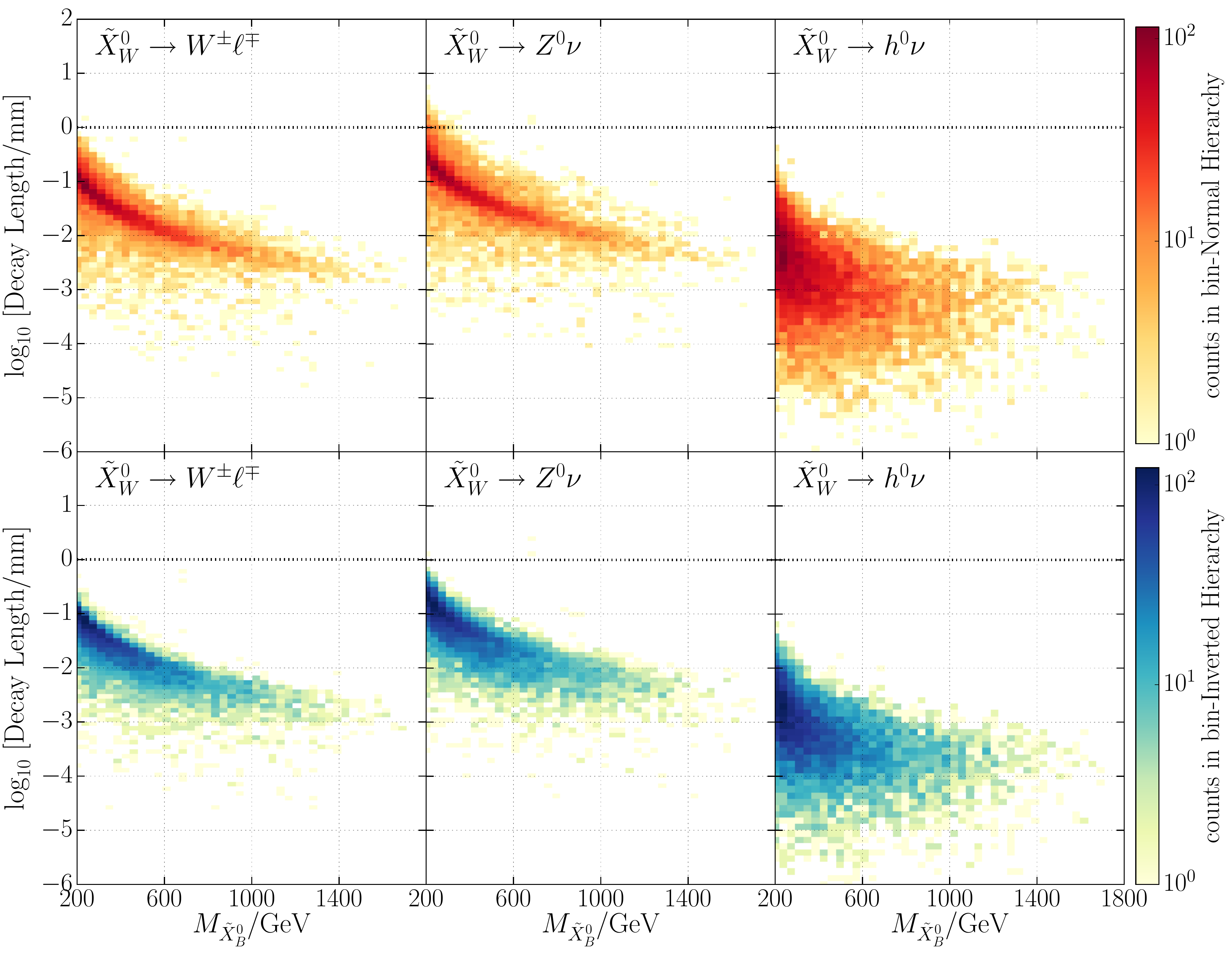}
\end{subfigure}
\caption{Wino neutralino LSP decay length in milimeters, for individual decay channels, for both normal and inverted hierarchies. We have chosen $\theta_{23}=0.597$ for the normal neutrino hierarchy and $\theta_{23}=0.529$ for the inverted hierarchy. The choice of $\theta_{23}$ has no impact on the decay length. The dotted line represents the 1mm line, below which all decays are considered prompt.}\label{fig:LSPprompt}
\end{figure}

In Figure \ref{fig:LSPprompt}, we study the decay lengths of the three decay channels separately.  We find that all three processes occur promptly in the detector. 

\subsection{Lepton family production}
 
We again study which of the three lepton families, if any, is favored within each of the three decay channels. For example, to quantify the probability to observe an electron $e^\mp$ in the ${\tilde X}^0_W\rightarrow W^\pm \ell^\mp$ process, over a muon $\mu^\mp$ or a tauon  $\tau^\mp$, we compute
\begin{equation}
\text{Br}_{{\tilde X}^0_W\rightarrow W^\pm e^\mp}=\frac{\Gamma_{{\tilde X}^0_W\rightarrow W^\pm e^\mp}}{\Gamma_{{\tilde X}^0_W\rightarrow W^\pm e^\mp}+\Gamma_{{\tilde X}^0_W\rightarrow W^\pm \mu^\mp} + \Gamma_{{\tilde X}^0_W\rightarrow W^\pm \tau^\mp} }
\end{equation}

\begin{figure}[t]
   \centering

   \begin{subfigure}[b]{0.47\textwidth}
\includegraphics[width=1.0\textwidth]{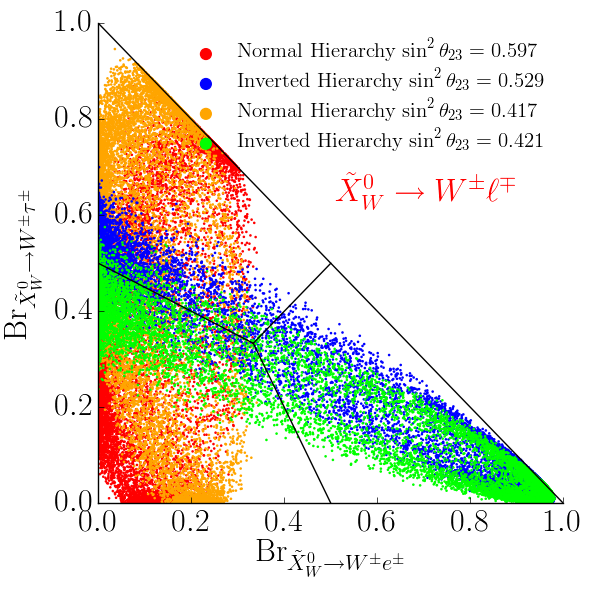}
\end{subfigure}
   \begin{subfigure}[b]{0.47\textwidth}
\includegraphics[width=1.0\textwidth]{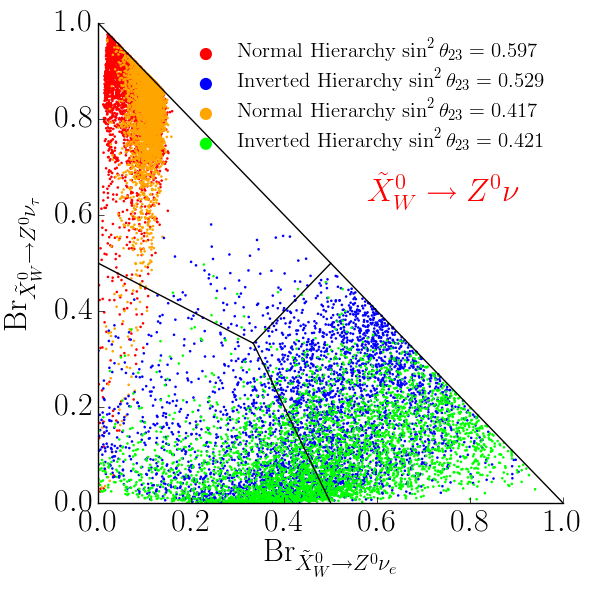}
\end{subfigure}\\
   \begin{subfigure}[b]{0.47\textwidth}
\includegraphics[width=1.0\textwidth]{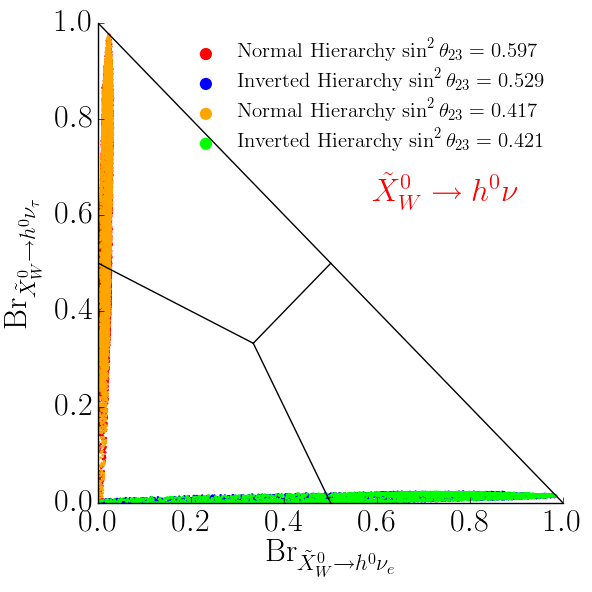} \ .
\end{subfigure}
    \caption{Branching ratios into the three lepton families,
for each of the three main decay channels of a Wino neutralino LSP. The associated neutrino hierarchy and the value of $\theta_{23}$ is specified by the color of the associated data point.}\label{fig:neutralino_lepton_family}
\end{figure}
\noindent Using this formalism, we proceed to quantify the branching ratios for each of the three decay processes ${\tilde X}^0_W\rightarrow W^{\pm} \ell^{\mp}$, ${\tilde X}^0_W\rightarrow Z^0 \nu_i$ and ${\tilde X}^0_W\rightarrow h^0 \nu_i$ into their individual lepton families. The results are shown in Figure \ref{fig:neutralino_lepton_family}.
\noindent In Figure \ref{fig:neutralino_lepton_family} we see that the ${\tilde X}^0_W\rightarrow W^\pm \ell^\mp$ process has an almost identical statistical distribution for lepton family production as does the chargino decay channel ${\tilde X}^\pm_W\rightarrow Z^0 \ell^\pm$. Additionaly, note that in a Wino neutralino decay via ${\tilde X}^0_W\rightarrow h^0 \nu_i$, the decay rate has a dominant term proportional to the square of $[V_{\text{PMNS}}^\dag]_{ij}\epsilon_j$. The combination leads to a branching ratio distribution as that observed in Figure \ref{fig:neutralino_lepton_family}--no $\nu_\tau$ neutrino is produced in the case of an inverted hierarchy and no $\nu_e$ is produced in the case of a normal hierarchy.

\section{Wino Neutralino NLSPs and Wino Chargino NLSPs }

\begin{figure}[t]
   \centering

   \begin{subfigure}[b]{0.49\textwidth}
 \centering
\includegraphics[width=1.0\textwidth]{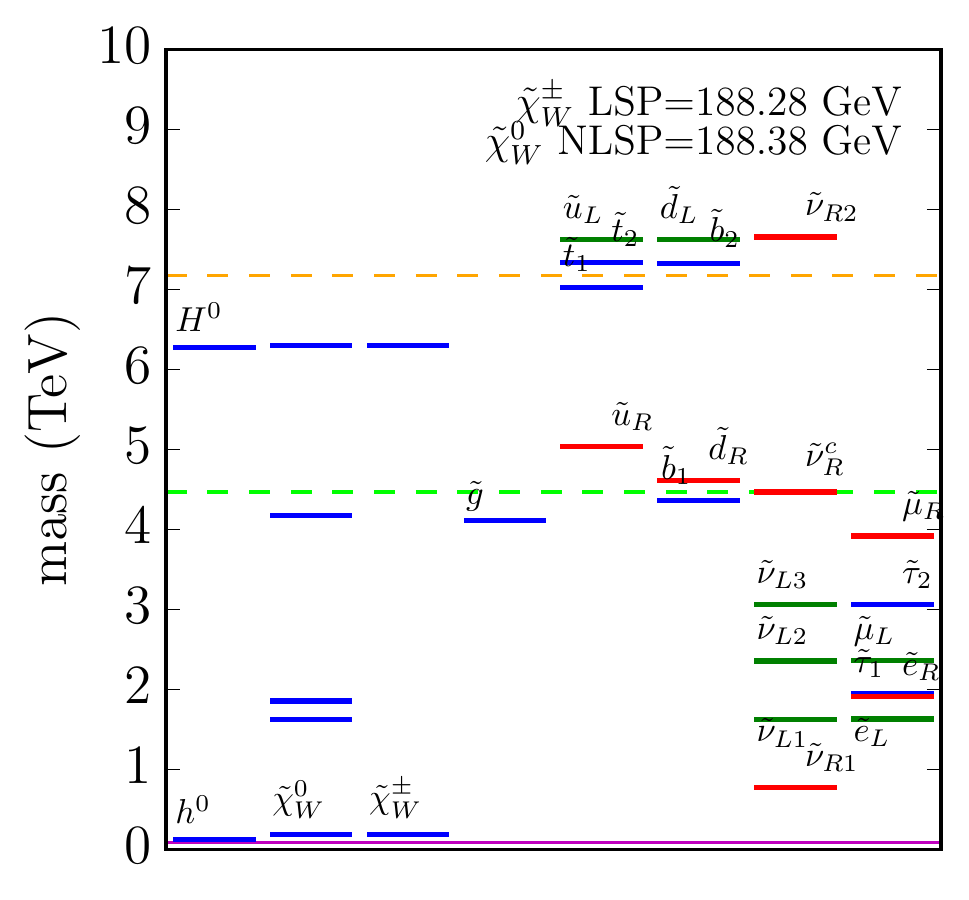}
\caption{}\label{fig:mass_spec1}
\end{subfigure}
   \begin{subfigure}[b]{0.49\textwidth}
 \centering
\includegraphics[width=1.0\textwidth]{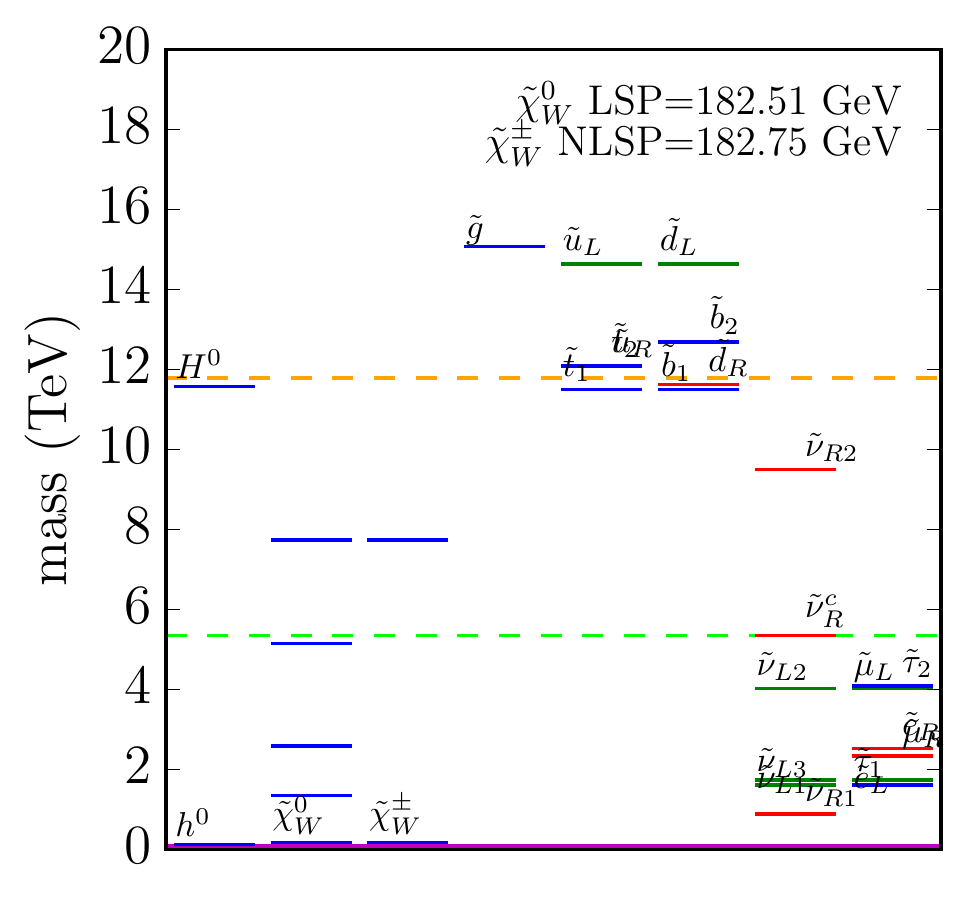}
\caption{}\label{fig:mass_spec2}
\end{subfigure}
\caption{a) A plot of the sparticle spectrum for a choice of one of the 4,858 valid black points associated with Wino chargino LSPs. The Wino neutralino NLSP is almost degenerate in mass with the LSP Wino chargino mass. b) A plot of the sparticle spectrum for a choice of one of the 4,869 valid black points associated with Wino neutralino LSPs. The Wino chargino NLSP is almost degenerate in mass with the LSP Wino neutralino mass.}
\end{figure}

Having analyzed the RPV decays of both Wino chargino LSPs and Wino neutralino LSPs, we now discuss the RPV decays of the NLSPs associated with each case. The reason this is important is the following. Let us begin with the Wino chargino LSPs associated with 4,858 valid black points. Now choose one of these black points. In Figure \ref{fig:mass_spec1}, we plot the entire sparticle spectrum of the theory for this fixed point. 

\begin{figure}[t]
   \centering

   \begin{subfigure}[b]{0.49\textwidth}
 \centering
\includegraphics[width=1.0\textwidth]{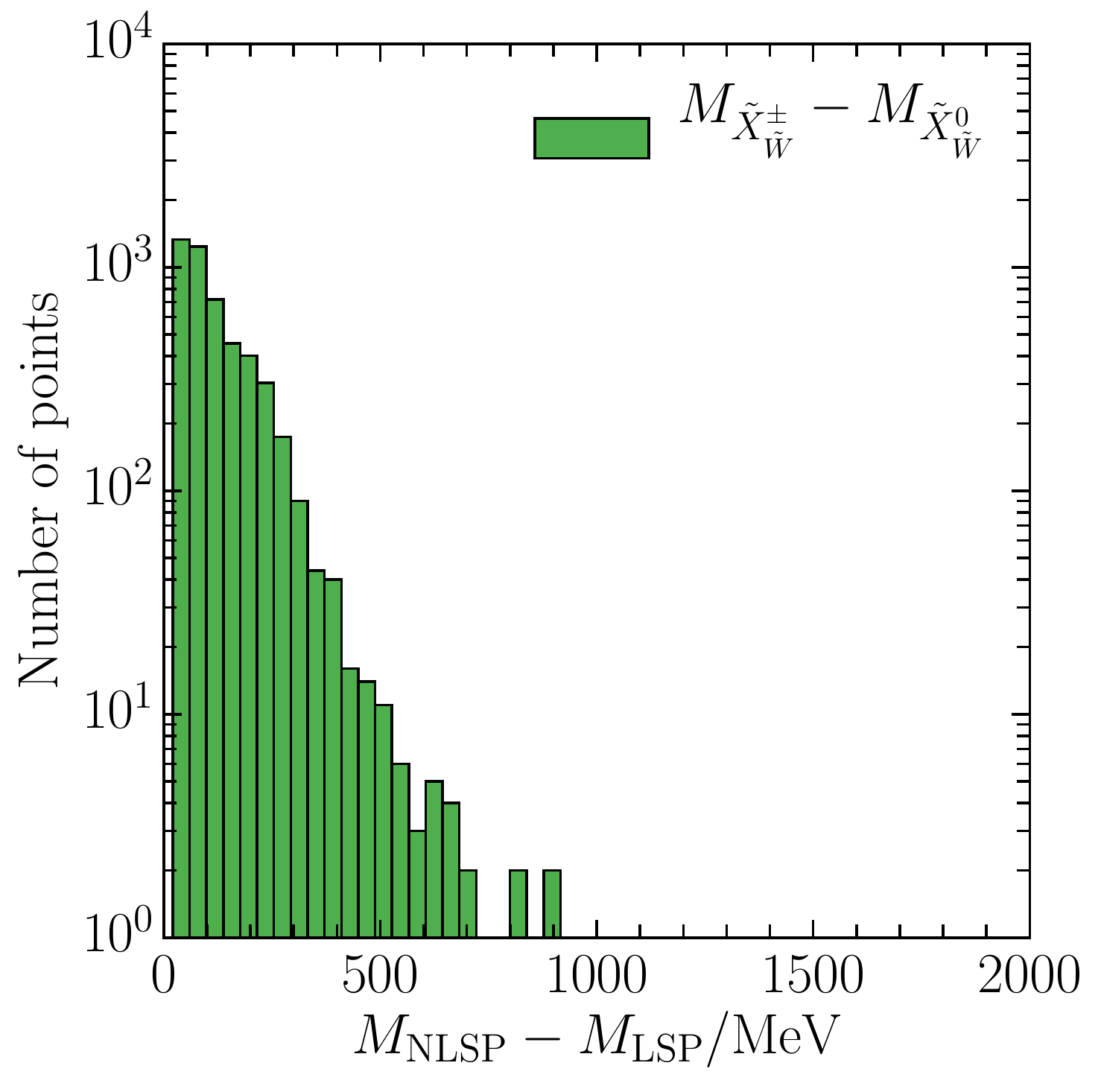}
\caption{}\label{fig:mass_diff1}
\end{subfigure}
   \begin{subfigure}[b]{0.49\textwidth}
 \centering
\includegraphics[width=1.0\textwidth]{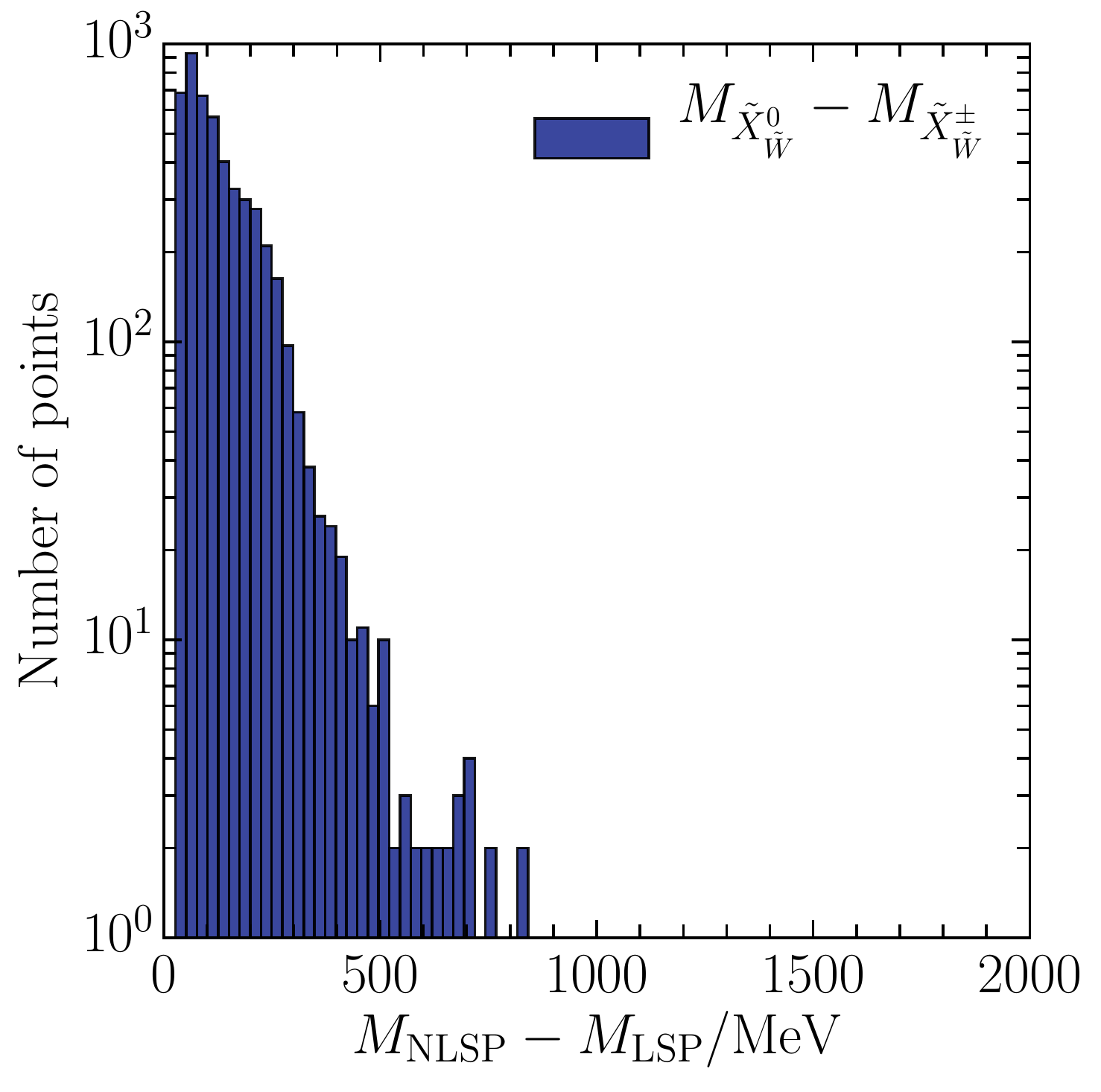}
\caption{}\label{fig:mass_diff2}
\end{subfigure}
\caption{a) The Wino neutralino NLSPs are all almost degenerate in mass with the LSPs, the Wino charginos. The mass difference is smaller than 200 MeV for most of the valid black points, as can be seen in the mass difference histogram. b) The Wino chargino NLSPs are all almost degenerate in mass with the LSPs, the Wino neutralinos. The mass difference is smaller than 200 MeV for most of the viable cases, as can be seen in the mass difference histogram}
\end{figure}

\noindent Of course, a Wino chargino is the LSP by construction.  
Interestingly, however, we see that the associated NLSP is, in fact, a Wino neutralino. This is not simply an accident of our specific choice of black point. In Figure \ref{fig:mass_diff1}, we plot the mass difference in MeV between the Wino neutralino NLSP and the Wino chargino LSP for all 4,858 black points. It is clear that for every Wino chargino LSP, the NLSP is a Wino neutralino whose mass is larger than, but very close to, the mass of the LSP-- as shown in Figure \ref{fig:mass_spec1} for a single such point. This is, perhaps, not surprising since the dominant contribution to the mass of both sparticles is given by the soft supersymmetry breaking  parameter $M_{2}$. See \cite{new} for details. 
Not surprisingly, we find that a similar, but reversed, situation occurs when the LSP is a Wino neutralino. Choosing one of the 4,869 associated valid black points, we find that the complete sparticle spectrum is given in Figure \ref{fig:mass_spec2}. \noindent Of course, a Wino neutralino is the LSP by construction.  
However, we now we find that the situation is reversed and that the associated NLSP is now a Wino chargino. Again, this is not simply an accident of our specific choice of black point. In Figure \ref{fig:mass_diff2}, we plot the mass difference in MeV between the Wino chargino and the Wino neutralino for all 4,869 Wino neutralino black points. It is clear that for every Wino neutralino LSP, the NLSP is a Wino chargino whose mass is larger than, but very close to, the mass of the LSP-- as in Figure \ref{fig:mass_spec2}. Once again, this is hardly surprising since the dominant contribution to the mass of both sparticles is given by the soft supersymetry breaking  parameter $M_{2}$.

Because the mass difference between the two species is so small, both the Wino chargino and the Wino neutralino will be produced at the LHC; assuming that one of them is the LSP and sufficiently light. We have already analyzed the decays of the LSP, both for the case in which the LSP is a Wino chargino and when the LSP is the Wino neutralino. These particles can decay into SM particles due to the RPV couplings in the B-L MSSM model we are studying. The NLSPs, however, as with any other sparticle in the mass spectrum that is not the LSP, can decay via channels that either violate R-parity or channels which conserve it. In general, the RPC couplings are much stronger than the RPV couplings introduced in our theory, since the latter need to be small enough to be consistent with the observed neutrino masses and not lead to unobserved effects such as proton decays. Therefore, the RPC decays of sparticles that are not the LSP are, in general, expected to have much higher branching ratio than the RPV decays. However, in the cases that we focus on, the NLSP is almost degenerate in mass with the LSP. The mass difference is so small that an RPC decay of a Wino neutralino NLSP into a Wino chargino LSP (or vice versa) might prove highly suppressed. Therefore, the NLSP would behave as though it was an LSP which decays via RPV decays. In the remainder of this section, we analyze both the RPV and the RPC decays of the Wino chargino and the Wino neutralino NSLPs and provide a quantitative comparison of the decay rates of those channels.

\subsection{RPV decays of the NLSPs}

\begin{figure}[t]
   \centering

   \begin{subfigure}[b]{0.8\textwidth}
 \centering
\includegraphics[width=1.0\textwidth]{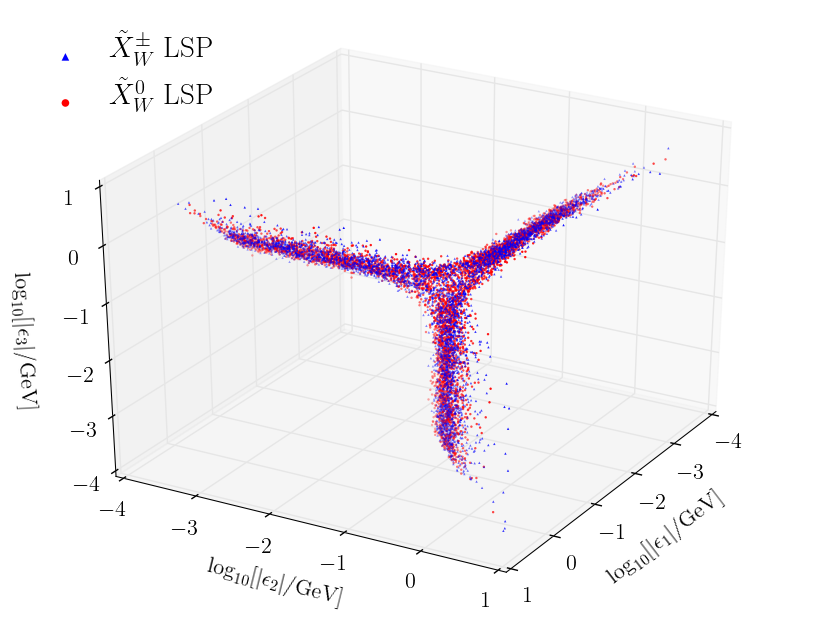}
\caption{}
\end{subfigure}

    \caption{Absolute values of the $\epsilon_1$, $\epsilon_2$ and $\epsilon_3$ parameters associated with the 4,858 black points with Wino chargino LSP (red) and with the 4,869 black points with Wino neutralino LSP (blue). We assume a normal hierarchy with $\theta_{23}$=0.597. We find that these RPV parameters lie in the same statistical regions, regardless of LSP species. }
    \label{fig:scatter_eps}
\end{figure}

We begin our discussion by analyzing the RPV decay channels of both the Wino chargino NLSP and the Wino neutralino NLSP. A Wino chargino and Wino neutralino LSP will, indeed, decay exactly as discussed in Sections 3 and 4 of this paper respectively. However, are the NLSP RPV decay rates and branching ratios the same as though they were actual LSPs? Does a Wino chargino NLSP, associated with the initial conditions for a Wino neutralino LSP, decay in the same way as an actual Wino chargino LSP? The same question arises for the Wino neutralino NLSP. Even though, in these cases, the LSP and NLSP masses are very close, the answer is not immediately obvious, since the decay rates and branching ratios do not depend only on these masses. The decay rates for charginos and neutralinos given in Appendix \ref{appendix:A} and \ref{appendix:B} are completely general, and apply for any chargino and neutralino species, regardless of if they are the LSP or just another particle in the spectrum. Those equations depend on a large number of parameters of the theory, such as $M_{BL}$, $v_R$, $\tan \beta$, $M_R$, as well as on the RPV couplings $\epsilon_{i}, v_{L_{i}}$, $i=1,2,3$. A Wino chargino LSP and a Wino chargino NLSP (associated with a Wino neutralino LSP) have the same RPV decay patterns only if all these parameters are contained within similar statistical intervals.

For example, let us consider the $\epsilon$ parameters. In Figure \ref{fig:scatter_eps} we plot the absolute values of the $\epsilon_1$, $\epsilon_2$  and $\epsilon_3$ couplings, associated with the 4,858 black points with Wino chargino LSP and with the 4,869 black points with Wino neutralino LSP, respectively. We assumed a normal hierarchy, with $\theta_{23}$=0.597. We find that these RPV parameters are statistically similar, whether associated with a Wino chargino LSP, or with a Wino neutralino LSP. This is clear from  Figure \ref{fig:scatter_eps}, where the points lie substantially on top of each other. Plotting $\epsilon_1$ against $\epsilon_2$ and $\epsilon_3$ for both initial conditions with Wino chargino and Wino neutralino LSP is only one of the tests one can make, since other parameters could be relevant. However, this scatter plot is a particularly pertinent one, since the RPV couplings depend on the neutrino masses and neutrino mixing angles, as well as on numerous mass terms from the B-L MSSM Lagrangian of our theory. Indeed, further analysis quickly concludes that other initial parameters have negligable effect. We conclude that, when analyzing Wino chargino LSP decays, one should simultaneously look for the RPV decays of the Wino neutralino-- as though it were the LSP -- and, vice versa.

\subsection{RPC decays of NLSPs}

\begin{figure}[t]
\includegraphics[width=1.0\linewidth]{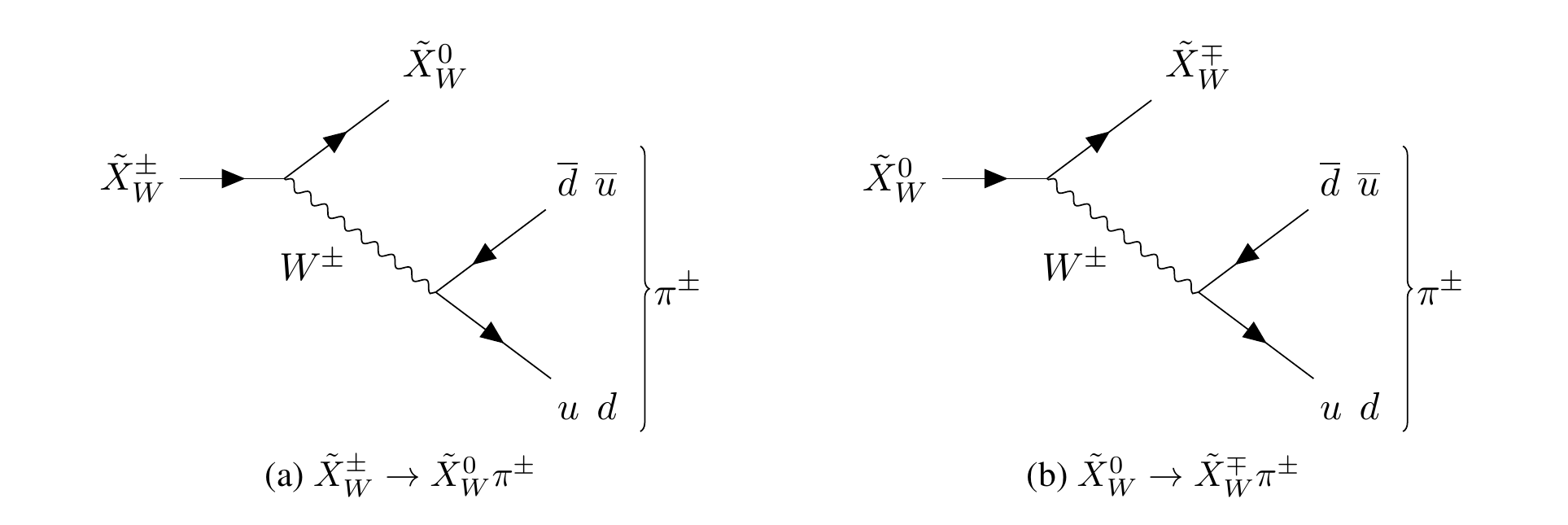}
\caption{Dominant RPC decay modes of (a) a Wino chargino NLSP and (b) a Wino neutralino NLSP. This decay mode dominates for NSLP-LSP mass difference $\delta M$ larger than the mass of the charged pions $\pi^\pm$; that is, $\delta M>m_{\pi^\pm}$}\label{fig:NLSPdecay}
\end{figure}

Figures \ref{fig:mass_diff1} and \ref{fig:mass_diff2} show that the mass differences between the Wino chargino LSP and the Wino neutralino NLSP, or between the Wino neutralino LSP and the Wino chargino NLSP, are generally smaller than 400 MeV. 
For this small mass splitting, there are only a limited number of possible RPC decay channels for the NLSP. If the mass splitting is larger than the charged pion mass, 
$m_{\pi}^{\pm} \sim140$~MeV, 
then the dominant RPC decay of the NLSP is into the LSP and a charged pion $\pi^\pm$. These processes, which involve the on-shell bosons $W^\pm$, are shown in Figure \ref{fig:NLSPdecay}.

\begin{figure}[t]
 \begin{minipage}{1.0\textwidth}
     \centering
        \begin{subfigure}[b]{0.48\linewidth}
   \centering
\includegraphics[width=0.8\textwidth]{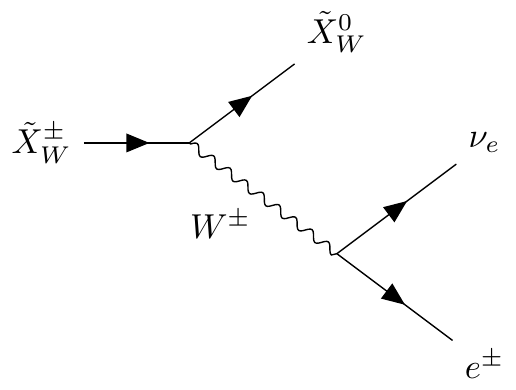}
\caption{${\tilde X}^\pm_W\rightarrow \tilde X_W^0 e^\pm \nu_e$}
       \label{fig:table2}
   \end{subfigure} 
   \begin{subfigure}[b]{0.48\linewidth}
   \centering
\includegraphics[width=0.8\textwidth]{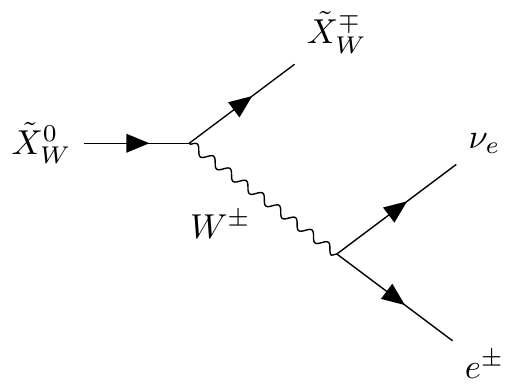}
\caption{${\tilde X}^0_W\rightarrow \tilde X_W^\mp e^\pm\nu_e$}
       \label{fig:table2}
   \end{subfigure} \\
\end{minipage}
\caption{Second most dominant RPC decay modes of (a) a Wino chargino NLSP and (b) a Wino neutralino NLSP. This decay mode dominates for NSLP-LSP mass difference $\delta M$ smaller than the mass of the charged pions $\pi^\pm$; that is, $\delta M<m_{\pi^\pm}$}\label{fig:NLSPdecay2}
\end{figure}
At leading order, the decay rate of the Wino chargino NLSP into a charged pion and the Wino neutralino LSP can be expressed in terms 
of the decay rate of the charged pion,
\begin{equation}\label{RPCrate}
\Gamma(\tilde X_W^\pm \rightarrow \tilde X_W^0 \pi^\pm)=\Gamma(\pi^\pm \rightarrow
\mu^\pm \nu_\mu)\times \frac{16\delta M^3}{m_\pi m_\mu^2}
\left(  1-\frac{m_\pi^2}{\delta M^2}   \right)^{1/2}\left( 1-\frac{m_\mu^2}{m_\pi^2}  \right)^{-2},
\end{equation}

\noindent where $\delta M=M_{\tilde X_W^\pm}-M_{\tilde X_W^0}$ is the mass difference between the NLSP and the LSP, and $m_{\pi}$ and $m_{\mu}$ denote the masses of the charged pion and the muon respectively. Conversely, in the case in which the Wino chargino is the LSP, the main RPC channel of the Wino neutralino NLSP is into a Wino chargino LSP and a charged pion. The decay rate is given by eq. \eqref{RPCrate}, but  now with $ \delta M=M_{\tilde X_W^0}-M_{\tilde X_W^\pm}$. 

Note that the processes shown in Figure \ref{fig:NLSPdecay} can only happen if the mass splitting between the LSP and the NLSP is larger than the charged pion mass $m_{\pi^\pm}$. For smaller mass differences, the decay modes shown in Figure \ref{fig:NLSPdecay2} then dominate. The decay rate of a Wino chargino NLSP decaying into a Wino neutralino LSP, an electron and a neutrino is given by

\begin{equation}\label{RPCrate2}
\Gamma(\tilde X_W^\pm \rightarrow \tilde X_W^0 e^\pm \nu_e)=\frac{2G_F^2}{15\pi^2}\delta M^5,
\end{equation}
where $\delta M=M_{\tilde X_W^\pm}-M_{\tilde X_W^0}$  and $G_F$ is the Fermi constant. Conversely,  the decay rate of a Wino neutralino NLSP into a Wino chargino LSP, an electon and a neutrino  is given by eq. \eqref{RPCrate2}, with $ \delta M=M_{\tilde X_W^0}-M_{\tilde X_W^\pm}$. Finally, we note that there is a similar RPC decay channel involving the muon. However, since the mass of the muon is much larger than that of the electron, this decay rate is greatly suppressed relative to \eqref{RPCrate2} and, hence, is irrelevant.

\subsection{RPV vs RPC}

In Section 5.1, we have shown that the RPV decays of the Wino chargino and the Wino neutralino NLSPs occur as if they were the both LSPs. Based on the analysis carried out in Section 3.3 and 4.2, we expect these RPV processes to allow for prompt decays of the NLSPs in the detector. In this section, we analyze whether the RPC processes of the NLSP can produce observable traces in the detector as well.

\begin{figure}[t]
   \centering

   \begin{subfigure}[b]{0.49\textwidth}
 \centering
\includegraphics[width=1.0\textwidth]{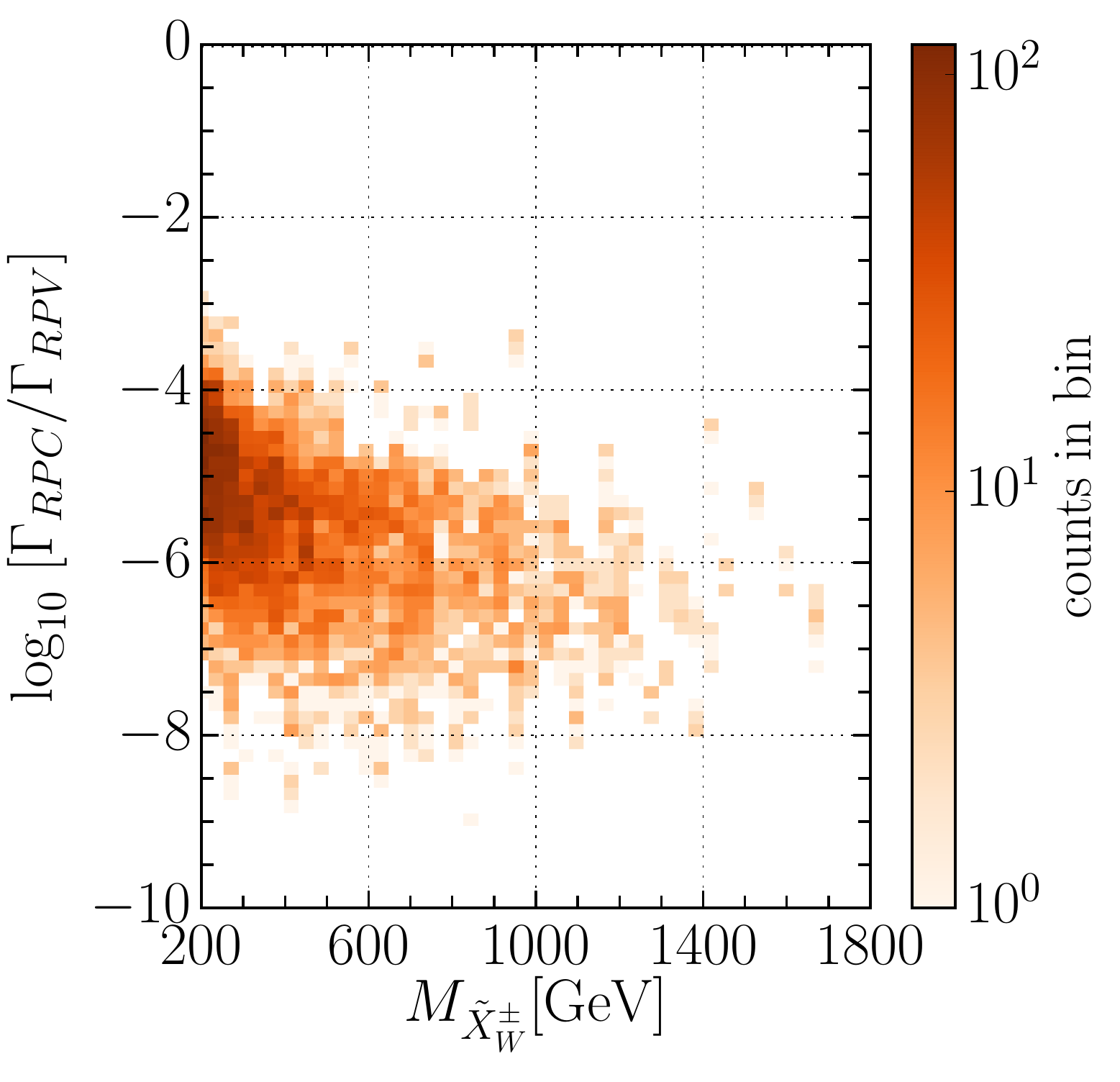}
\caption{}\label{fig:ratio1}
\end{subfigure}
   \begin{subfigure}[b]{0.49\textwidth}
 \centering
\includegraphics[width=1.0\textwidth]{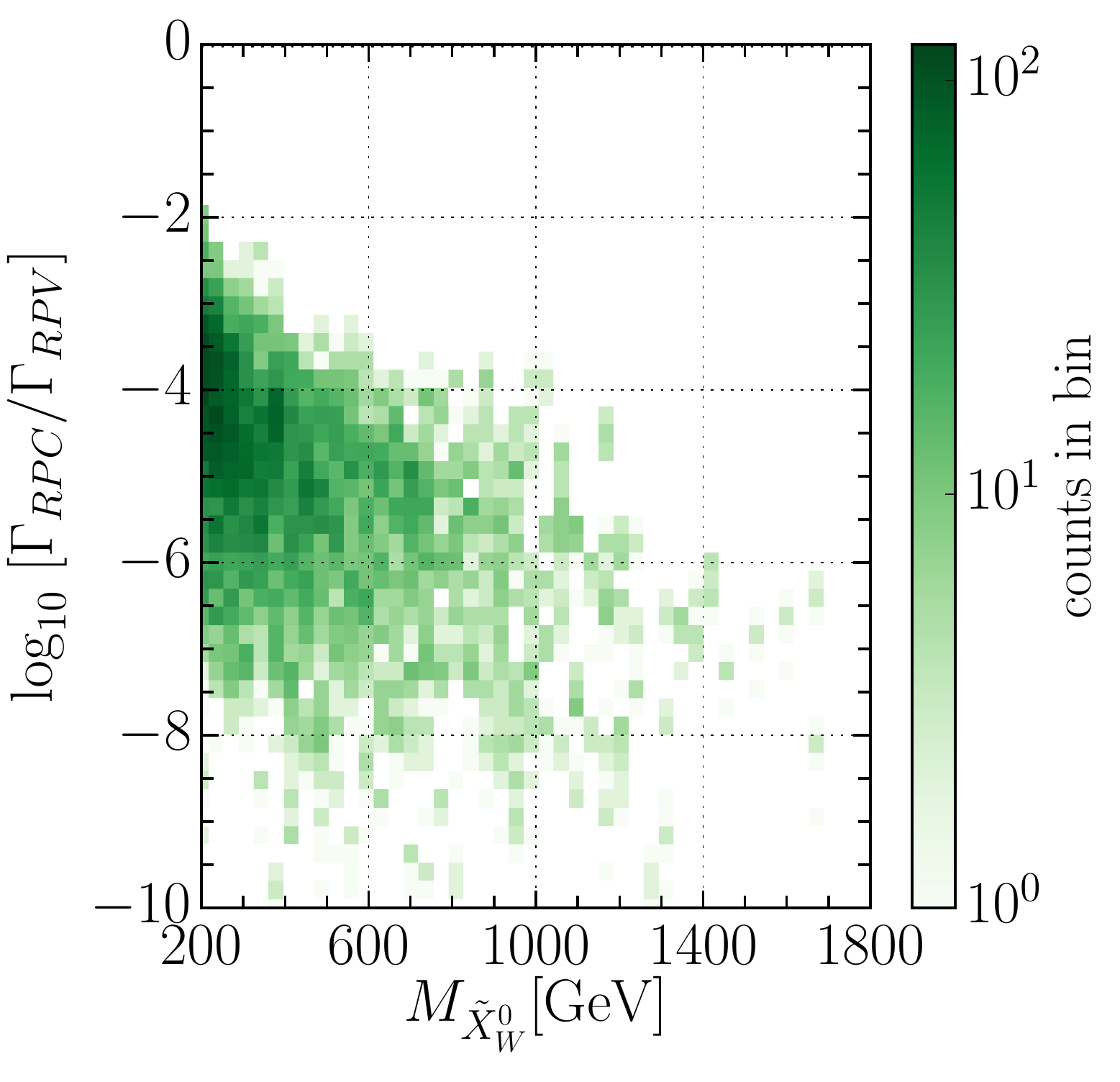}
\caption{}\label{fig:ratio2}
\end{subfigure}
\caption{Ratios between the decay rates of the RPC and the RPV channels. On the horizontal axis, we show the NLSP mass. We study both scenarios with (a) Wino chargino NLSPs and (b) Wino neutralino NLSPs. In all cases, the RPV processes are strongly dominant.}
\label{fig:ratios}
\end{figure}

 In Figure \ref{fig:ratios}, we present the ratios $\Gamma_{\text{RPC}}/\Gamma_{\text{RPV}}$ for all simulated NLSPs, where $\Gamma_{\rm RPC}$ is computed by summing over all RPC channels discussed in Section 5.2. We find that, {\it in all cases}, the RPC processes are strongly suppressed compared with the RPV ones. This suppression of the RPC decays is due to the near degeneracy in mass between the Wino chargino LSP and the Wino neutralino NLSP shown in Figure \ref{fig:mass_diff1}, and the similar mass degeneracy between a Wino neutralino LSP and its Wino chargino NLSP displayed in Figure \ref{fig:mass_diff2}-- with mass splittings ranging between 20 MeV and 500 MeV. Note that most mass splittings are $\lesssim 200$ MeV.
 Therefore, the RPC decays of Winos charginos and Wino neutralinos NLSPs are not expected to produce any visible traces in the detector. Hence, in both cases, the only decay mode of the LSP and the dominant decay mode of the NLSP are precisely the RPV decays discussed in Sections 3 and 4
above.

Since this final conclusion rests on having a reliable computation of the mass splitting between the Wino NLSP and the Wino LSP, we end this Section by discussing the role that higher-loop contributions might have. As stated in the Introduction, all calculations performed in this paper are done using one-loop RG $\beta$ and $\gamma$ functions ignoring finite loop corrections. For the values of the parameters required to produce realistic low-energy phenomenology, such higher loop effects are not expected to have a large effect as regards the Wino NLSP and Wino LSP mass splitting--on the order of several hundred MeV. Indeed, such loop contributions to NLSP and LSP mass splitting have been explicitly computed to the one-loop and two-loop level in \cite{Gherghetta:1999sw, Cheng:1998hc, Feng1999xx}  and \cite{Ibe:2012sx} respectively--albeit in different theoretical contexts than our own, without the B-L extension, and without the RPV couplings which mix charginos and neutralinos with leptons. These papers found that, indeed, the higher loop corrections are small--on the order of 100-200 MeV. We conclude, therefore, that our expectation that the higher loop contributions to the mass splitting of the Wino NLSP and the Wino LSP in our theory would only be, at most, on the order of several hundred MeV, is indeed correct. Note that adding several hundred MeV might raise our maximum possible mass splitting to $\sim$ 700 MeV--with most other mass splittings being much smaller. Even at 700 MeV, the decay ratio $\Gamma_{\rm RPC}/\Gamma_{\rm RPV}\ll 1$ and, hence, our conclusions in the previous paragraph remain unaltered.

\section{Conclusion}

In this paper, we have systematically derived the RPV decay channels, decay rates, branching ratios, and the relationship of these results to the neutrino mass hierarchy and the $\theta_{23}$ neutrino mixing angle, for both a Wino Chargino LSP and a Wino neutralino LSP-- all within the context of the explicit $B-L$ MSSM theory. It is shown that the Wino neutralino is the NLSP for a Wino chargino LSP and vice versa--with the mass splitting between them being small. Hence, while the Wino NLSP RPC decays are suppressed, its RPV decays should be observable at the LHC in addition to those of the Wino LSP. Since the $B-L$ MSSM is completely compatible with all low energy phenomenological data, the results of this paper are explicit physical predictions that are of interest to the CERN SUSY ATLAS experimentalists. Run 1 and early run 2 data have already been used in the search for these processes--see \cite{Aaboud:2017opj,ATLAS:2017hbw,Jackson:2015lmj,ATLAS:2015jla}--and newer run 2 data is presently being analyzed.

\section*{Acknowledgments}
The authors would like to thank Evelyn Thomson, Elliot Lipeles, Jeff Dandoy, Christopher Mauger, Nuno Barros, Leigh Schaefer, and Christian Herwig for helpful suggestions. Ovrut and Purves would also like to acknowledge many informative conversations with Zachary Marshall and Sogee Spinner concerning RPV decays of a stop LSP. Burt Ovrut and Sebastian Dumitru are supported in part by DOE No. DE-SC0007901 and SAS Account 020-0188-2-010202-6603-0338.

\begin{appendices}

\section {Notation}\label{appendix:notation}

In this Appendix, we present, for clarity, all the notation used throughout the paper.
\subsection{Gauge Eigenstates}

\begin{itemize}
\item{Bosons}

\underline{{\it vector gauge bosons}}

~~~ $SU(2)_L-\quad\> W^1_\mu\>, W^2_\mu\>, W^3_\mu$, $\quad $   
coupling parameter $g_2$

~~~ $U(1)_{B-L}-\quad \> B^'_\mu \> $, $\quad $   
coupling parameter $g_{BL}$

~~~ $U(1)_{3R}-\quad \> {W_R}_\mu \> $, $\quad $   
coupling parameter $g_R$

~~~ $U(1)_{Y}-\quad \> {B}_\mu \> $, $\quad $   
coupling parameter $g'$

~~~ $U(1)_{EM}-\quad \> {\gamma}^0_\mu \> $, $\quad $   
coupling parameter $e$

~~~B-L Breaking:    $U(1)_{3R}\otimes U(1)_{B-L}\rightarrow U(1)_Y,\quad$    massive boson ${Z_R}_\mu$,$\>\>$ coupling $g_{Z_R}$

~~~EW Breaking:    $SU(2)_L\otimes U(1)_Y \rightarrow U(1)_{EM},\quad$    massive bosons $Z^0_\mu
,\>W^\pm_\mu\quad$

\underline{{\it Higgs scalars}}

~~~$H_u^0\>, H_u^+\>, H_d^0\>, H_d^-\quad $

\item{Weyl Spinors}

\underline{ {\it gauginos}}

~~~ $SU(2)_L-\> \tilde W^0\>, \tilde W^\pm$,\quad
$U(1)_{B-L}- \>\tilde B^' , \quad$ $U(1)_{3R}-\> {\tilde W_R} \> $,\quad
$U(1)_{Y}-\> \tilde {B},\quad$ $U(1)_{EM}- \> \tilde {\gamma}^0 \> $

\vspace{1mm}
\underline{{\it Higgsinos}}

~~~$\tilde H_u^0\>, \tilde H_u^+\>, \tilde H_d^0\>, \tilde H_d^-$

\underline{{\it leptons}}

 ~~~left chiral\quad  $e_i,\>\nu_i, \>\> i=1,2,3 \quad \text{where} \quad e_1=e,\>e_2=\mu,\>e_3=\tau$
          
 ~~~right chiral\quad  $e^c_i,\>\nu^c_i, \>\> i=1,2,3 \quad \text{where} \quad e^c_1=e^c,\>e_2^c=\mu^c,\>e_3^c=\tau^c$

\underline{{\it sleptons}}

 ~~~left chiral\quad  $\tilde e_i,\>\tilde \nu_i, \>\> i=1,2,3 \quad \text{where} \quad \tilde e_1=\tilde e,\>\tilde e_2=\tilde \mu,\>\tilde e_3=\tilde \tau$
          
 ~~~right chiral\quad  $\tilde e^c_i,\>\tilde \nu^c_i, \>\> i=1,2,3 \quad \text{where} \quad \tilde e^c_1=\tilde e^c,\>\tilde e_2^c=\tilde \mu^c,\>\tilde e_3^c=\tilde \tau^c$

\end{itemize}

\subsection{Mass terms}

\indent \quad \quad \underline{ {\it gauginos}}

~~~ $\tilde W^0\>, \tilde W^\pm \rightarrow M_2,\quad$
$\tilde B^'\rightarrow M_{BL} , \quad$ $ {\tilde W_R} \rightarrow M_R,\quad $ $\tilde {B} \rightarrow M_1,\quad$ 

\vspace{1mm}

\underline{{\it Higgsinos}}

~~~$\tilde H_u^0\>, \tilde H_u^+\>, \tilde H_d^0\>, \tilde H_d^- \rightarrow \mu$

\vspace{1mm}

\underline{{\it left chiral charged leptons}}

~~~$e_i \rightarrow m_{e_i}, \>$ for $i=1,2,3$

\vspace{1mm}

\underline{{\it right chiral charged leptons}}

~~~$e_i^c \rightarrow m_{e_i^c}, \>$ for $i=1,2,3$

\subsection{Mass Eigenstates}

\begin{itemize}

\item{Weyl Spinors}

\underline{{\it leptons}}\\
 \quad  $e_i,\>\nu_i, \>\> i=1,2,3 \quad \text{where} \quad e_1=e,\>e_2=\mu,\>e_3=\tau$

\underline{{\it charginos and neutralinos}}\\
  \quad $\tilde \chi^\pm_1, \quad \tilde \chi^\pm_2, \quad \tilde \chi_n^0,\quad n=1,2,3,4,5,6$

\item{4-component Spinors}

\underline{{\it  leptons}}\\
~~~$\ell_i^-=\left(\begin{matrix}e_i\\ {e_i^c}^\dag\end{matrix}\right),\quad
\ell_i^+=\left(\begin{matrix}{e_i^c}\\ e_i^\dag\end{matrix}\right), \quad
\nu_i=\left(\begin{matrix}\nu_i\\ {\nu_i}^\dag\end{matrix}\right)\quad i=1,2,3$

\underline{{\it charginos and neutralinos}}

$\tilde X^-_1=\left(\begin{matrix}\tilde \chi^-_1\\ \tilde {\chi}^{+\dag}_1\end{matrix}\right),\quad
\tilde X^+_1=\left(\begin{matrix}\tilde \chi^+_1\\ \tilde {\chi}^{-\dag}_1\end{matrix}\right), \quad
\tilde X^0_n=\left(\begin{matrix}\tilde \chi^0_n\\ \tilde {\chi}^{0\dag}_n\end{matrix}\right)$

\end{itemize}

\subsection{VEV's}

\begin{itemize}
\item {sneutrino VEV's \\
\quad $\left<\tilde \nu^c_{3}\right> \equiv \frac{1}{\sqrt 2} {v_R} \quad 
\epsilon_i=\frac{1}{2}Y_{\nu i3}v_R \quad \left<\tilde \nu_{i}\right> \equiv \frac{1}{\sqrt 2} {v_L}_i, \quad i=1,2,3$}

\item {Higgs VEV's \\
$\left< H_u^0\right> \equiv \frac{1}{\sqrt 2}v_u, \ \ \left< H_d^0\right> \equiv \frac{1}{\sqrt 2}v_d, \quad \tan \beta=v_u/v_d$}

\end{itemize}

\subsection{Relevant angles}

\begin{itemize}

\item{$\beta$ - Higgs VEVs ratio}
\begin{equation}
 \tan \beta=v_u/v_d
 \end{equation}

\item{$\theta_W$ - Weinberg angle}
\begin{equation}
\sin^2 \theta_W=0.22  \quad s_W=\sin \theta_W \quad c_W=\cos \theta_W
\end{equation}

\item{$\theta_R$ - $U_{B-L}$, $U_{3R}$ couplings ratio}

\begin{equation}
\cos \theta_R = \frac{g_R}{\sqrt{g_R^2+g_{BL}^2}} \ .
\end{equation}

\item{$\alpha$ - Higgs bosons rotation matrix}

\begin{equation}
\left(\begin{matrix}H_u^0\\H_d^0\end{matrix}\right)=
\left(\begin{matrix}v_u\\v_d\end{matrix}\right)+
\frac{1}{\sqrt{2}}R_{\alpha}\left(\begin{matrix}h^0\\H^0\end{matrix}\right)+
\frac{i}{\sqrt{2}}R_{\beta_0}\left(\begin{matrix}G^0\\\Gamma^0\end{matrix}\right) \ ,
\end{equation}

\begin{equation}
R_{\alpha}=\left( \begin{matrix}
\cos{\alpha}&\sin{\alpha}\\ -\sin{\alpha}&\cos{\alpha}
\end{matrix}\right),
\end{equation}

\item{$\phi_\pm$ - Chargino rotation matrix}

\begin{equation}
\tan 2\phi_-=2\sqrt{2}M_{W^\pm}\frac{\mu \cos \beta +M_2 \sin \beta}{\mu^2-M_2^2-2M_{W^\pm}^2
\cos 2\beta}
\label{bernard1}
\end{equation}
\begin{equation}
\tan 2\phi_+=2\sqrt{2}M_{W^\pm}\frac{\mu \sin \beta +M_2 \cos \beta}{\mu^2-M_2^2+2M_{W^\pm}^2
\cos 2\beta}
\label{bernard2}
\end{equation}

\item{Neutrino rotation matrix $V_{\text{PMNS}}$}

 The $3 \times 3$ Pontecorvo-Maki-Nakagawa-Sakata matrix is
\begin{eqnarray}
	V_\pmns &=& 
	\begin{pmatrix}
		c_{12} c_{13}
		&
		s_{12} c_{13}
		&
		s_{13} e^{-i \delta}
		\\
		-s_{12} c_{23} - c_{12} s_{23} s_{13} e^{i \delta}
		&
		c_{12} c_{23} - s_{12} s_{23} s_{13} e^{i \delta}
		&
		c_{13} s_{23}
		\\
		s_{12} s_{23} - c_{12}  c_{23} s_{13} e^{i \delta}
		&
		-c_{12} s_{23} - s_{12}  c_{23} s_{13} e^{i \delta}
		&
		c_{13} c_{23}
	\end{pmatrix}\nonumber\\ &&\times \text{diag}(1, e^{i \mathcal{A}/2}, 1) \ , \label{eq:32}
\end{eqnarray}
Values for the matrix terms can be found in \cite{Capozzi:2018ubv}.
\end{itemize}

\section{Chargino states}\label{appendix:Charginos}

The chargino mass eigenstates in Weyl notation are denoted by $\tilde \chi_{1}^{\pm}$ and  $\tilde \chi_{2}^{\pm}$. These states are labelled so that $\tilde \chi_{1}^{\pm}$ is always the lighter of the two. It follows that $\tilde \chi_{2}^{\pm}$ can never be either the LSP or the NLSP and, hence, we ignore it in this paper.
The lightest chargino mass eigenstate $\tilde \chi_{1}^{\pm}$ is a linear combination of the charged Wino, ${\tilde{W}}^{\pm}$, the charged Higgsino, ${\tilde{H}}^{\pm}$ and three charged lepton components, either left-chiral $e_i$ or right-chiral $e^c_i$. It was shown in \cite{new} that the number of viable initial conditions consistent with all low-energy phenomenology satisfying the inequality $|M_{2}|<\mu$ is {\it vastly} larger than those with $|\mu|<|M_{2}|$. The reason for this is the following. First, the value of $\mu$ is naturally restricted to be of the order of a few TeV in order to solve the ``little hierarchy problem''. This has been discussed in detail in \cite{new}. Furthermore, we study the cases in which the chargino eigenstates are the LSPs and light enough to be detected at LHC in the near future. Hence, the mass term $M_2$, which gives the dominant contribution in the expression for the chargino mass, is naturally constrained to be on the order of a few hundred GeVs.
Therefore, in this paper, we will restrict the discussion to the first mass inequality. When $|M_{2}|<\mu$,  the mass eigenstate $\tilde \chi_{1}^{\pm}$ is given by
\begin{equation}
\tilde \chi_1^+=\cos \phi_+ \tilde W^+ +\sin \phi_+ H_u^+ +\mathcal{V}_{1\> 2+i} e_i^c
\end{equation}
and
\begin{equation}
\tilde \chi_1^-=\cos \phi_-\tilde W^- +\sin \phi_- H_d^- +\mathcal{U}_{1\> 2+i} e_i,
\end{equation}
where $\phi_\pm$ are chargino mixing angles presented in \eqref{bernard1}, \eqref{bernard2} above and $\mathcal{V}_{1\>2+i}$, $\mathcal{U}_{1\>2+i}$ are matrix elements of the extended chargino mixing matrices \cite{new}. These elements couple the SM charged leptons to the charged Winos and Higgsinos. 
They are given by
\begin{equation}
\mathcal{V}_{1\>2+i}=-\cos \phi_+ \frac{g_2 \tan \beta m_{e_i}}{\sqrt{2}M_2\mu}v_{L_i}+\sin \phi_+\frac{m_{e_i}}{\mu v_d}v_{L_i} \ .
\label{finish3}
\end{equation}
and
\begin{equation}
\mathcal{U}_{1\>2+i}=-\cos \phi_- \frac{g_2 v_d}{\sqrt{2}M_2\mu}\epsilon_i^*+\sin \phi_-\frac{\epsilon_i^*}{\mu} \ .
\label{finish3}
\end{equation}
$\mathcal{V}_{1\>2+i}$ and $\mathcal{U}_{1\>2+i}$ are proportional to the RPV couplings $\epsilon_i$ and $v_{L_i}$ and are therefore are very small; $\mathcal{V}_{1\>2+i}, \>\mathcal{U}_{1\>2+i} \ll 1$. Hence, one can ignore the charged lepton terms in equations (B.1) and (B.2).
It follows that
\begin{equation}
\tilde \chi_1^+ \simeq \cos \phi_+ \tilde W^+ +\sin \phi_+ H_u^+
\end{equation}
and
\begin{equation}
\tilde \chi_1^-\simeq \cos \phi_-\tilde W^- +\sin \phi_- H_d^- \ .
\end{equation}
When $|\cos \phi_\pm |> |\sin \phi_\pm|$, the Wino components of the chargino states $\tilde \chi_1^\pm$ are the most dominant. Therefore, we call these states Wino charginos and denote them by $\tilde \chi_W^\pm$. In 4-component spinor notation they are

\begin{equation}
\tilde X^\pm_W=\left(\begin{matrix}\tilde \chi^\pm_W\\ \tilde {\chi}^{\mp\dag}_W\end{matrix}\right) \ .
\end{equation}
Indeed, we demonstrate in Section 2 of this paper that the requirement that the initial conditions be consistent with all low energy phenomenology constrains the vast majority of valid black points to satisfy $|\cos \phi_\pm |> |\sin \phi_\pm|$. Wino chargino RPV decays comprise the first subject that we study in this paper. 

\section{Neutralino states}\label{appendix:Neutralinos}

 After diagonalizing the neutralino mass matrix, one obtains six neutralino mass eigenstates, $\tilde \chi_n^0$ with $n=1,2,3,4,5,6$. Unlike for charginos, the label $n$ does not automatically imply any mass ordering (i.e. the $\tilde \chi_1^0$ neutralino is not necessarily the lightest). Each of the six neutralinos $\tilde \chi_n^0$ is a superposition of a Rino $\tilde W_R$, a Wino $\tilde W_2$, two neutral Higgsinos $\tilde H_d^0$, $\tilde H_u^0$, a Blino $\tilde B^'$, a third generation right handed neutrino $\nu_3^c$ and three generations of left handed neutrinos $\nu_i$, for $i=1,2,3$.  

In the theoretical context we work in,  the off-diagonal terms are much smaller than the diagonal ones, which allows to determine which component dominates in each of the neutralino states $\tilde \chi_n^0$. That is, $\tilde \chi_1^0$ has a dominant Bino $\tilde B$ component, $\tilde \chi_2^0$ has a dominant Wino $\tilde W$ component, $\tilde \chi_{3,4}^0$ have dominant Higgsino $\tilde H_u^0, \> \tilde H_d^0$ component, $\tilde \chi_{5,6}^0$ have a dominant right-handed neutrino $\nu_3^c$ component. Therefore we use the notation
\begin{equation}
{\tilde \chi}_1^0={\tilde \chi}_B^0,\quad  {\tilde \chi}_2^0={\tilde \chi}_W^0, \quad {\tilde \chi}_3^0={\tilde \chi}_{H_d}^0,
\quad {\tilde \chi}_4^0={\tilde \chi}_{H_u}^0, \quad {\tilde \chi}_5^0={\tilde \chi}_{\nu_{3a}}^0, \quad {\tilde \chi}_6^0={\tilde \chi}_{\nu_{3b}}^0,
\end{equation}
to express which component dominates in each neutralino state. In our studies, the exact content of each neutralino state  is calculated numerically, after diagonalizing the neutralino mass mixing matrix. 
The second objects of study in this paper are the Wino neutralinos, $\tilde \chi_W^0$, which, as the notation implies, have a dominant Wino component. In 4-component spinor notation they are
\begin{equation}
\tilde X^0_W=\left(\begin{matrix}\tilde \chi^0_W\\ \tilde {\chi}^{0\dag}_W\end{matrix}\right).
\end{equation}

\section{Chargino decay rates}\label{appendix:A}

In \cite{new}, we computed the RPV decay rates of a general light chargino state $\tilde X_1^{\pm}$, without looking into its Wino and Higgsino content. We reproduce the results here, for reference. 

\begin{enumerate}
\item{\boldmath{ ${\tilde X}^\pm_1\rightarrow W^\pm \nu$}}
\begin{equation}\label{eq:Chargino_Decay1}
\Gamma_{{\tilde X}_1^\pm\rightarrow W^\pm \nu_{i}}=\frac{\left(|{G_L}|^2_{{\tilde X}^\pm_1\rightarrow W^\pm \nu_i}+|{G_R}|^2_{{\tilde X}^\pm_1\rightarrow W^\pm \nu_i} 
\right)}{64\pi}
\frac{M_{{\tilde \chi}_1^\pm}^3}{M_{W^\pm}^2}\left(1-\frac{M_{W^\pm}^2}{M_{{\tilde \chi}_1^\pm}^2}\right)^2\left(1+2\frac{M_{W^\pm}^2}{M_{{\tilde \chi}_1^\pm}^2}\right),
\end{equation}
where
\begin{multline}\label{eq:72}
 {G_L}_{{\tilde X}^+_1\rightarrow W^+ \nu_i}= -{G^*_R}_{{\tilde X}^-_1\rightarrow W^- \nu_i} =\frac{g_2}{\sqrt 2}\Bigg[
\left( -\cos \phi_- \frac{g_2 v_d}{\sqrt{2}M_2\mu}\epsilon_j^*+\sin \phi_-\frac{\epsilon_j^*}{\mu}\right)\\
+\sin \phi_- \frac{1}{16d_{\tilde \chi^0}}\left( M_{\tilde \gamma}v_R^2v_u(v_d\epsilon_j-\mu v_{L_j}^*)-4M_2\mu(M_{\tilde Y}v_R^2+g_R^2M_{BL}v_u^2)\epsilon_j \right)\\
-\sqrt{2}\cos \phi_- \frac{g_2\mu}{8d_{\tilde \chi_0}}\left( 2g_R^2M_{BL}v_dv_u^2\epsilon_j+M_{\tilde Y}v_R^2(v_d\epsilon_j+\mu v_{L_j}^*) \right)\Bigg]\left[V_{\text{PMNS}}\right]_{ji}
\end{multline}
and
\begin{multline}\label{eq:73}
 {G_R}_{{\tilde X}^+_1\rightarrow W^+ \nu_i}= -{G^*_L}_{{\tilde X}^-_1\rightarrow W^- \nu_i} = \\
\frac{g_2}{\sqrt 2}\Bigg[-\sin \phi_+ \frac{1}{16d_{{\tilde \chi}^0}}
[ M_{\tilde \gamma} v_R^2v_u (v_d\epsilon_j^*+\mu v_{L_j})
-4g_R^2\mu M_2M_{BL}v_d v_u\epsilon_j^*]  
 \\+\sqrt{2} \cos \phi_+ 
\frac{g_2\mu}{8d_{{\tilde \chi}^0}}
[2g_R^2M_{BL}v_dv_u^2\epsilon_j+M_{\tilde Y}v_R^2(v_d\epsilon_j^*+\mu v_{L_j})]
\Bigg] \left[V_{\text{PMNS}}^\dag \right]_{ij} \ .
 \end{multline}
 
\item{\boldmath{ ${\tilde X}^\pm_1\rightarrow Z^0 \ell_i^\pm$}}
\begin{equation}\label{eq:Chargino_Decay2}
\Gamma_{{\tilde X}_1^\pm\rightarrow Z^0 \ell_i^\pm}=\frac{\Big( |{G_L}|_{{\tilde X}^\pm_1\rightarrow Z^0 \ell^\pm_i}^2+|{G_R}|_{{\tilde X}^\pm_1\rightarrow Z^0 \ell^\pm_i}^2\Big)}{64\pi}
\frac{M_{{\tilde \chi}_1^\pm}^3}{M_{Z^0}^2}\left(1-\frac{M_{Z^0}^2}{M_{{\tilde \chi}_1}^2}\right)^2
\left(1+2\frac{M_{Z^0}^2}{M_{{\tilde \chi}_1^\pm}^2}\right).
\end{equation}
where
\begin{multline}
{G_L}_{{\tilde X}^+_1\rightarrow Z^0 \ell^+_i}=-{G_R}^*_{{\tilde X}^-_1\rightarrow Z^0 \ell^-_i}=g_2c_W\Big(\frac{g_2}{\sqrt{2}M_2\mu}(v_d\epsilon_i+\mu v_{L_i}^*)\Big)
\cos \phi_- +\\ +\frac{g_2}{c_W}
\left(\frac{1}{2}-s_W^2\right)
\Big(-\cos \phi_- \frac{g_2 v_d}{\sqrt{2}M_2\mu}\epsilon_i+\sin \phi_-\frac{\epsilon_i}{\mu}\Big)-\frac{g_2}{c_W}\left(\frac{1}{2}-s_W^2\right)
\Big(\frac{\epsilon_i}{\mu} \Big)\sin \phi_-
\end{multline}
and
\begin{multline}
{G_R}_{{\tilde X}^+_1\rightarrow Z^0 \ell^+_i}=-{G_L}^*_{{\tilde X}^-_1\rightarrow Z^0 \ell^-_i}=g_2c_W\cos \phi_+ \Big(-\frac{1}{\sqrt{2}M_2\mu}g_2\tan \beta m_{e_i}v_{L_i} \Big)-\\+
\frac{g_2}{c_W}s_W^2\Big( -\cos \phi_+ \frac{g_2 \tan \beta m_{e_i}}{\sqrt{2}M_2\mu}v_{L_i}+\sin \phi_+\frac{m_{e_i}}{\mu v_d}v_{L_i}\Big)
-\frac{g_2}{c_W}\left(\frac{1}{2}-s_W^2\right)\sin \phi_+\Big(\frac{m_{e_i}}{ v_d \mu}v_{L_i} \Big) \ .
\end{multline}
\noindent There is no sum over the $i$ in $v_{L_i}m_{e_i}$.

\item{\boldmath{ ${\tilde X}^\pm_1\rightarrow h^0 \ell_i^\pm$}}

\begin{equation}\label{eq:Chargino_Decay4}
\Gamma_{{\tilde X}_1^\pm\rightarrow h^0 \ell_i^\pm}=\frac{\Big(|{G_L}|_{{\tilde X}^\pm_1\rightarrow h^0 \ell_i^\pm}^2+|{G_R}|_{{\tilde X}^\pm_1\rightarrow h^0 \ell_i^\pm}^2\Big)}{64\pi}
M_{{\tilde X}_1^\pm}\left(1-\frac{M_{h^0}^2}{M_{{\tilde X}_1^\pm}^2}\right)^2.
\end{equation}
where
\begin{multline}
{G_L}_{{\tilde X}^+_1\rightarrow h^0 \ell_i^-}=-{G_R}^*_{{\tilde X}^+_1\rightarrow h^0 \ell_i^+} =-\frac{1}{\sqrt 2}Y_{e_i}\sin \alpha 
\Big(-\cos \phi_+ \frac{g_2 \tan \beta m_{e_i}}{\sqrt{2}M_2\mu}v_{L_i}^*+\sin \phi_+\frac{m_{e_i}}{\mu v_d}v_{L_i}^* \Big)+\\-
\frac{1}{ 2}g_2 \sin \alpha \cos \phi_+ \Big( \frac{\epsilon_i}{\mu} \Big)-
\frac{1}{2}g_2\cos \alpha \sin \phi_+\Big( \frac{g_2}{\sqrt{2}M_2\mu}(v_d\epsilon_i+\mu v_{L_i}^*) \Big)
\end{multline}
and
\begin{multline}
{G_R}_{{\tilde X}^-_1\rightarrow h^0 \ell_i^-}=-{G_L}^*_{{\tilde X}^+_1\rightarrow h^0 \ell_i^+}=-\frac{1}{\sqrt 2}Y_{e_i}\sin \alpha 
\Big(-\cos \phi_- \frac{g_2v_d}{\sqrt{2}M_2\mu}\epsilon_i^*+\sin \phi_-\frac{\epsilon_i^*}{\mu}\Big)\\
+\frac{1}{2}g_2 \sin \alpha \sin \phi_-\Big(-\cos \phi_+\frac{1}{\sqrt{2}M_2\mu}g_2\tan \beta m_{e_i}v_{L_i} -\sin \phi_+ \frac{m_{e_i}}{\mu v_d}v_{L_i}\Big)\\
-\frac{1}{2}g_2\cos \alpha \cos \phi_-\Big( \frac{m_{e_i}}{ v_d \mu}v_{L_i} \Big) \ .
\end{multline}
\noindent There is no sum over $i$ in either of these expressions. 
\end{enumerate}

\section{Neutralino decay rates}\label{appendix:B}

In \cite{new}, we computed the RPV decay rates of a general neutralino state $\tilde X_n^0$. the index $n$ indicates the neutralino species as follows:
\begin{equation}
{\tilde X}_1^0={\tilde X}_B^0,\quad  {\tilde X}_2^0={\tilde X}_W^0, \quad {\tilde X}_3^0={\tilde X}_{H_d}^0,
\quad {\tilde X}_4^0={\tilde X}_{H_u}^0, \quad {\tilde X}_5^0={\tilde X}_{\nu_{3a}}^0, \quad {\tilde X}_6^0={\tilde X}_{\nu_{3b}}^0.
\end{equation}
We reproduce the results here, for reference. 

\begin{enumerate}

\item{\boldmath{$\tilde X^0_n\rightarrow Z^0 \nu$}}

\begin{equation}\label{Neutralino_Decay_Rate1}
\Gamma_{{\tilde X}^0_n\rightarrow Z^0\nu_{i}}=
\frac{\Big(|{G_L}|_{{\tilde X}^0_n\rightarrow Z^0\nu_{i}}^2
+|{G_R}|_{{\tilde X}^0_n\rightarrow Z^0\nu_{i}}^2 \Big)
}{64\pi}
\frac{M_{{\tilde \chi}_n^0}^3}{M_{Z^0}^2}\left(1-\frac{M_{Z^0}^2}{M_{{\tilde \chi}_n^0}^2}\right)^2
\left(1+2\frac{M_{Z^0}^2}{M_{{\tilde \chi}^0_n}^2}\right),
\end{equation}
where
\begin{multline}
 {G_L}_{{\tilde X}^0_n\rightarrow Z^0 \nu_{i}}=
g_2\Big(\frac{1}{2c_W}\mathcal{N}_{n\>6+j}\mathcal{N}^*_{6+j\>6+i}-\frac{1}{c_W}\left(\frac{1}{2}+s_W^2\right)\mathcal{N}_{n\>4}\mathcal{N}^*_{6+i\>4} \Big)\\
+g_2\Big( \frac{1}{c_W}\left(\frac{1}{2}+s_W^2\right) \mathcal{N}^*_{n\>3}\mathcal{N}_{6+i\>3}\Big)
\end{multline}
and
\begin{multline}
{G_R}_{{\tilde X}^0_n\rightarrow Z^0 \nu_{i}}=g_2\Big( -\frac{1}{c_W}\left(\frac{1}{2}+s_W^2\right) \mathcal{N}_{n\>3}\mathcal{N}^*_{6+i\>3}\Big)\\
-{g_2}\Big[
\Big(-\frac{1}{2c_W}\mathcal{N}^*_{n\>6+j}\mathcal{N}_{6+j\>6+i}-\frac{1}{c_W}\left(\frac{1}{2}+s_W^2\right)\mathcal{N}^*_{n\>4}\mathcal{N}_{6+i\>4} \Big)
\end{multline}

\item{\boldmath{$\tilde X^0_n\rightarrow W^\mp \ell^\pm$}}

\begin{equation}\label{Neutralino_Decay_Rate2}
\Gamma_{{\tilde X}^0_n\rightarrow W^\mp \ell_i^\pm}=\frac{\Big(|{G_L}|_{{\tilde X}^0_n\rightarrow W^\pm \ell_i^\mp}^2+|{G_R}|_{{\tilde X}^0_n\rightarrow W^\pm \ell_i^\mp}^2\Big)}{64\pi}
\frac{M_{{\tilde \chi}_1^\pm}^3}{M_{W^\pm}^2}\left(1-\frac{M_{W^\pm}^2}{M_{{\tilde \chi}_n^0}^2}\right)^2
\left(1+2\frac{M_{W^\pm}^2}{M_{{\tilde \chi}_n^0}^2}\right),
\end{equation}
where
\begin{equation}
 {G_L}_{{\tilde X}^0_n\rightarrow W^- \ell_i^+}=-{G_R}_{{\tilde X}^0_n\rightarrow W^+ \ell_i^-}=\frac{g_2}{\sqrt{2}}\Big[\mathcal{N}_{n\>4}\mathcal{V}^*_{2+i\>2}+\sqrt{2}\mathcal{V}^*_{2+i\>1}\mathcal{N}_{n\>2}\Big]
\end{equation}
and
\begin{equation}
{G_R}_{{\tilde X}^0_n\rightarrow W^- \ell_i^+}=-{G_L}_{{\tilde X}^0_n\rightarrow W^+ \ell_i^-}=\frac{g_2}{\sqrt{2}}\Big[-\mathcal{U}_{2+i\>2+j}\mathcal{N}^*_{n\>6+j}-\mathcal{U}_{2+i\>2}\mathcal{N}^*_{n\>3}+\sqrt{2}\mathcal{N}^*_{n\>2}\mathcal{U}_{2+i\>1}\Big]
\end{equation}

\item{\boldmath{$\tilde X^0_n\rightarrow h^0 \nu$}}
\begin{equation}\label{Neutralino_Decay_Rate3}
\Gamma_{{\tilde X}^0_n\rightarrow h^0\nu_{i}}=\frac{\Big(|{G_L}|_{{\tilde X}^0_n\rightarrow h^0\nu_{i}}^2+|{G_R}|_{{\tilde X}^0_n\rightarrow h^0\nu_{i}}^2\Big)}{64\pi}
M_{{\tilde \chi}_n^0}\left(1-\frac{M_{h^0}^2}{M_{{\tilde \chi}_n^0}^2}\right)^2
\end{equation}
where
\begin{multline}
{G_L}_{{\tilde X}^0_n\rightarrow h^0 \nu_{i} }=\frac{g_2}{{2}}\Big(
\cos \alpha (\mathcal{N}^*_{n\>4}\mathcal{N}^*_{6+i\>2}+\mathcal{N}^*_{6+i\>4}\mathcal{N}_{n\>2}^*)+\sin \alpha (\mathcal{N}^*_{n\>3}\mathcal{N}^*_{6+i\>2}+\mathcal{N}^*_{6+i\>3}\mathcal{N}_{n\>2}^*)\Big)\\
-\frac{g'}{{2}}\Big(
\cos\alpha \left(\sin \theta_R(\mathcal{N}^*_{n\>4}\mathcal{N}^*_{6+i\>1}+\mathcal{N}^*_{6+i\>4}\mathcal{N}^*_{n\>1})+\cos \theta_R(\mathcal{N}^*_{n\>4}\mathcal{N}^*_{6+i\>5}+\mathcal{N}^*_{6+i\>4}\mathcal{N}^*_{n\>5})   \right)\\
+\sin \alpha \left( \sin \theta_R(\mathcal{N}^*_{n\>3}\mathcal{N}^*_{6+i\>1}+\mathcal{N}^*_{6+i\>3}\mathcal{N}^*_{n\>1})+\cos \theta_R(\mathcal{N}^*_{n\>3}\mathcal{N}^*_{6+i\>5}+\mathcal{N}^*_{6+i\>3}\mathcal{N}^*_{n\>5}) \right)\Big)\\
+\frac{1}{\sqrt 2}Y_{\nu i3}\cos\alpha \Big(\mathcal{N}^*_{n\>6+j}\mathcal{N}^*_{6+i\>6}
+\mathcal{N}^*_{6+i\>6+j}
\mathcal{N}^*_{n\>6}\Big)
\end{multline}
and
\begin{multline}
{G_R}_{{\tilde X}^0_n\rightarrow h^0 \nu_{i} }=\frac{g_2}{{2}}\Big(
\cos \alpha (\mathcal{N}_{n\>4}\mathcal{N}_{6+i\>2}+\mathcal{N}_{6+i\>4}\mathcal{N}_{n\>2})+\sin \alpha (\mathcal{N}_{n\>3}\mathcal{N}_{6+i\>2}+\mathcal{N}_{6+i\>3}\mathcal{N}_{n\>2})\Big)\\
+\frac{g'}{{2}}\Big(\cos\alpha \left(\sin \theta_R(\mathcal{N}_{n\>4}\mathcal{N}_{1\>6+i}+\mathcal{N}_{6+i\>4}\mathcal{N}_{n\>1})+\cos \theta_R(\mathcal{N}_{n\>4}\mathcal{N}_{6+i\>5}+\mathcal{N}_{6+i\>4}\mathcal{N}_{n\>5})   \right)\\
+\sin \alpha \left( \sin \theta_R(\mathcal{N}_{n\>3}\mathcal{N}_{6+i\>1}+\mathcal{N}_{6+i\>3}\mathcal{N}_{n\>1})+\cos \theta_R(\mathcal{N}_{n\>3}\mathcal{N}_{6+i\>5}+\mathcal{N}_{6+i\>3}\mathcal{N}_{n\>5}) \right)\Big)\\
+\Big(\mathcal{N}_{n\>6+j}\mathcal{N}_{6+i\>6}
+\frac{1}{\sqrt 2}Y_{\nu i3}\cos\alpha \Big(\mathcal{N}_{6+i\>6+j}
\mathcal{N}_{n\>6}\Big)
\end{multline}

\end{enumerate}
The matrices $\mathcal{U}$, $\mathcal{V}$ and $\mathcal{N}$ matrices rotate the gaugino eigenstates into the neutralino and chargino mass eigenstates. They are presented in Appendices B.1 and B.2 of \cite {new}. 

Note that in all cases in Appendix A and B above, we sum over $j=1,2,3$.

\end{appendices}

\end{document}